\documentclass{vldb}
\usepackage{verbatim}
\usepackage{graphicx}
\usepackage{multirow}
\usepackage{algorithm}
\PassOptionsToPackage{noend}{algpseudocode}
\usepackage{algpseudocode}
\usepackage[labelfont=bf,textfont=bf]{caption}
\usepackage[labelfont=bf,textfont=bf,singlelinecheck=off,justification=raggedright]{subcaption}

\usepackage[normalem]{ulem}
\usepackage{longtable}
\usepackage{array}
\usepackage{url}
\usepackage{balance}
\usepackage{color}
\usepackage[table]{xcolor}
\usepackage{hhline}
\usepackage{paralist}
\usepackage{amsmath}

\usepackage{amsthm}
\usepackage{bm}
\usepackage[normalem]{ulem} 
\newtheorem{defn}{Definition}
\newtheorem{thm}{Theorem}[section]
\newtheorem{lem}[thm]{Lemma}
\graphicspath{{.}}
\definecolor{myred}{RGB}{189, 52, 67}
\definecolor{mygreen}{RGB}{19, 136, 8}
\definecolor{myblue}{RGB}{16, 52, 166}

\newif\ifJournal
\newif\ifFuture
\newif\ifGrey
\newif\ifColor

\Greytrue

\vldbTitle{Return of the Lernaean Hydra: Experimental Evaluation of Data Series Approximate Similarity Search}
\vldbAuthors{Karima Echihabi, Kostas Zoumpatianos, Themis Palpanas, and Houda Benbrahim}
\vldbDOI{https://doi.org/10.14778/3368289.3368303}
\vldbVolume{13}
\vldbNumber{3}
\vldbYear{2019}

\begin{document}
	
	\title{Return of the Lernaean Hydra: Experimental Evaluation of\\ Data Series Approximate Similarity Search}		

	\numberofauthors{4}
	
	\author{
		%
		%
		\alignauthor
		\affaddr{
			Karima Echihabi\\
			IRDA, Rabat IT Center, \\
			ENSIAS, Mohammed V Univ.}
		\affaddr{{karima.echihabi@gmail.com}}\\
		\alignauthor 
		\affaddr{
			Kostas Zoumpatianos\\
			Harvard University} 
		\affaddr{{kostas@seas.harvard.edu}}
		\alignauthor
		\affaddr{
			Themis Palpanas\\
			Universit{\'e} de Paris}\\ 
		\affaddr{{themis@mi.parisdescartes.fr}}\\
		\and
		\alignauthor 
		\affaddr{
			Houda Benbrahim\\
			IRDA, Rabat IT Center, \\
			ENSIAS, Mohammed V Univ.} 
		\affaddr{{houda.benbrahim@um5.ac.ma}}
	}
	
 \maketitle
	
\begin{abstract}
Data series are a special type of multidimensional data present in numerous domains, where similarity search is a key operation that
has been extensively studied in the data series literature. 
In parallel, the multidimensional community 
has studied approximate similarity search techniques. 
We propose a 
taxonomy of similarity search techniques that reconciles the terminology used in these two domains, we describe modifications to data series indexing techniques enabling them to answer approximate similarity queries with quality guarantees, and we conduct a thorough experimental evaluation to compare approximate similarity search techniques under a unified framework, on synthetic and real datasets in memory and on disk. 
Although data series differ from generic multidimensional vectors (series usually exhibit correlation between neighboring values), our results show that data series techniques answer approximate 
queries with strong guarantees and an excellent empirical performance, on data series and vectors alike. 
These techniques outperform the state-of-the-art approximate techniques for vectors when operating on disk, and remain competitive in memory. 
\end{abstract}

\section{Introduction}
\label{sec:introduction}
\noindent{\bf Motivation.} 
A data series is a sequence of ordered real values\footnote{The order attribute can be angle, mass, time, etc.~\cite{conf/sofsem/Palpanas2016}. When the order is time, the series is called a \emph{time series}. We use \emph{data series}, \emph{time series} and \emph{sequence} interchangeably.}. 
Data series are ubiquitous, appearing in nearly every domain including science and engineering, medicine, business, finance and economics~\cite{KashinoSM99,Shasha99,humanbehaviorpatterns,volker,DBLP:conf/edbt/MirylenkaCPPM16,HuijseEPPZ14,percomJournal,windturbines,spikesorting,VALMOD,journal/jte/Williams2003,conf/compstats/Hebrail2000}. 
The increasing presence of IoT technologies is making collections of data series grow to multiple terabytes~\cite{DBLP:journal/sigmod/Palpanas15}. 
These data series collections need to be analyzed in order to extract knowledge and produce value~\cite{Palpanas2019}. 
The process of retrieving similar data series (i.e., similarity search), forms the backbone of most analytical tasks, including outlier detection~\cite{journal/csur/Chandola2009,conf/icde/boniol20}, frequent pattern mining~\cite{conf/kdd/Mueen2012}, clustering~\cite{conf/kdd/Keogh1998,conf/sdm/Rodrigues2006,conf/icdm/Keogh2011,journal/pattrecog/Warren2005}, and classification~\cite{journal/jmlr/Chen2009}. 
Thus, to render data analysis algorithms and pipelines scalable, we need to make similarity search more efficient.

\noindent{\bf Similarity Search.}
A large number of data series similarity search methods has been studied, supporting exact search~\cite{conf/fodo/Agrawal1993,conf/kdd/shieh1998,conf/icde/Rafiei99,conf/kdd/Karras2011,conf/kdd/Mueen2012,code/Mueen2017}, approximate search~\cite{conf/icde/shatkay1996,conf/kdd/Keogh1997,conf/ssdm/Wang2000,conf/kdd/ColeSZ05,journal/vldb/Dallachiesa2014}, or both~\cite{conf/icdm/Camerra2010,journal/edbt/Schafer2012,conf/vldb/Wang2013,journal/kais/Camerra2014,journal/vldb/Zoumpatianos2016,dpisax,journal/pvldb/kondylakis18,ulisse,journal/vldb/linardi19,conf/bigdata/peng18,dpisaxjournal,coconutjournal,conf/icde/peng20}. 
In parallel, the research community has also developed exact~\cite{journal/cacm/bentley1975,conf/sigmod/Guttman1984,conf/icmd/Beckmann1990,conf/vldb/bertchold1996, conf/vldb/Ciaccia1997,conf/vldb/Weber1998,conf/cikm/Hakan2000} and approximate~\cite{conf/stoc/indyk1998} similarity search techniques geared towards generic multidimensional vector data\footnote{A comprehensive survey of techniques for multidimensional vectors can be found elsewhere~\cite{book/multiD/samet2005}.}. 
In the past few years though, we are witnessing a renewed interest in the development of approximate methods~\cite{journal/tpami/jegou2011,journal/iccv/xia2013,conf/vldb/sun14,journal/pami/babenko15,journal/corr/malkov16}. 

This study is the first experimental comparison of the efficiency and accuracy of data series approximate similarity search methods ever conducted. Specifically, we evaluate the accuracy of both data series specific approximate similarity search methods, as well as that of approximate similarity search algorithms that operate on general multidimensional vectors. 
Moreover, we propose modifications to data series techniques in order to support approximate query answering with theoretical guarantees, following~\cite{conf/icde/Ciaccia2000}. 

Our experimental evaluation covers in-memory and out-of-core experiments, the largest publicly available datasets, extensive comparison criteria, and a new set of methods that have never been compared before. 
We thus differ from other experimental studies, which focused on the efficiency of exact search~\cite{journal/pvldb/echihabi2018}, the accuracy of dimensionality reduction techniques and similarity measures for classification tasks~\cite{journal/dmkd/Keogh2003,conf/vldb/Ding2008,DBLP:journals/datamine/BagnallLBLK17}, or in-memory high-dimensional methods~\cite{journal/tkde/li19, conf/sisap/martin17, journal/pvld/naidan2015}. 
In this study, we focus on the problem of \emph{approximate whole matching similarity search in collections with a very large number of data series}, i.e., similarity search that produces approximate (not exact) results, by calculating distances on the whole (not a sub-) sequence.
This problem represents a common use case across many domains~\cite{journal/tpami/ge2014,conf/vldb/sun14,journal/pami/babenko15,journal/corr/malkov16,DBLP:conf/kdd/ColeSZ05,DBLP:conf/sigmod/ManninoA18,DBLP:conf/edbt/GogolouTPB19,Palpanas2019}. 



\noindent{\bf Contributions.}
Our key contributions are as follows:

1. We present a similarity search taxonomy that classifies methods based on the quality guarantees they provide for the search results, and that unifies the varied nomenclature used in the literature. 
Following this taxonomy, we include a brief survey of similarity search approaches supporting approximate search, bringing together works from the data series and multidimensional data research communities.

2. We propose a new set of approximate approaches with theoretical guarantees on accuracy and excellent empirical performance, based on modifications to the current data series exact methods.

3. We evaluate all methods under a unified framework to prevent implementation bias. We used the most efficient C/C++ implementations available for all approaches, and developed from scratch in C the ones that were only implemented in other programming languages. Our new implementations are considerably faster than the original ones.

4. We conduct the first comprehensive experimental evaluation of the efficiency and accuracy of data series approximate similarity search approaches, using synthetic and real series and vector datasets from different domains, including the two largest vector datasets publicly available.
The results unveil the strengths and weaknesses of each method, and lead to recommendations as to which approach to use. 

5. Our results show that the methods derived from the exact data series indexing approaches generally surpass the state-of-the-art techniques for approximate search in vector spaces. 
This observation had not been made in the past, and it paves the way for exciting new developments in the field of approximate similarity search for data series and multidimensional data at large. 
 

6. We share all source codes, datasets, and queries~\cite{url/DSSeval2}. 




\section{Definitions and Terminology}
\label{sec:problem}


Similarity search represents a common problem in various areas of computer science.
In the case of data series, several different flavors have been studied in the literature, often times using overloaded and conflicting terms.
We summarize here these variations, and provide definitions, thus setting a common language (for more details, see~\cite{journal/pvldb/echihabi2018}).

\noindent{\bf On Sequences.}
A \textit{\textbf{data series}} $S(p_1,p_2,...,p_n)$ is an ordered sequence of points, $p_i$, $1 \leq i \leq n$.
The number of points, $|S|=n$, is the length of the series.
We denote the $i$-th point in $S$ by $S[i]$; then $S[i:j]$ denotes the \textit{\textbf{subsequence}} $S(p_i,p_{i+1},...,p_{j-1},p_j)$, where $1 \leq i \leq j \leq n$.
We use $\mathbb{S}$ to represent all the series in a collection (dataset).
Each point in the series may represent the value of a single variable, i.e., \textit{\textbf{univariate series}}, or of multiple variables, i.e., \textit{\textbf{multivariate series}}.
If these values encode errors, or imprecisions, we talk about uncertain data  series~\cite{DBLP:conf/ssdbm/AssfalgKKR09,DBLP:conf/edbt/YehWYC09,DBLP:conf/kdd/SarangiM10,DBLP:journals/pvldb/DallachiesaNMP12,journal/vldb/Dallachiesa2014}.

Note that in the context of similarity search, a data series of length $n$ can be represented as a single point in an $n$-dimensional space. 
Then the values and length of $S$ are referred to as \emph{dimensions} and \emph{dimensionality}, respectively.

%



\noindent{\bf On Distance Measures.}
A data series \textit{\textbf{distance}} is a function that measures the (dis)similarity of two data series~\cite{berndt1994using,das1997finding,DBLP:conf/edbt/AssfalgKKKPR06,DBLP:conf/icde/ChenNOT07,journal/dmkd/Wang2013,DBLP:conf/ssdbm/MirylenkaDP17}.
The distance between a query series, $S_Q$, and a candidate series, $S_C$, is denoted by $d(S_Q,S_C)$.
The Euclidean distance is the most widely used, and one of the most effective for large series collections~\cite{conf/vldb/Ding2008}.
Some similarity search methods also rely on the \textit{lower-bounding distance} (distances in the reduced dimensionality space are guaranteed to be smaller than or equal to distances in the original space)~\cite{journal/kais/Camerra2014,journal/vldb/Zoumpatianos2016,journal/edbt/Schafer2012,conf/vldb/Wang2013,dpisax,ulisse,conf/vldb/Ciaccia1997,conf/kdd/Karras2011} and \textit{upper-bounding distance} (distances in the reduced space are larger than the distances in the original space)~\cite{conf/vldb/Wang2013,conf/kdd/Karras2011}. 

\noindent{\bf On Similarity Search Queries.}
We assume a data series collection, $\mathbb{S}$, a query series, $S_Q$, and a distance function $d(\cdot,\cdot)$.
%
A \textit{\textbf{k-Nearest-Neighbor (k-NN) query}} identifies the $k$ series in the collection with the smallest distances to the query series, while an \textit{\textbf{r-range query}} identifies all the series in the collection within range $r$ {\color{black}from} the query series.

\vspace*{-0.1cm}
\begin{defn}~\cite{journal/pvldb/echihabi2018} \label{def:knnquery}
Given an integer $k$, a \textit{\textbf{k-NN query}} retrieves the set of series $\mathbb{A} = \{ \{S_{C_1},...,S_{C_k}\} \subseteq \mathbb{S} | \forall \ S_C \in \mathbb{A} \ and \ \forall \ S_{C'} \notin \mathbb{A}, \ d(S_Q,S_C) \leq d(S_Q,S_{C'})\}$.
\end{defn}
\vspace*{-0.3cm}
\begin{defn}~\cite{journal/pvldb/echihabi2018} \label{def:rquery}
Given a distance $r$, an \textit{\textbf{r-range query}} retrieves the set of series $\mathbb{A} = \{S_C \in \mathbb{S} | d(S_Q,S_C) \leq r\}$.
\end{defn}
\vspace*{-0.1cm}

We additionally identify 
\textit{\textbf{whole matching (WM)}} queries (similarity between an entire query series and an entire candidate series), and 
\textit{\textbf{subsequence matching (SM)}} queries (similarity between an entire query series and all subsequences of a candidate series).

\vspace*{-0.1cm}
\begin{defn}~\cite{journal/pvldb/echihabi2018} \label{def:wholematch}
	A \textit{\textbf{WM query}} finds the candidate data series $S \in \mathbb{S}$ that matches $S_Q$, where $|S|=|S_Q|$. 
\end{defn}
\vspace*{-0.3cm}
\begin{defn}~\cite{journal/pvldb/echihabi2018} \label{def:submatch}
	A \textit{\textbf{SM query}} finds the subsequence $S[i:j]$ of a candidate data series $S \in \mathbb{S}$ that matches $S_Q$, where $|S[i:j]| = |S_Q| < |S|$.
\end{defn}
\vspace*{-0.1cm}



In practice, we have WM queries on large collections of short series~\cite{SENTINEL-2,url:sds}, SM queries on large collections of short series~\cite{url:adhd}, and SM queries on collections of long series~\cite{url/data/seismic}.
Note that a SM query can be converted to WM~\cite{ulisse,journal/vldb/linardi19}.

\noindent{\bf On Similarity Search Methods.}
The similarity search algorithms (k-NN or range) that always produce correct and complete answers are called \textit{\textbf{exact}}.
Algorithms that do not satisfy this property are called 
\textit{\textbf{approximate}}.
%
%
An {\textit{\bf$\bm{\epsilon}$-approximate}} algorithm guarantees that its distance results have a relative error no more than $\epsilon$, i.e., the approximate distance is at most $(1+\epsilon)$ times the exact one. 
A {\bf $\bm{\delta}$-$\bm{\epsilon}$-approximate} algorithm, guarantees that its distance results will have a relative error no more than $\epsilon$ (i.e., the approximate distance is at most $(1+\epsilon)$ times the exact distance), with a probability of at least $\delta$.
An \textit{\textbf{ng-approximate}} (no-guarantees approximate) algorithm does not provide any guarantees (deterministic, or probabilistic) on the error bounds of its distance results.

\vspace*{-0.2cm}
\begin{defn}~\cite{journal/pvldb/echihabi2018} \label{def:epsmatch}
Given a query $S_Q$, and $\epsilon \geq 0$, an \textit{\textbf{$\bm{\epsilon}$-approximate}} algorithm guarantees that all results, $S_C$, are at a distance $d(S_Q,S_C) \leq (1+\epsilon)\ d(S_Q,[\text{k-th NN of }S_Q])$ in the case of a $k$-NN query, and distance $d(S_Q,S_C) \leq (1+\epsilon)r$ in the case of an r-range query.
\end{defn}

\vspace*{-0.4cm}
\begin{defn}~\cite{journal/pvldb/echihabi2018} \label{def:probmatch}
Given a query $S_Q$, $\epsilon \geq 0$, and $\delta \in [0,1]$, a \textit{\textbf{$\bm{\delta}$-$\bm{\epsilon}$-approximate}} algorithm produces results, $S_C$, for which $Pr[d(S_Q,S_C)$ $\leq (1+\epsilon)\ d(S_Q,[\text{k-th NN of }S_Q])] \geq \delta$ in the case of a $k$-NN query, and $Pr[d(S_Q,S_C) \leq (1+\epsilon)r] \geq \delta$) in the case of an r-range query.
\end{defn}
\vspace*{-0.4cm}
\begin{defn}~\cite{journal/pvldb/echihabi2018} \label{def:appmatch}
Given a query $S_Q$, an \textit{\textbf{ng-approximate}} algorithm produces results, $S_C$, that are at a distance $d(S_Q,S_C) \leq (1+\theta)\ d(S_Q,[\text{k-th NN of }S_Q])$ in the case of a $k$-NN query, and distance $d(S_Q,S_C) \leq (1+\theta)r$ in the case of an r-range query, for an arbitrary value $\theta \in \mathbb{R}_{>0}$.
\end{defn}

In the data series literature, \textit{ng-approximate} algorithms have been referred to as \emph{approximate}, or \emph{heuristic} search~\cite{journal/kais/Camerra2014,journal/vldb/Zoumpatianos2016,journal/edbt/Schafer2012,conf/vldb/Wang2013,dpisax,ulisse}.
Unless otherwise specified, 
we will refer to \textit{ng-approximate} algorithms simply as approximate. Approximate matching in the data series literature
consists of pruning the search space, by traversing one path of an index structure representing the data, visiting at most one leaf, to get a baseline best-so-far (bsf) match.
In the multidimensional literature, ng-approximate similarity search is also called \textit{Approximate Nearest Neighbor (ANN)}~\cite{journal/tpami/jegou2011}, $\epsilon$-approximate 1-NN search is called \textit{c-ANN}~\cite{conf/vldb/sun14}, and $\epsilon$-approximate k-NN search is called \textit{c-k-ANN}~\cite{qalsh}, where $c$ stands for the approximation error and corresponds to $1+\epsilon$. 

Observe that when $\delta = 1$, a $\delta$-$\epsilon$-approximate method becomes $\epsilon$-approximate, and when $\epsilon=0$, an $\epsilon$-approximate method becomes exact~\cite{conf/icde/Ciaccia2000}.
It it also possible that the same approach implements both approximate and exact algorithms~\cite{conf/kdd/shieh1998,conf/vldb/Wang2013,journal/kais/Camerra2014,journal/vldb/Zoumpatianos2016,journal/edbt/Schafer2012}. 

\noindent{\bf Scope.}
In this study, we focus on \emph{univariate} series with \emph{no uncertainty}, where each point is drawn from the domain of real values, $\mathbb{R}$, and we evaluate \emph{approximate} methods for \emph{whole matching} in datasets containing a \emph{very large number of series}, using \emph{$k$-NN queries} and the \emph{Euclidean distance}. 
This scenario is key to numerous 
analysis pipelines in 
practice~\cite{journal/pattrecog/Warren2005,conf/kdd/Zoumpatianos2015,conf/sofsem/Palpanas2016,Palpanas2019}, in fields as varied as neuroscience~\cite{golay1998new}, seismology~\cite{kakizawa1998discrimination}, retail data~\cite{DBLP:conf/kdd/KumarPW02}, and energy~\cite{kovsmelj1990cross}. 

\vspace*{-0.3cm}

\section{Similarity Search Primer}
\label{sec:approaches}

Similarity search methods aim at answering a query efficiently by limiting the number of data points accessed, 
while minimizing the I/O cost of accessing raw data on disk and the CPU cost 
when comparing raw data to the query (e.g., Euclidean distance calculations). 
These goals are achieved by exploiting summarization techniques, and using efficient data structures (e.g., an index) and search algorithms. 
Note that solutions based on sequential scans are geared to exact similarity search~\cite{conf/kdd/Mueen2012,code/Mueen2017}, and cannot support efficient approximate search, since all candidates are always read.


Answering a similarity query using an index typically involves two steps: a filtering step where the pre-built index is used to prune candidates and a refinement step where the surviving candidates are compared to the query in the original high dimensional space~\cite{conf/sigmod/Guttman1984,conf/vldb/Weber1998,conf/cikm/Hakan2000,journal/kais/Camerra2014,journal/vldb/Zoumpatianos2016,journal/edbt/Schafer2012,conf/vldb/Wang2013,conf/icmd/Beckmann1990,dpisax,ulisse}. Some exact~\cite{conf/icmd/Beckmann1990,journal/edbt/Schafer2012,conf/vldb/Weber1998,conf/cikm/Hakan2000} and approximate methods~\cite{conf/vldb/sun14,journal/pami/babenko15} first summarize the original data and then index these summarizations, while others tie together data reduction and indexing~\cite{journal/kais/Camerra2014,journal/vldb/Zoumpatianos2016,conf/vldb/Wang2013}.  Some approximate methods return the candidates obtained in the filtering step~\cite{journal/pami/babenko15}. There also exist exact~\cite{conf/vldb/Ciaccia1997} and approximate~\cite{journal/corr/malkov16} methods that index high dimensional data directly.

A variety of data structures exist for similarity search indexes, including trees~\cite{conf/sigmod/Guttman1984,conf/icmd/Beckmann1990,journal/kais/Camerra2014,journal/vldb/Zoumpatianos2016,conf/vldb/Wang2013,dpisax,ulisse,conf/vldb/sun14,journal/edbt/Schafer2012}, inverted indexes~\cite{journal/tpami/jegou2011,conf/icassp/jegou2011,journal/iccv/xia2013,journal/pami/babenko15}, filter files~\cite{conf/vldb/Weber1998,conf/cikm/Hakan2000,journal/vldb/Zoumpatianos2016}, hash tables~\cite{conf/stoc/indyk1998,conf/poccs/broder1997,conf/sigcg/datar2004,conf/stoc/charikar02,journal/nips/liu2004,conf/soda/panigrahy2006,journal/siamdm/motwani2007,conf/vldb/lv2007,conf/sigmod/gan2012,journal/atct/odonnell2014,qalsh} and graphs~\cite{conf/siam/arya1993,journal/apr/chavez2010,conf/sigkdd/aoyama2011,conf/iccv/wang2013,journal/is/malkov2014,conf/sisap/chavez2015,conf/icmm/jiang2016,journal/corr/malkov16}. 
%
%
%
There also exist multi-step approaches, e.g., Stepwise~\cite{conf/kdd/Karras2011}, that transform and organize data according to a hierarchy of resolutions. 

Next, we outline the \emph{approximate} similarity search methods (refer also to Table~\ref{tab:multiprogram}) and their summarization techniques. 
(\emph{Exact} methods are detailed in~\cite{journal/pvldb/echihabi2018}).

\begin{table*}[tb]
		\caption{{\color{black}Similarity search methods used in this study} 
		("$\bullet$" indicates our modifications to original methods). {\color{black}All methods support in-memory data, but only methods ticked in last column support disk-resident data.}}
	{\small
		\centering
		\begin{tabular*}{\linewidth}{|*{11}{c|}} 
			\cline{3-11} 
			\multicolumn{1}{c}{}& & \multicolumn{4}{c|}{Matching Accuracy}  & 
			 \multicolumn{2}{|c|}{Representation} & \multicolumn{3}{c|}{Implementation}\\    		
			\cline{3-11} 
			\multicolumn{1}{c}{}& & exact & ng-appr. & $\epsilon$-appr. & $\delta$-$\epsilon$-appr. & Raw & Reduced & Original  & New & Disk-resident Data \\    		
			\cline{1-11}			 		 
			\cline{2-11}			 
			\multicolumn{1}{|c|}{\multirow{2}{*}{{Graphs}}}
			& \multicolumn{1}{c|}{HNSW} & 
			& \cite{journal/corr/malkov16} &   &  & \checkmark & & C++ &  &\\	
			\cline{2-11}			 	 
			& \multicolumn{1}{c|}{NSG} & 
			& \cite{nsg} &   &  & \checkmark & & C++ &  &\\	
			\cline{1-11}			 
			\multicolumn{1}{|c|}{\multirow{1}{*}{{Inv. Indexes}}}
			& \multicolumn{1}{c|}{IMI} &  & \cite{journal/pami/babenko15,journal/tpami/ge2014}  &  &  &  & OPQ & C++  & &\checkmark\\	
			\cline{1-11}			 	 
			\multicolumn{1}{|c|}{\multirow{2}{*}{{LSH}}}
			& \multicolumn{1}{c|}{QALSH} & &  & & \cite{qalsh} &   & Signatures & C++ & &\\	
			\cline{2-11}			 		 			
			& \multicolumn{1}{c|}{SRS} & & &  & \cite{conf/vldb/sun14} &  & Signatures & C++ & &\\				
			\cline{1-11}			 
			\multicolumn{1}{|c|}{\multirow{1}{*}{{Scans}}}
			& \multicolumn{1}{c|}{VA+file} & {\cite{conf/cikm/Hakan2000}} & $\bullet$&$\bullet$&$\bullet$ & & {DFT} & {MATLAB} & {C} & \checkmark\\	
			\cline{2-11}			 		 			
			\cline{1-11}			 
			\multicolumn{1}{|c|}{\multirow{4}{*}{{Trees}}}
			& \multicolumn{1}{c|}{Flann} & & \cite{flann} & &  & \checkmark  &  & C++ & &\\	
			\cline{2-11}			 
			& \multicolumn{1}{c|}{DSTree} &\cite{conf/vldb/Wang2013} & \cite{conf/vldb/Wang2013} &$\bullet$&$\bullet$&  & EAPCA  & Java & C & \checkmark\\	
			\cline{2-11}			 
			& \multicolumn{1}{c|}{HD-index} & & \cite{hdindex} & &  &   & Hilbert keys & C++ & & \checkmark\\	
			\cline{2-11}			 
			& \multicolumn{1}{c|}{iSAX2+} & \cite{journal/kais/Camerra2014} & \cite{journal/kais/Camerra2014} &$\bullet$&$\bullet$&  &  iSAX &  C\# & C & \checkmark\\	
			\cline{1-11}			 
			\cline{1-11}			 		 
		\end{tabular*}
	} 
	\label{tab:multiprogram}
\end{table*}

\subsection{Summarization Techniques}

\noindent\textbf{Random projections} (used by SRS~\cite{conf/vldb/sun14}) reduce the original high dimensional data into a lower dimensional space by multiplying it with a random matrix. 
The Johnson-Lindenstrauss (JL) Lemma~\cite{conf/map/johnson84} guarantees that if the projected space has a large enough number of dimensions, there is a high probability that the pairwise distances are preserved, with a distortion not exceeding $(1+\epsilon)$.

\noindent{\bf Piecewise Aggregate Approximation} (PAA)~\cite{journal/kais/Keogh2001} and {\it Adaptive Piecewise Constant Approximation} (APCA)~\cite{journal/acds/Chakrabarti2002} are segmentation techniques that approximate a data series $S$ using $l$ segments (of equal/arbitrary length, respectively). The approximation represents each segment with the mean value of its points. 
The {\it Extended APCA} (EAPCA)~\cite{conf/vldb/Wang2013} technique extends APCA by representing each segment with both the mean and the standard deviation.

\noindent\textbf{Quantization} 
is a lossy compression process that maps a set of infinite numbers to a finite set of codewords that together constitute the codebook. 
A \emph{scalar} quantizer operates on the individual dimensions of a vector independently, whereas a \emph{vector} quantizer considers the vector as a whole (leveraging the correlation between dimensions~\cite{journal/tit/gray1998}). 
The size $k$ of a codebook increases exponentially with the number of bits allocated for each code. 
A \emph{product} quantizer~\cite{journal/tpami/jegou2011} 
splits
the original vector of dimension $d$ into $m$ smaller subvectors, on which a lower-complexity vector quantization is performed. 
The codebook then consists of the cartesian product of the codebooks of the $m$ subquantizers. 
Scalar and vector quantization are special cases of product quantization, where $m$ is equal to $d$ and 1, respectively.

\noindent(i) 
{\it Optimized Product Quantization} (OPQ) (used by IMI~\cite{journal/tpami/ge2014}) improves the accuracy of the original product quantizer~\cite{journal/tpami/jegou2011} by adding a preprocessing step consisting of a linear transformation of the original vectors, which decorrelates the dimensions and optimizes space decomposition. {\color{black} A similar quantization technique, CK-Means, was proposed in~\cite{ck-means} but OPQ is considered the state-of-the-art~\cite{conf/CVPR/kalantidis2014,journal/ite/matsui2018}}. 

\noindent(ii) 
The {\it Symbolic Aggregate Approximation} (SAX)~\cite{conf/dmkd/LinKLC03} technique starts by transforming the data series into $l$ real values using PAA, and then applies a \emph{scalar} quantization technique to represent the PAA values 
using discrete symbols forming an alphabet of size $a$, called the cardinality of SAX. 
The $l$ symbols form the SAX representation. 
The $i$SAX~\cite{conf/kdd/Shieh2008} technique 
allows comparisons of SAX representations of different cardinalities, which makes SAX indexable. 


\noindent(iii) \textit{The Karhunen-Lo\`{e}ve transform (KLT).} 
The original VA+file method~\cite{conf/cikm/Hakan2000} first converts a data series $S$ of length $n$ using KLT into $n$ real values to de-correlate the data, then applies a \emph{scalar} quantizer to encode the real values as discrete symbols.
As {\color{black}we} will explain in the next subsection, for efficiency considerations, we altered the VA+file to use the {\it Discrete Fourier Transform} (DFT) instead of KLT. DFT~\cite{conf/fodo/Agrawal1993,conf/sigmod/Faloutsos1994,conf/sigmod/Rafiei1997,journal/corr/Rafiei1998} 
approximates a data series using $l$ frequency coefficients, 
and can be efficiently implemented with Fast Fourier Transform (FFT), which is 
optimal for whole matching (alternatively, the MFT algorithm~\cite{conf/icdsp/Albrecht1997} is adapted to subsequence matching 
since it uses sliding windows).

\vspace*{-0.3cm}

\subsection{Approximate Similarity Search Methods}

There exist several techniques for approximate similarity search~\cite{conf/stoc/indyk1998,conf/vldb/Gionis1999, journal/jda/bustos2004,conf/icde/houle2005,conf/kdd/ColeSZ05,journal/tpami/chavez2008,conf/icsis/amato2008,conf/sisap/tellez2011,conf/vldb/sun14,journal/tpami/ge2014,journal/corr/malkov16,conf/cvpr/yandex16}  {\color{black}\cite{conf/sigmod/berchtold1998,conf/pods/ooi2000,conf/vldb/yu2001}}. 
We focus on the {\color{black}7} most prominent techniques designed for multidimensional data, and we also describe the approximate search algorithms designed specifically for data series. 
We also propose a new set of techniques that can answer $\delta$-$\epsilon$-approximate queries based on modifications to existing exact similarity methods for data series.



\subsubsection{State-of-the-Art for Multidimensional Vectors}

{\color {black} \noindent{\bf Flann}~\cite{flann} is an in-memory ensemble technique for $ng$-approximate nearest neighbor search in high-dimensional spaces. Given a dataset and a desired search accuracy, Flann selects and auto-tunes the most appropriate algorithm among randomized kd-trees~\cite{random-kd-trees} and a new proposed approach based on hierarchical k-means trees~\cite{flann}.}

{\color {black} \noindent{\bf HD-index}~\cite{hdindex} is an $ng$-approximate nearest neighbor technique that partitions the original space into disjoint partitions of lower dimensionality, then represents each partition by an RBD tree (modified B+tree with leaves containing distances of data objects to reference objects) built on the Hilbert keys of data objects. A query $Q$ is partitioned according to the same scheme, 
searching the hilbert key of $Q$ in the RDB tree of each partition, then refining the candidates first using approximate distances based on triangular and Ptolemaic inequalities then using the real distances.}

\noindent{\bf HNSW}. 
HNSW~\cite{journal/corr/malkov16} is an {\color{black} in-memory} $ng$-approximate method that belongs to the class of proximity graphs that exploit two fundamental geometric structures: the Voronoi Diagram (VD) and the Delaunay Triangulation (DT). 
A VD is obtained when a given space is decomposed using a finite number of points, called \emph{sites}, into regions such that each site is associated with a region consisting of all points that are closer to it than to any other site. 
The DT is the dual of the VD.
It is constructed by connecting sites with an edge if their regions share a side. 
Since constructing a DT for a generic metric space is not always possible (except if the DT is the complete graph)~\cite{journal/vldbj/navarro2002}, proximity graphs, which approximate the DT by conserving only certain edges, have been proposed~\cite{conf/siam/arya1993,journal/apr/chavez2010,conf/sigkdd/aoyama2011,conf/iccv/wang2013,journal/is/malkov2014,conf/sisap/chavez2015,conf/icmm/jiang2016,journal/corr/malkov16}. 
A k-NN graph is a proximity graph, where only the links to the closest neighbors are preserved. 
Such graphs suffer from two limitations: (i) the curse of dimensionality; and (ii) the poor performance on clustered data (the graph has a high probability of being disconnected). 
To address these limitations, the Navigable Small World (NSW) method~\cite{journal/is/malkov2014} proposed to heuristically augment the approximation of the DT with long range links to satisfy 
the small world navigation properties~\cite{conf/stoc/kleinberg2000}.  
The HNSW graph~\cite{journal/corr/malkov16} improves the search efficiency of NSW 
by organizing the links in hierarchical layers according to their lengths. 
Search starts at the top layer, which contains only the longest links, and proceeds down the hierarchy. 
HNSW 
is considered the state-of-the-art~\cite{conf/sisap/martin17}. 

{\color {black} \noindent{\bf NSG}~\cite{nsg} is a recent in-memory proximity graph approach that approximates a graph structure called MRNG~\cite{nsg} which belongs to the class of Monotonic Search Networks (MSNET). 
Building an MRNG graph for large datasets becomes impractical; that is why the state-of-the-art techniques approximate it. NSG approximates the MRNG graph by relaxing the monotonicity requirement and edge selection strategy, and dropping the longest edges in the graph.}

\noindent{\bf IMI}. Among the different quantization-based inverted indexes proposed in the literature~\cite{journal/tpami/jegou2011,conf/icassp/jegou2011,journal/iccv/xia2013,journal/pami/babenko15}, IMI~\cite{journal/tpami/ge2014,journal/pami/babenko15} is considered the state-of-the-art~\cite{journal/ite/matsui2018}. 
This class of techniques builds an inverted index storing the list of data points that lie in the proximity of each codeword. 
The codebook
is the set of representative points obtained by performing clustering on the original data. When a query arrives, the $ng$-approximate search algorithm returns the list of all points corresponding to the closest codeword (or list of codewords).

{\color{black} \noindent{\bf LSH.} The LSH family}~\cite{journal/corr/andoni2018} encompasses a class of randomized algorithms that solve the $\delta$-$\epsilon$-approximate nearest neighbor problem in sub-linear time, for $\delta < 1 $. 
The main intuition 
is that two points that are nearby in a high dimensional space, will remain nearby when projected to a lower dimensional space~\cite{conf/stoc/indyk1998}. 
LSH techniques {\color{black}partition points into buckets} using hash functions, which guarantee that only nearby points are likely to be mapped to the same bucket. 
Given a dataset $\mathbb{S}$ and a query $S_Q$, $L$ hash functions are applied to all points in $\mathbb{S}$ and to the query $S_Q$. Only points that fall at least once in the same bucket as $S_Q$, in each of the $L$ hash tables, are further processed in a linear scan to find the $\delta$-$\epsilon$-approximate nearest-neighbor. 
There exist many variants of LSH, either proposing different hash functions to support particular similarity measures~\cite{conf/poccs/broder1997,conf/sigcg/datar2004,conf/stoc/charikar02,conf/sigmod/gan2012}, 
or improving the theoretical bounds on query accuracy (i.e., $\delta$ or $\epsilon$), query efficiency or the index size~\cite{journal/nips/liu2004,conf/soda/panigrahy2006,journal/siamdm/motwani2007,conf/vldb/lv2007,conf/sigmod/gan2012,journal/atct/odonnell2014,conf/vldb/sun14,qalsh} {\color{black}~\cite{sk-lsh}}. 
{\color{black}In this work, we select SRS~\cite{conf/vldb/sun14} and QALSH~\cite{qalsh} to represent the class of LSH techniques because they are considered the state-of-the-art in terms of footprint and accuracy, respectively~\cite{hdindex}}. SRS answers $\delta$-$\epsilon$-approximate queries using size linear to the dataset size, while empirically outperforming other LSH methods (with size super-linear to the dataset size~\cite{conf/poccs/broder1997}). 
{\color{black} QALSH is a query-aware LSH technique that partitions points into buckets using the query as anchor. Other LSH methods typically partition data points before a query arrives, using a random projection followed by a random shift. QALSH, does not perform the second step until a query arrives, thus improving the likelihood that points similar to the query are mapped to the same bucket.}

\subsubsection{State-of-the-Art for Data Series}
While a number of data series methods support approximate similarity search ~\cite{conf/icde/shatkay1996,conf/kdd/Keogh1997,conf/ssdm/Wang2000,conf/kdd/ColeSZ05,conf/icdm/Camerra2010,journal/edbt/Schafer2012,conf/vldb/Wang2013,journal/kais/Camerra2014,journal/vldb/Zoumpatianos2016}, we focus on those that fit the scope of this study, i.e., methods that support out-of-core k-NN queries with Euclidean distance. 
In particular, we examine DSTree~\cite{conf/vldb/Wang2013}, iSAX2+~\cite{journal/kais/Camerra2014}, and VA+file~\cite{conf/cikm/Hakan2000}, the three data series methods that perform the best in terms of exact search~\cite{journal/pvldb/echihabi2018}, and also inherently support ng-approximate search.

\noindent{\bf DSTree}~\cite{conf/vldb/Wang2013} is a tree index based on the EAPCA summarization technique and supports ng-approximate and exact query answering. 
Its dynamic segmentation algorithm 
allows tree nodes to split vertically and horizontally, unlike the other data series indexes which allow either one or the other. 
The DSTree supports a lower and upper bounding distance and uses them to calculate a QoS measure that determines the optimal way to split any given node. 
We significantly improved the efficiency of the original DSTree Java implementation by developing it from scratch in C and optimizing its buffering and memory management, 
making it 4 times faster across datasets ranging between 25-250GB.

\noindent{\bf SAX-based indexes} include different flavors of tree indexes based on SAX summarization. 
The original iSAX index~\cite{conf/kdd/shieh1998} was enhanced with a better spliting policy and bulk-loading support in iSAX 2.0~\cite{conf/icdm/Camerra2010}, while
iSAX2+~\cite{journal/kais/Camerra2014} further optimized bulk-loading. 
ADS+~\cite{journal/vldb/Zoumpatianos2016} then improved upon iSAX2+ by making it adaptive, 
Coconut~\cite{journal/pvldb/kondylakis18,DBLP:conf/sigmod/KondylakisDZP19,coconutjournal} by constructing a compact and contiguous data layout, and DPiSAX~\cite{dpisax,dpisaxjournal}, ParIS~\cite{conf/bigdata/peng18} and MESSI~\cite{conf/icde/peng20} by exploiting parallelization.
Here, we use iSAX2+, because of its excellent performance~\cite{journal/pvldb/echihabi2018} and the fact that the SIMS query answering strategy~\cite{journal/vldb/Zoumpatianos2016} of ADS+, Coconut, and ParIS is not immediately amenable to approximate search with guarantees (we plan to extend these methods in our future work). 
We do not include DPiSAX and MESSI, because they are distributed, and in-memory only, algorithms, respectively. 
\ifJournal
\begin{figure}[t]
	\captionsetup{justification=centering}
	\captionsetup[subfigure]{justification=centering}
	\begin{subfigure}{0.49\columnwidth}
		\centering
		\includegraphics[width=\columnwidth]{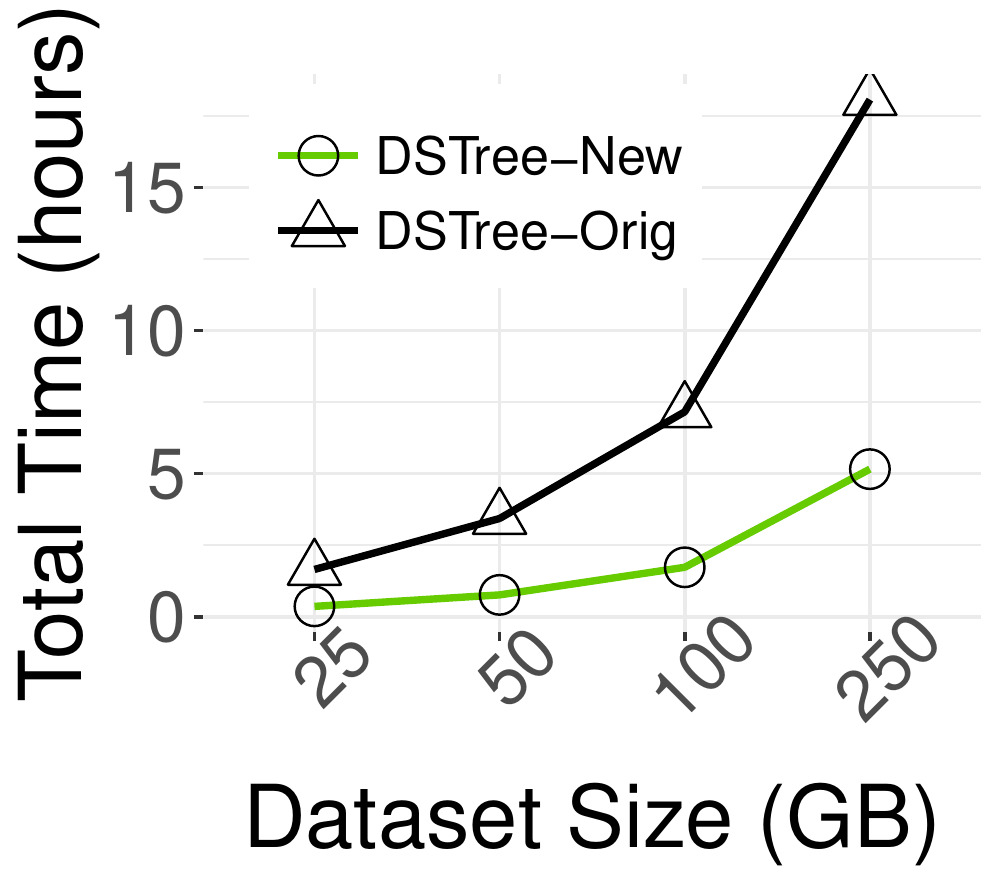}
		\caption{Total Time (Indexing and Answering 100 Exact Queries)}
		\label{fig:dstree:orig:new:combined}
	\end{subfigure}u
	\begin{subfigure}{0.49\columnwidth}
		\centering
		\includegraphics[width=\columnwidth]{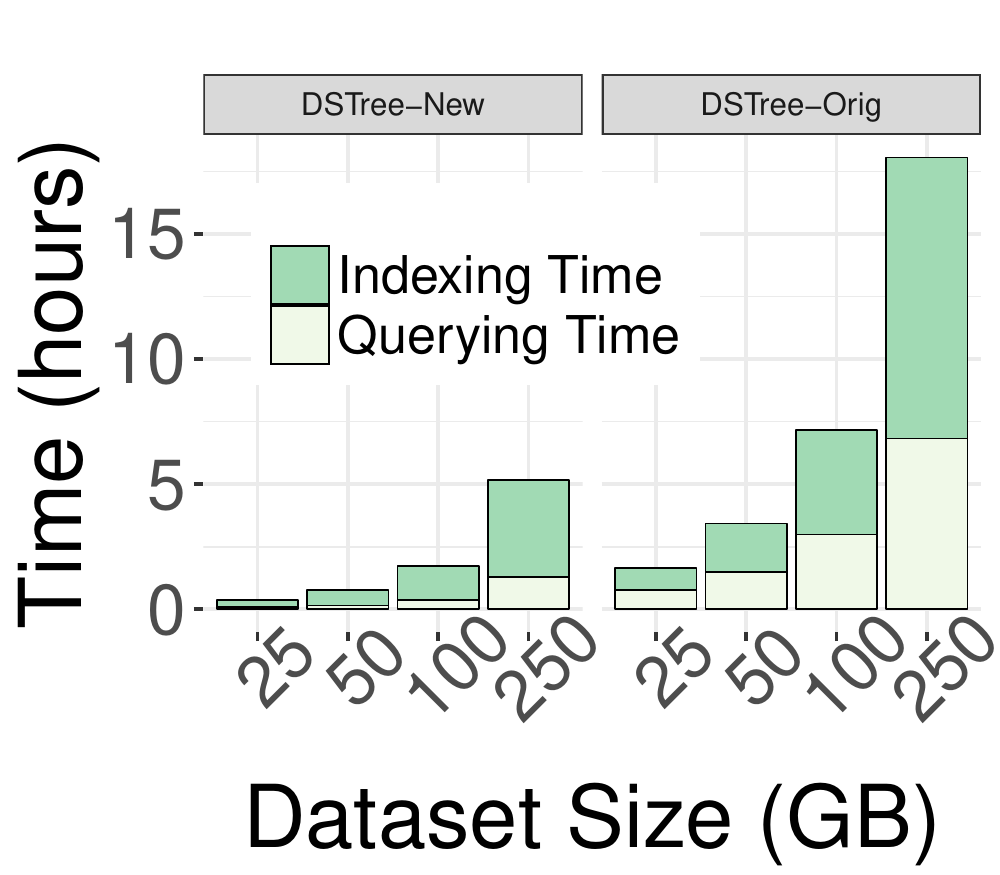}
		\caption{Detailed Times for Indexing and Answering 100 Exact Queries}
		\label{fig:dstree:orig:new:detailed}
	\end{subfigure}
	\caption{DSTree Implementation Optimization}
	\label{fig:dstree:orig:new}
}
\end{figure}
\fi

{\color {black} \noindent{\bf TARDIS}~\cite{conf/icde/zhang2019} is a distributed indexing method that supports exact and $ng$-approximate kNN queries. It improves the efficiency and accuracy of iSAX by building a more compact, k-ary tree index, exploiting word-level (instead of character-level) cardinality, and using a novel conversion scheme between SAX representations. We do not include TARDIS in the experimental evaluation since it is a distributed algorithm (built in Scala for Spark). 
}

\noindent{\bf VA+file}~\cite{conf/cikm/Hakan2000} is a skip-sequential method that improves the accuracy and efficiency of the VA-file~\cite{conf/vldb/Weber1998}. 
Both techniques create a file that contains quantization-based summarizations of the original multidimensional data. 
Search proceeds by sequentially reading each summarization, 
calculating its lower bounding distance to the query, and accessing the original multidimensional vector only if the lower bounding distance is less than the current \emph{best-so-far (bsf)} answer. 
We greatly improved the performance of the original VA+file by approximating KLT with DFT~\cite{conf/cikm/Hakan2000,journal/acta/maccone2007} and implementing it in C instead of Matlab. 
In the rest of the text, whenever we mention the VA+file, we refer to the modified version.

\vspace*{-0.2cm}

\subsubsection{Extensions of Data Series Methods}
\label{sec:dataseriesextensions}

We now propose extensions to the data series methods described above, that will allow them to support $\epsilon$-approximate and $\delta$-$\epsilon$-approximate search (in addition to ng-approximate that they already support).
Due to space limitations, we only discuss the tree-based methods (such as iSAX2+ and DSTree); skip-sequential techniques (such as VA+file) can be modified following the same ideas.

The exact 1-NN search algorithms of DSTree and iSAX2+ are based on an optimal exact NN algorithm first proposed for PMR-Quadtree~\cite{conf/isasd/samet1995}, which was then generalized for any hierarchical index structure that is constructed using a conservative and recursive partitioning of the data~\cite{conf/pods/berchtold1997}. 

\begin{algorithm}[tb]
	{\scriptsize
		\caption{exactNN({$\bm{S_Q}$},{$\bm{idx}$})}
		\begin{algorithmic}[1]
			\\{$\bm{bsf.dist}$} $\gets$ $\infty$ ; {$\bm{bsf.node}$} $\gets$ $NULL$;		
			\For {each {{$\bm{rootNode}$} in {$\bm{idx}$} }}         
			\State{$\bm{result.node}$} $\gets$ {$\bm{rootNode}$};
			\State{$\bm{result.dist}$} $\gets$ calcMinDist({$\bm{S_Q}$},{$\bm{rootNode}$});			
			\State{push $\bm{result}$ to $\bm{pqueue}$} 
			\EndFor
			\\{$\bm{bsf}$} $\gets$ {\color{mygreen}\dashuline{ng-approxNN}}({{$\bm{S_Q}$},$\bm{idx}$}); 		
			\\add {$\bm{bsf}$} to {$\bm{pqueue}$};
			\While{ {$\bm{result}$} $\gets$ pop next node from {$\bm{pqueue}$ } } 
			\State{$\bm{n}$} $\gets$ {$\bm{result.node}$};
			\If { {$\bm{n.dist}$} $>$ {$\bm{bsf.dist}$}} 
			break;
			\EndIf                 
			\If {{$\bm{n}$} is a leaf}    \Comment{a leaf node}     
			\For {each {{$\bm{S_C}$} in {$\bm{n}$} }}         
			\State {$\bm{realDist}$} $\gets$ calcRealDist({$\bm{S_Q}$},{$\bm{S_C}$});
			\If { {$\bm{realDist}$} $<$ {$\bm{bsf.dist}$}} 
			\State {$\bm{bsf.dist}$} $\gets$ {$\bm{realDist}$} ;
			\State {$\bm{bsf.node}$} $\gets$ {$\bm{n}$};		               
			\EndIf                  
			\EndFor        
			\Else  \Comment{an internal node}
			\For {each {{$\bm{childNode}$} in {$\bm{n}$} }}         
			\State {$\bm{minDist}$} $\gets$ calcMinDist({$\bm{S_Q}$},{$\bm{childNode}$});
			\If { {$\bm{minDist}$} $<$ {$\bm{bsf.dist}$}} add {$\bm{childNode}$} to
			\State {$\bm{pqueue}$ } with priority {$\bm{minDist}$}; 
			\EndIf                  
			\EndFor        
			\EndIf 
			\EndWhile\label{euclidendwhile}
			\State \Return {$\bm{bsf}$}
		\end{algorithmic}
		\label{alg:exactNN}
	} 
\end{algorithm}

Algorithm~\ref{alg:exactNN} describes an index-invariant algorithm for exact 1-NN search. It takes as arguments a query $S_Q$ and an index $idx$.
Lines 1-5 initialize the \emph{best-so-far (bsf)} answer and a priority queue with the root node(s) of the index in increasing order of lower bounding ($lb$) distances (the $lb$ distance is calculated by the function $calcMinDist$). 
In line 6, the $ng$-approxNN function traverses one path of the index tree visiting one leaf to return an $ng$-approximate bsf answer, {\color{black} which} is added to the queue (line 7). 
In line 8, the algorithm pops nodes from the queue, terminating in line 10 if the $lb$ distance of the current node is greater than the current \emph{bsf} distance (the $lb$ distances of all remaining nodes in the queue are also greater than the \emph{bsf}). 
Otherwise, if the node is a leaf, the \emph{bsf} is updated if a better answer is found (lines 11-16); if the node is an internal node, its children are added to the queue provided their $lb$ distances are greater than the \emph{bsf} distance (lines 18-21).

\begin{algorithm}[tb]
	{\scriptsize
		\caption{{\color{myred}\underline{\underline{delta}}}{\color{myblue}\underline{Epsilon}}NN({$\bm{S_Q}$},{$\bm{idx}$},{$\bm{\delta}$},{$\bm{\epsilon}$}, {$\bm{F_Q(.)}$})}
		\begin{algorithmic}[1]
			\\{$\bm{bsf.dist}$} $\gets$ $\infty$ ; {$\bm{bsf.node}$} $\gets$ $NULL$;			
			\Statex	 {\color{myred}\underline{\underline{${\bm {r_\delta(Q)}}$  $\gets$ calcDeltaRadius({$\bm{S_Q}$},{$\bm{\delta}$}, {$\bm{F_Q(.)}$})}}}; 
			\\{$\bm{bsf}$} $\gets$ {\color{mygreen}\dashuline{ng-approxNN}}({{$\bm{S_Q}$},$\bm{idx}$}); 		
			\\add {$\bm{bsf}$} to {$\bm{pqueue}$};
			\For {each {{$\bm{rootNode}$} in {$\bm{idx}$} }}         
			\State{$\bm{result.node}$} $\gets$ {$\bm{rootNode}$};
			\State{$\bm{result.dist}$} $\gets$ calcMinDist({$\bm{S_Q}$},{$\bm{rootNode}$});			
			\State{push $\bm{result}$ to $\bm{pqueue}$} 
			\EndFor
			\While{ {$\bm{result}$} $\gets$ pop next node from {$\bm{pqueue}$ } } 
			\State{$\bm{n}$} $\gets$ {$\bm{result.node}$};
			\If { {$\bm{n.dist}$} $>$ {$\bm{bsf.dist}{\color{myblue}\underline{/(1+\epsilon)}}$}} 
			break;
			\EndIf                 
			\If {{$\bm{n}$} is a leaf}    \Comment{a leaf node}     
			\For {each {{$\bm{S_C}$} in {$\bm{n}$} }}         
			\State {$\bm{realDist}$} $\gets$ calcRealDist({$\bm{S_Q}$},{$\bm{S_C}$});
			\If { {$\bm{realDist}$} $<$ {$\bm{bsf.dist}$}} 
			\State {$\bm{bsf.dist}$} $\gets$ {$\bm{realDist}$} ;
			\State {$\bm{bsf.node}$} $\gets$ {$\bm{n}$};		               
			\Statex\hspace{2cm}{{\color{myred} \underline{\underline{if { {$\bm{bsf.dist}$} $\leq$ $(1+\epsilon)$ {$\bm {r_\delta(Q)}$}} then exit;}}}}
			\EndIf                  
			\EndFor        
			\Else  \Comment{an internal node}
			\For {each {{$\bm{childNode}$} in {$\bm{n}$} }}         
			\State {$\bm{minDist}$} $\gets$ calcMinDist({$\bm{S_Q}$},{$\bm{childNode}$});
			\If { {$\bm{minDist}$} $<$ {$\bm{bsf.dist}{\color{myblue}/\underline{(1+\epsilon)}}$}} add
			\State {$\bm{childNode}$} to {$\bm{pqueue}$ } with priority {$\bm{minDist}$}; 
			\EndIf                  
			\EndFor        
			\EndIf 
			\EndWhile\label{euclidendwhile}
			\State \Return {$\bm{bsf}$}
		\end{algorithmic}
		\label{alg:deltaepsilonNN}
	} 
\end{algorithm}

We can use Algorithm~\ref{alg:exactNN} for $ng$-approximate search, by visiting one leaf and returning the first \emph{bsf}. 
This $ng$-approximate answer can be anywhere in the data space

We extend approximate search in Algorithm~\ref{alg:exactNN} by introducing two changes: (i) allow the index to visit up to $nprobe$ leaves (user parameter); and (ii) apply the modifications suggested in~\cite{conf/icde/Ciaccia2000} to support $\delta$-$\epsilon$-approximate NN search. 
The first change is straightforward, so we only describe the second change in Algorithm~\ref{alg:deltaepsilonNN}. 
%
%
To return the $\epsilon$-approximate NN of $S_Q$, $S_\epsilon$, {\bf\emph{bsf.dist}} is replaced with {\bf\emph{bsf.dist$/(1+\epsilon)$}} in lines 10 and 20.  
To return the $\delta$-$\epsilon$-approximate NN of $S_Q$, $S_{\delta\epsilon}$, we also modify lines 1 and 16.

The distance $r_\delta(Q)$ is initialized in line 1 using $F_Q(\cdot)$, $S_Q$ and $\delta$. $F_Q(\cdot)$ represents the relative distance distribution of $S_Q$. 
Intuitively, $r_\delta(Q)$ is the maximum distance from $S_Q$, such that the sphere with center $S_Q$ and radius $r_\delta(Q)$ is empty with probability $\delta$. 
As proposed in~\cite{conf/pods/Ciaccia1998}, we use $F(\cdot)$, the overall distance distribution, instead of $F_Q(\cdot)$ to estimate $r_\delta(Q)$. 
The delta radius $r_\delta(Q)$ is then used in line 16 as a stopping condition.
When  $\delta = 1$, Algorithm~\ref{alg:deltaepsilonNN} returns $S_{\delta\epsilon}$, the $\epsilon$-approximate NN of $S_Q$, and when $\delta = 1$ and $\epsilon=0$, Algorithm~\ref{alg:deltaepsilonNN} becomes equivalent to Algorithm~\ref{alg:exactNN}, i.e., it returns $S_x$, the exact NN of $S_Q$. 
Our implementations generalize Algorithm~\ref{alg:deltaepsilonNN} to the case of $k \ge 1$. 
These modifications are straightforward and omitted for the sake of brevity. 
A proof of correctness for Algorithm~\ref{alg:deltaepsilonNN} can be found in~\cite{conf/icde/Ciaccia2000,conf/sisap/ciaccia17} for $k = 1$ and $k \ge 1$, respectively.

\begin{figure}[tb]
	\centering
	\captionsetup{justification=centering}
	\includegraphics[width=\columnwidth]{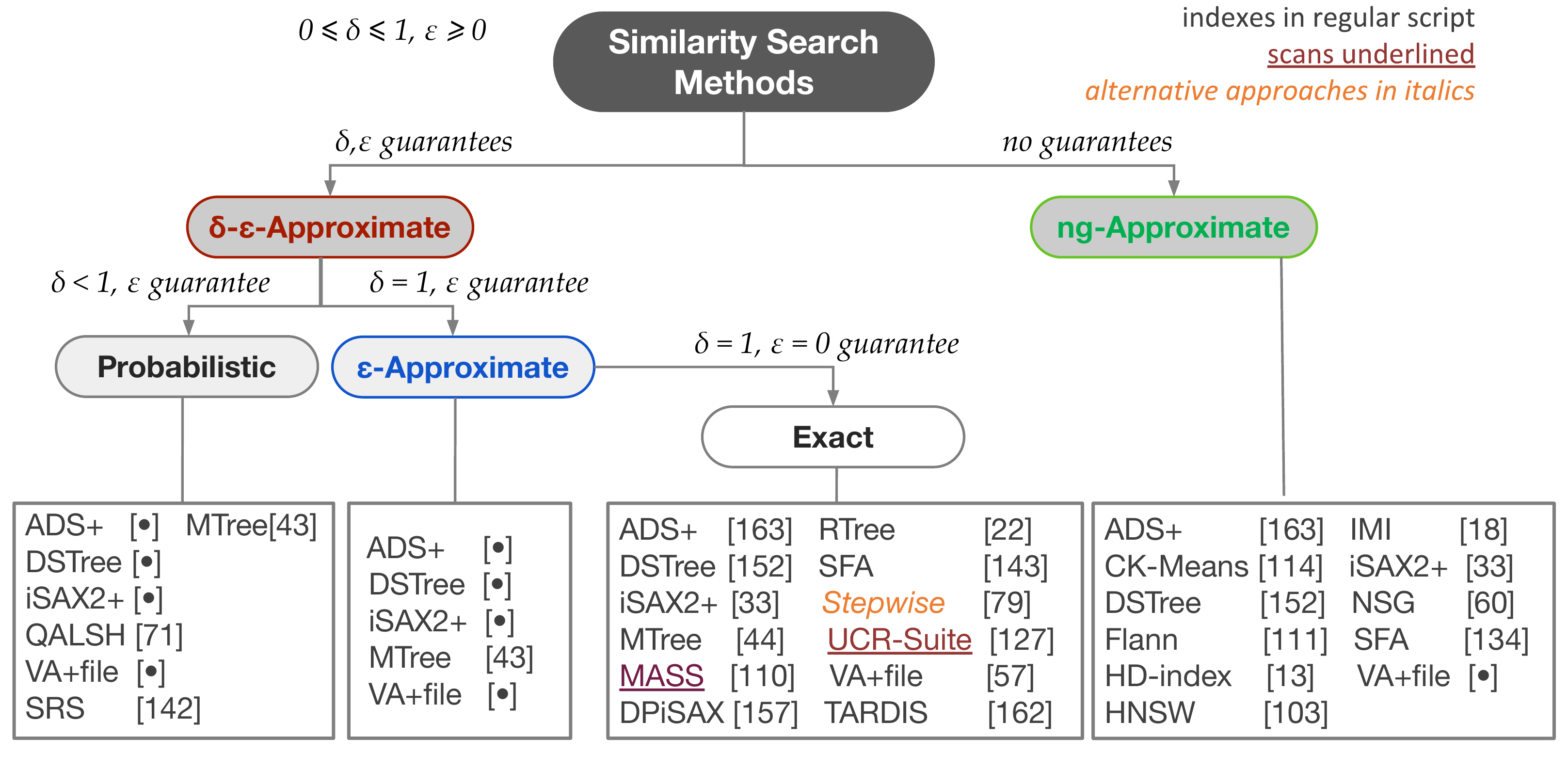}
	\caption{{\color{black} Taxonomy of similarity search methods.}}
	\vspace*{-0.2cm}
	\label{fig:taxonomy}
\end{figure}

\subsection{Taxonomy of Similarity Search Methods}
Figure~\ref{fig:taxonomy} presents a taxonomy of similarity search methods based on the type of guarantees they provide (methods with multiple types of guarantees are included in more than one leaf of the taxonomy).
We call probabilistic the general $\delta$-$\epsilon$-approximate methods. 
When $\delta =1$ we have the $\epsilon$-approximate methods.
Setting $\delta=1$ and $\epsilon=0$, we get the exact methods. 
Finally, methods that provide no guarantees are categorized under ng-approximate. 
Here, {\color{black} we cover 7 state-of-the-art methods from the high-dimensional literature, Flann, HD-index, HNSW, IMI, NSG, QALSH and SRS, as well as the 3 best methods from the data series community~\cite{journal/pvldb/echihabi2018}, iSAX2+, DSTree and VA+file. 
}

\section{Experimental Evaluation}
\label{sec:experiments}
{\color{black}
We assessed all methods on the same framework. 
Source code, datasets, queries, and all results are available in~\cite{url/DSSeval2}.

\vspace{1cm}

\subsection{Experimental Setup}
\label{subsec:framework}
\label{subsec:environment}

\noindent{\bf Environment.} 
All methods were compiled with GCC 6.2.0 under Ubuntu Linux 16.04.2 with their default compilation flags; optimization level was set to 2. 
Experiments were run on a server with two Intel Xeon E5-2650 v4 2.2GHz CPUs,
75GB\footnote{We used GRUB to limit the amount of RAM, so that all methods are forced to use the disk. Note that GRUB prevents the operating system from using the rest of the RAM as a file cache, which is what we wanted for our experiments.} of RAM, 
and 10.8TB (6 x 1.8TB) 10K RPM SAS hard drives 
in RAID0 with a throughput of 1290 MB/sec.


\noindent{\textbf{Algorithms.}}
We use the most efficient C/C++ implementation available for each method: 
iSAX2+~\cite{url/DSSeval}, DSTree~\cite{url/DSSeval} and VA+file~\cite{url/DSSeval} representing exact data series methods with support for approximate queries; and HNSW~\cite{url/hnsw}, Faiss IMI~\cite{url/faiss}, SRS~\cite{url/srs}, {\color{black} FLANN~\cite{flann}, and QALSH~\cite{qalsh} representing strictly approximate methods for vectors. We ran experiments with the HD-index~\cite{hdindex} and NSG~\cite{nsg}, but since they could not scale for our smallest 25GB dataset, we do not report results for them.}
We extended DSTree, iSAX2+ and VA+file with Algorithm~\ref{alg:deltaepsilonNN}, approximating $r_{\delta}$ with density histograms on a 100K data series sample, following the C++ implementation of~\cite{conf/icde/Ciaccia2000}. 
All methods are single core implementations, except for HNSW and IMI that make use of multi-threading and SIMD vectorization. 
Data series points are 
represented using single precision values and methods based on fixed summarizations use 16 dimensions. 

\noindent{\textbf{Datasets.}}
We use synthetic and real datasets. Synthetic datasets, called $Rand$, were generated as random-walks using a summing process with steps following a Gaussian distribution (0,1). 
Such data model financial time series~\cite{conf/sigmod/Faloutsos1994} and have been widely used in the literature~\cite{conf/sigmod/Faloutsos1994,journal/kais/Camerra2014,conf/kdd/Zoumpatianos2015}. 
Our four real datasets cover domains as varied as deep learning, computer vision, seismology, and neuroscience. 
\emph{Deep1B}~\cite{url/data/deep1b} comprises 1 billion vectors of size 96 extracted from the last layers of a convolutional neural network. 
\emph{Sift1B}~\cite{conf/icassp/jegou2011,url/data/sift} consists of 1 billion SIFT vectors of size 128 representing image feature descriptions. 
To the best of our knowledge, these two vector datasets are the largest publicly available real datasets. 
\emph{Seismic100GB}~\cite{url/data/seismic}, contains 100 million data series of size 256 representing earthquake recordings at seismic stations worldwide. 
\emph{Sald100GB}~\cite{url/data/eeg} contains neuroscience MRI data and includes 200 million data series of size 128. 
In our experiments, we vary the size of the datasets from 25GB to 250GB. 
The name of each dataset is suffixed with its size. 
We do not use other real datasets that have appeared in the literature~\cite{UCRArchive,conf/sisap/martin17}, because they are very small, not exceeding 1GB in size.

\noindent{\textbf{Queries.}}
All our query workloads consist of 100 query series run asynchronously, i.e., not in batch mode. 
Synthetic queries were generated using the same random-walk generator as the $Rand$ dataset (with a different seed, reported in~\cite{url/DSSeval2}). 
For the Deep1B and Sift1B datasets, we randomly select 100 queries from the real workloads that come with the datasets archives. For the other real datasets, query workloads were generated by adding progressively larger amounts of noise to data series extracted from the raw data, so as to produce queries having different levels of difficulty, following the ideas in~\cite{johannesjoural2018}. 
Our experiments cover $ng$-approximate and $\delta$-$\epsilon$-approximate k-NN queries, where k $\in [1,100]$. We also include results for exact queries to serve as a yardstick. 

\noindent{\textbf{Scenarios.}}
{\color{black}Our experimental evaluation proceeds in four main steps: 
(i) we tune methods to their optimal parameters (\S\ref{ssec:parametrization}); (ii) we evaluate the indexing scalability of the methods
(\S\ref{ssec:indexing_efficiency}); (iii) we compare in-memory and out-of-core scalability and  accuracy of all methods (\S\ref{ssec:query_efficiency_mem}-\S\ref{ssec:query_efficiency_disk}); and (iv) we perform additional experiments on the best performing methods for disk-resident data (\S\ref{ssec:query_efficiency_disk})}.

\noindent{\textbf{Measures.}} We assess methods using the following criteria:

\noindent(1) Scalability and search efficiency using: \emph{wall clock time} (input, output, CPU  and total time), \emph{throughput} (\# of queries answered per minute), and two implementation-independent measures: the \emph{number of random disk accesses} (\# of disk seeks) and the \emph{percentage of data accessed}. 

\noindent(2) Search accuracy is assessed using: \emph{Avg\_Recall}, \emph{Mean Average Precision (MAP)}, and \emph{Mean Relative Error (MRE)}. Recall is the most commonly used accuracy metric in the approximate similarity search literature. However, since it does not consider rank accuracy, we also use MAP~\cite{conf/sigir/turpin2006} that is popular in information retrieval~\cite{book/manning2008,conf/sigir/buckley2000} {\color{black} and has been proposed recently in the high-dimensional community~\cite{hdindex} as an alternative accuracy measure to recall}. 
For a workload of queries $S_{Q_i} : i \in [1, N_Q]$, these are defined as follows.
\begin{compactitem}
\item $Avg\_Recall(workload) = \sum_{i=1}^{N_Q} Recall(S_{Q_i}) / N_Q $ 
{\color{black}\item $MAP(workload) = \sum_{i=1}^{N_Q} AP(S_{Q_i}) / N_Q $}
\item $MRE(workload) = \sum_{i=1}^{N_Q} RE(S_{Q_i}) / N_Q $
\end{compactitem}
where:  

\noindent$\bullet$  
$Recall(S_{Q_i}) = \frac{\textit{\# true neighbors returned by }{Q_i}}{k}$

\noindent$\bullet$  
$AP(S_{Q_i}) = \frac {\sum_{r=1}^{k} (P(S_{Q_i,r}) \times rel(r))} {k}, \forall i \in [1,N_Q]$ 

$-$  
$P({S_{Q_i}},r) = \frac {\text{\# true neighbors among the first $r$ elements}} {r}$.

$-$ $rel(r)$ is equal 1 if the neighbor returned at position $r$ 

is one of the $k$ exact neighbors of $S_{Q_i}$ and 0 otherwise.
	
\noindent$\bullet$  
$RE(S_{Q_i}) = \frac{1}{k} \times \sum_{r=1}^{k} \frac {d(S_{Q_i},S_{C_{r}}) - d(S_{Q_i},S_{C_i})} {d(S_{Q_i},S_{C_i})}$. 	
$S_{C_i}$ is the exact nearest neighbor of $S_{Q_i}$ and $S_{C_{r}}$ is the $r$-th NN retrieved\footnote{Note that in Definition~\ref{def:epsmatch}, $\epsilon$ is an upper bound on $RE(S_{Q_i})$.}.
Without loss of generality, we do not consider the case where $d(S_{Q_i},S_{C_i}) = 0$. 
(i.e., range queries with radius zero, or kNN queries where the 1-NN is the query itself\footnote{In these cases, the MRE definition can be extended to use the symmetric mean absolute percentage error~\cite{journal/omega/Flores1986}.}.) 

\noindent(3) Size, using the \emph{main memory} footprint of the algorithm.

\noindent{\textbf{Procedure.}}
Experiments involve two steps: index building and query answering. Caches are fully cleared before each step, and stay warm between consecutive queries.
For large datasets that do not fit in memory, the effect of caching is minimized for all methods. 
All experiments use workloads of 100 queries. 
Results reported for workloads of 10K queries are extrapolated: we discard the $5$ best and $5$ worst queries of the original 100 (in terms of total execution time), and multiply the average of the 90 remaining queries by 10K.

\subsection{Results}
\label{subsec:results}


\subsubsection{\textbf{Parametrization}}
\label{ssec:parametrization}

{\color{black}
We start by fine tuning each method (graphs omitted for brevity). 
In order to understand the speed/accuracy tradeoffs, we fix the total memory size available to 75GB. 
The optimal parameters for DSTree, iSAX2+ and VA+file are set according to~\cite{journal/pvldb/echihabi2018}. 
For indexing, the buffer and leaf sizes are set to 60GB and 100K, respectively, for both DSTree and iSAX2+. 
iSAX2+ is set to use 16 segments. 
VA+file uses a 20GB buffer and 16 DFT symbols. 
For SRS, we set M (the projected space dimensionality) to 16 so that the representations of all datasets fit in memory. 
The settings were the same for all datasets. 
The fine tuning for HNSW and IMI proved more tricky and involved many testing iterations since the index building parameters strongly affect the speed/accuracy of query answering and differ greatly across datasets. 
For this reason, different parameters were chosen for different datasets. 
For the in-memory method HNSW, we set efConstruction (the number of neighbors considered during index construction) to 500, and M (the number of bi-directional edges created for every new node during indexing) to 4 for the Rand25GB dataset. 
{\color{black} For Deep25GB and Sift25GB, we set efConstruction to 500 and M to 16}.
To tune the Faiss implementation of IMI, we followed the guidelines in~\cite{url/faiss}. 
For the in-memory datasets, we set the index factory key to PQ32\_128,IMI2x12,PQ32 and the training size to 1048576, while for disk based datasets, the index key is PQ32\_128,IMI2x14,PQ32 and the training size 4194304. 
To tune $\delta$-$\epsilon$-approximate search performance and accuracy, we vary $\delta$ and $\epsilon$ for SRS and $\epsilon$ for DSTree, iSAX2+ and VA+file (except in one experiment where we also vary $\delta$). 
For $ng$-approximate search, we vary the $nprobe$ parameter for DSTree/iSAX2+/IMI/VA+file ($nprobe$ represents the number of visited leaves for DSTree/iSAX2+, the number of visited raw series for VA+file, and the number of inverted lists for IMI), and the \emph{efs} parameter for HNSW (which represents the number of non-pruned candidates). 

\subsubsection{\textbf{\color{black}Indexing Efficiency}} 
\label{ssec:indexing_efficiency}
In this section, we evaluate the indexing scalability of each method by varying the dataset size. We used four synthetic datasets of sizes 25GB, 50GB, 100GB and 250GB, two of which fit in memory (total RAM was 75GB).
{\color{black}
Figure~\ref{fig:exact:datasize:time:indexing:cache} shows that iSAX2+ is the fastest method at index building in and out of memory, followed by VA+file, SRS, DSTree, {\color{black}FLANN}, QALSH, IMI and HNSW. 
Even though IMI and HNSW are the only parallel methods, they are the slowest at index building. {\color{black}Although FLANN is slow at indexing the 50GB dataset, we think this is more due to memory management issues in the code, which cause swapping.}
For HNSW, the major cost is building the graph structure, whereas IMI spends most of the time on determining the clusters and computing the product quantizers. We also measured the breakdown of the indexing time and found out that all methods can be significantly improved by parallelism except iSAX2+ and QALSH that are I/O bound.
In terms of footprint, the DSTree is the most memory-efficient, followed by iSAX2+. 
IMI, SRS, VA+file {\color{black}and FLANN} are two orders of magnitude larger, while {\color{black}QALSH} and HNSW are a further order of magnitude bigger (Figure~\ref{fig:exact:datasize:memory:indexing:cache}).}

\begin{figure}[!htb]
	\captionsetup{justification=centering}
	\captionsetup[subfigure]{justification=centering}
	\begin{subfigure}{\columnwidth }
		\centering
		\includegraphics[scale=0.14]{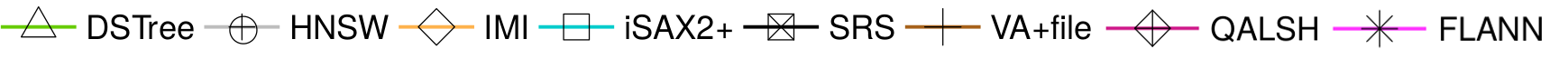}
	\end{subfigure}		
	\begin{subfigure}{0.49\columnwidth }
		\centering
		\includegraphics[width=\columnwidth ]{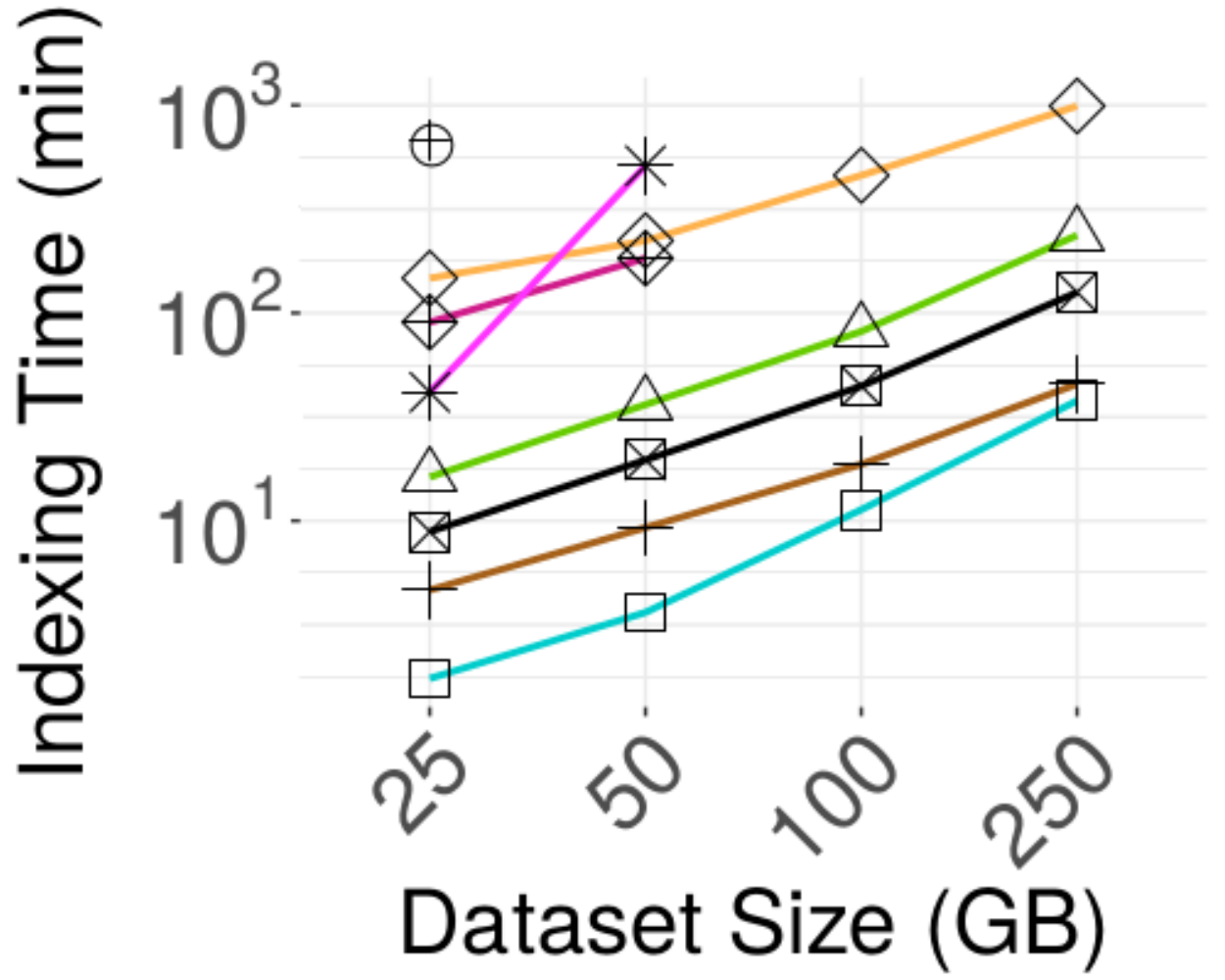}
		\caption{Indexing time}
		\label{fig:exact:datasize:time:indexing:cache}
	\end{subfigure}
	\begin{subfigure}{0.49\columnwidth }
		\centering
		\includegraphics[width=\columnwidth ]{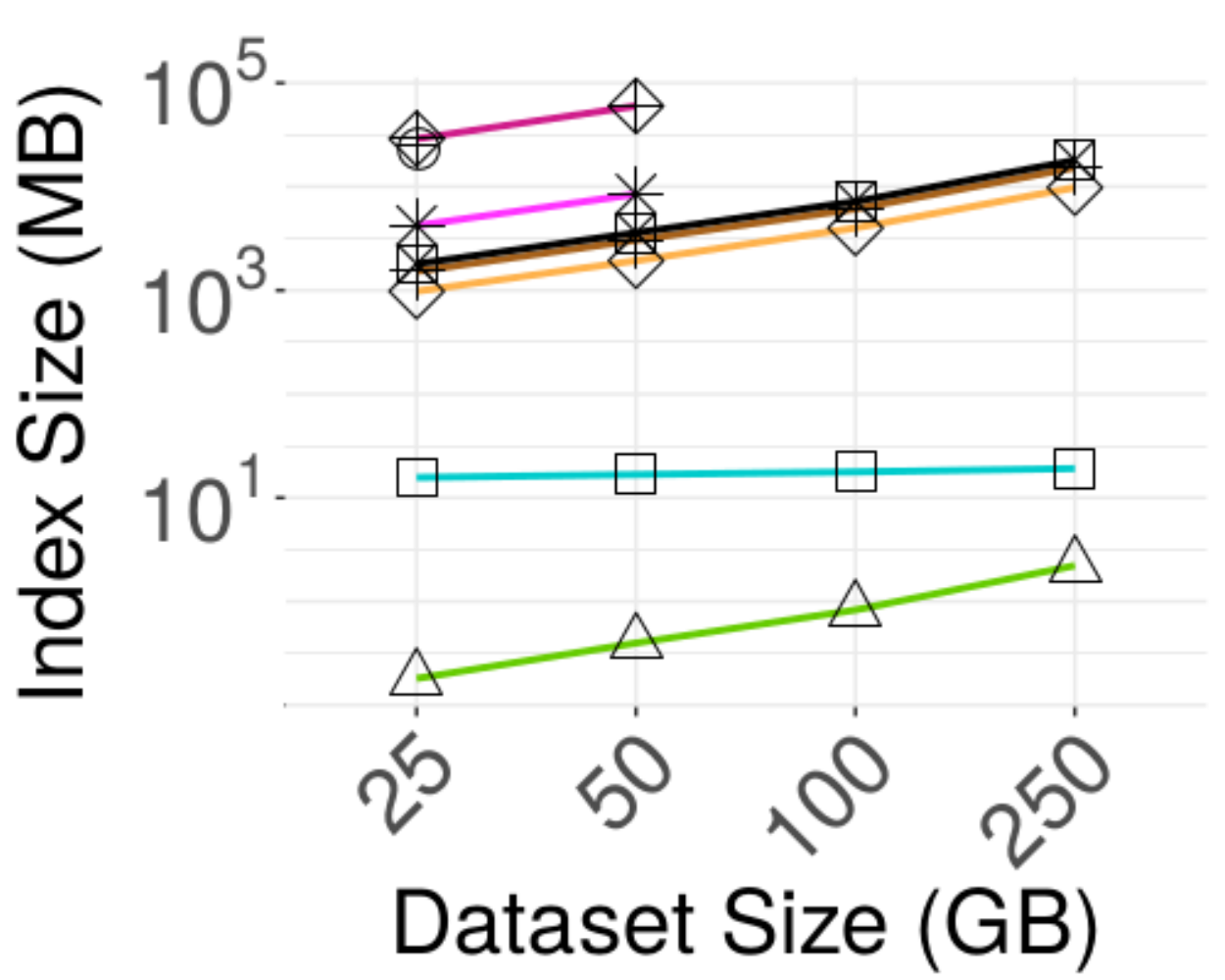}
		\caption{Size in memory}
		\label{fig:exact:datasize:memory:indexing:cache}
	\end{subfigure}
	\caption{{\color{black} Comparison of indexing scalability}}	
	\label{fig:exact:datasize:time:memory:indexing:combined:cache}
\end{figure}

\subsubsection{\textbf{Query Answering Efficiency and Accuracy: in-Memory Datasets}}
\label{ssec:query_efficiency_mem}

{\color{black}
	We now compare query answering efficiency and accuracy, in addition to the indexing time, thus, measuring how well each method amortizes index construction time over a large number of queries, and the level of accuracy achieved.
}

\noindent\textbf{Summary.} 
For our in-memory experiments, we used four datasets of 25GB each: two synthetic (with series of length 256 and 16384, respectively), and two real: Deep25GB and Sift25GB. 
We ran 1NN, 10NN and {\color{black} 100NN} queries on the four datasets and we observed that, while the running times increase with k, the relative performance of the methods stays the same. 
Due to lack of space, Figure~\ref{fig:approx:accuracy:efficiency:synthetic:25GB:inmemory:hdd} shows the 100NN query results only (full results are in~\cite{url/DSSeval2}), which we discuss below. 
{\color{black} Note that  HNSW, QALSH and FLANN store all raw data in-memory, 
while all other approaches use the memory to store their data structures, but read the raw data from disk; IMI does not access the raw data at all (it only uses the in-memory summaries).}

\noindent\textbf{Short Series}. 
For $ng$-approximate queries of length $256$ on the Rand25GB dataset, HNSW has the largest throughput for any given accuracy, followed by {\color{black}FLANN}, IMI, DSTree and iSAX2+ (Figure~\ref{fig:approx:accuracy:qefficiency:synthetic:25GB:256:hdd:ng:100NN:100:nocache}). 
However, HNSW does not reach a MAP of 1, which is only obtained by the data series indexes (DSTree, iSAX2+, VA+file). The skip-sequential method VA+file performs poorly on approximate search since it prunes per series and not per cluster like the tree-based methods do. 
When indexing time is also considered, iSAX2+ wins for the workload consisting of 100 queries (Figure~\ref{fig:approx:accuracy:efficiency:synthetic:25GB:256:hdd:ng:100NN:100:nocache}), and DSTree for the 10K queries (Figure~\ref{fig:approx:accuracy:efficiency:synthetic:25GB:256:hdd:ng:100NN:10K:nocache}). 

Regarding $\delta$-$\epsilon$-approximate search, DSTree offers the best throughput/accuracy tradeoff, followed by iSAX2+, SRS, {\color{black}VA+file and finally QALSH}. 
SRS does not achieve a MAP higher than 0.5, while DSTree and iSAX2+ are at least 3 times faster than SRS for a similar accuracy (Figure~\ref{fig:approx:accuracy:qefficiency:synthetic:25GB:256:hdd:de:100NN:100:nocache}). 
When we consider the combined indexing and querying times, iSAX2+ wins over all methods for 100 queries (Figure~\ref{fig:approx:accuracy:efficiency:synthetic:25GB:256:hdd:de:100NN:100:nocache}), and DSTree wins for 10K queries (Figure~\ref{fig:approx:accuracy:efficiency:synthetic:25GB:256:hdd:de:100NN:10K:nocache}).

\begin{figure*}[!htb]
	\captionsetup{justification=centering}
	\captionsetup[subfigure]{justification=centering}
	\begin{subfigure}{\textwidth}
		\centering
		\includegraphics[scale=0.18]{{full_epsilon_legend_25GB}}\\
	\end{subfigure}	
	\begin{subfigure}{0.16\textwidth}
		\centering
		\includegraphics[width=\textwidth]{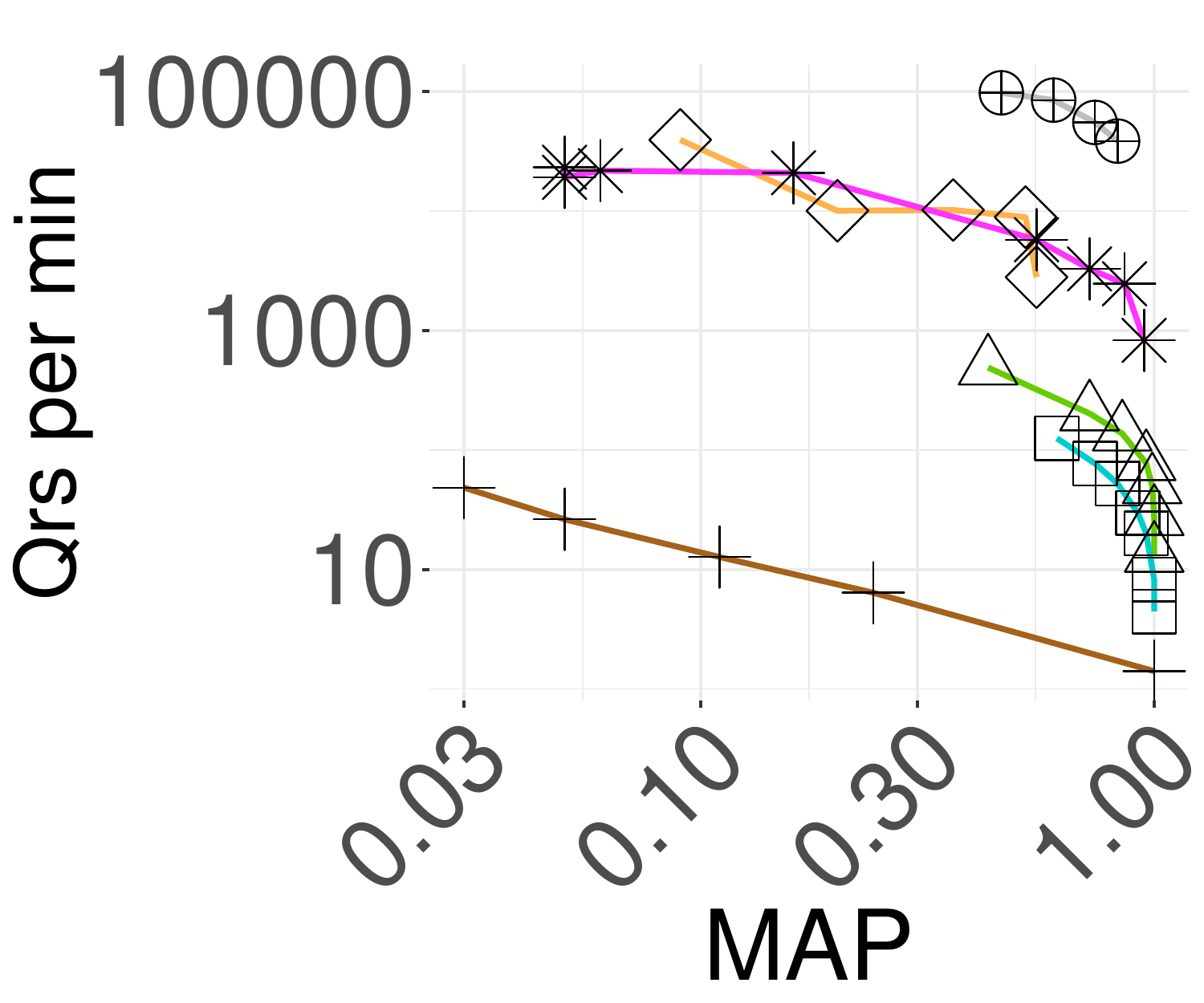}
		\scriptsize \caption{Rand25GB\\256 (ng)} 
		\label{fig:approx:accuracy:qefficiency:synthetic:25GB:256:hdd:ng:100NN:100:nocache}
	\end{subfigure}
	\begin{subfigure}{0.16\textwidth}
		\centering
		\includegraphics[width=\textwidth]{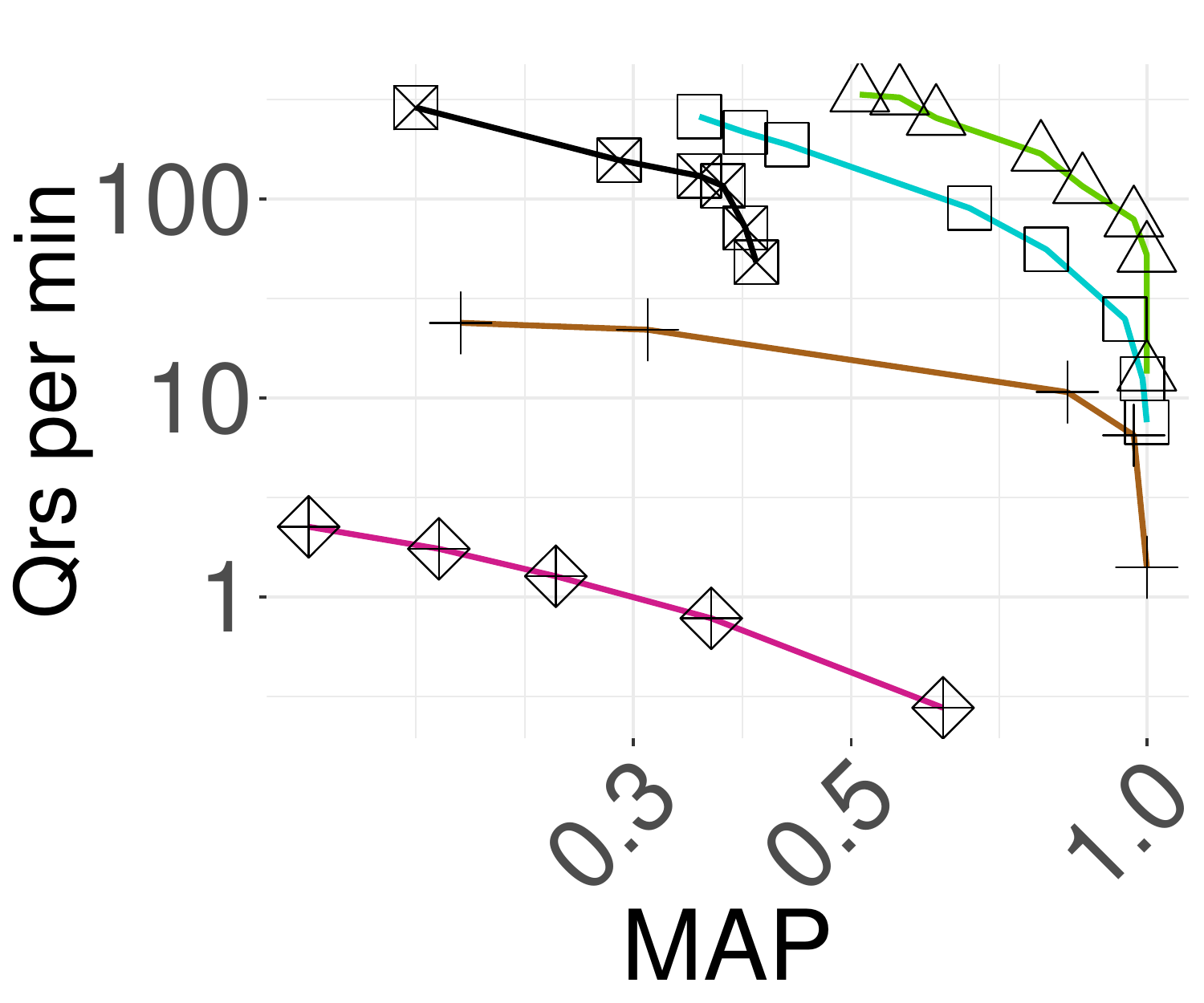}
		\scriptsize \caption{Rand25GB\\256 ($\bm{\delta\epsilon}$)} 
		\label{fig:approx:accuracy:qefficiency:synthetic:25GB:256:hdd:de:100NN:100:nocache}
	\end{subfigure}
	\begin{subfigure}{0.16\textwidth}
		\centering
		\includegraphics[width=\textwidth]{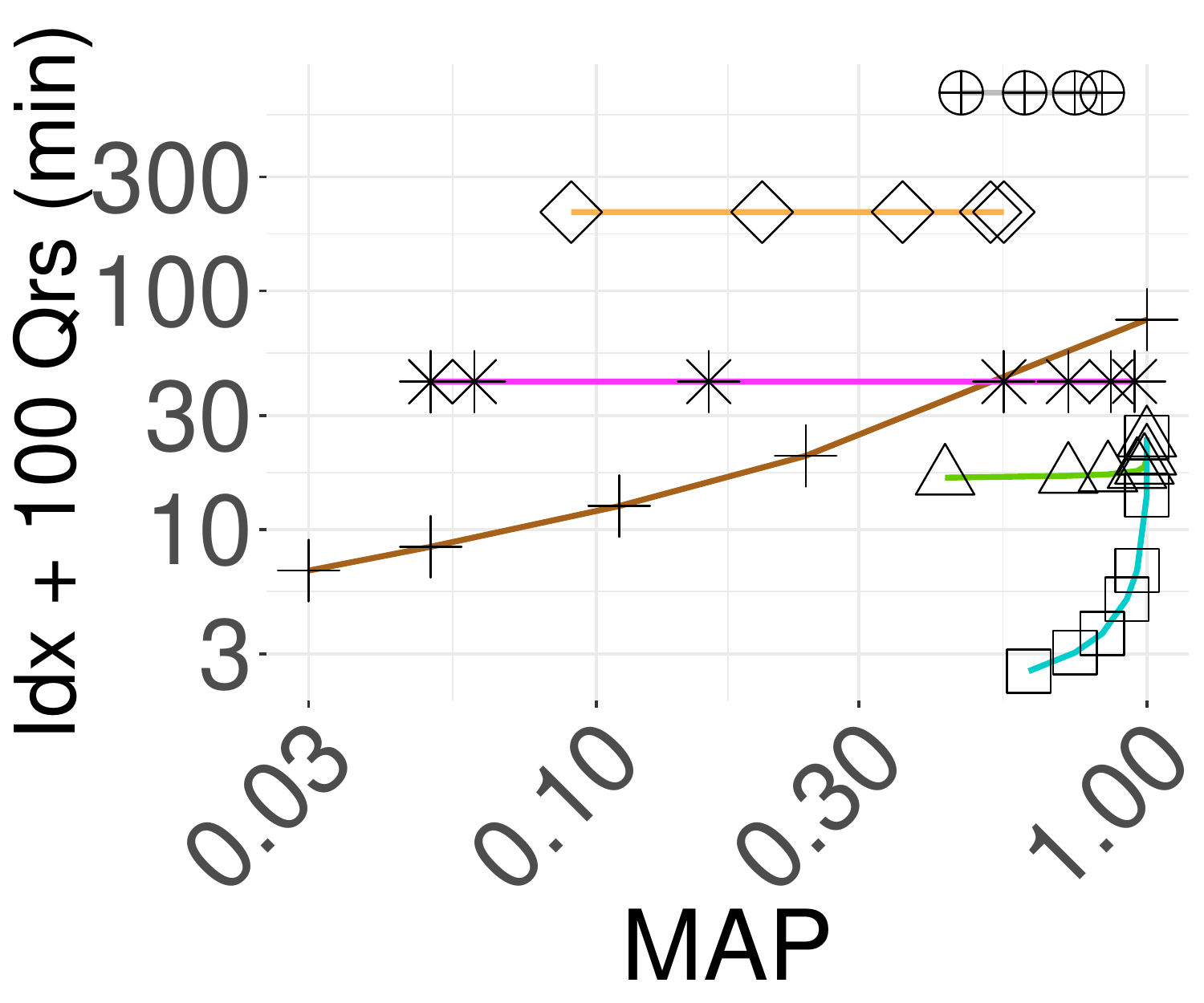}
		\scriptsize \caption{Rand25GB\\256 (ng)} 
		\label{fig:approx:accuracy:efficiency:synthetic:25GB:256:hdd:ng:100NN:100:nocache}
	\end{subfigure}
	\begin{subfigure}{0.16\textwidth}
		\centering
		\includegraphics[width=\textwidth]{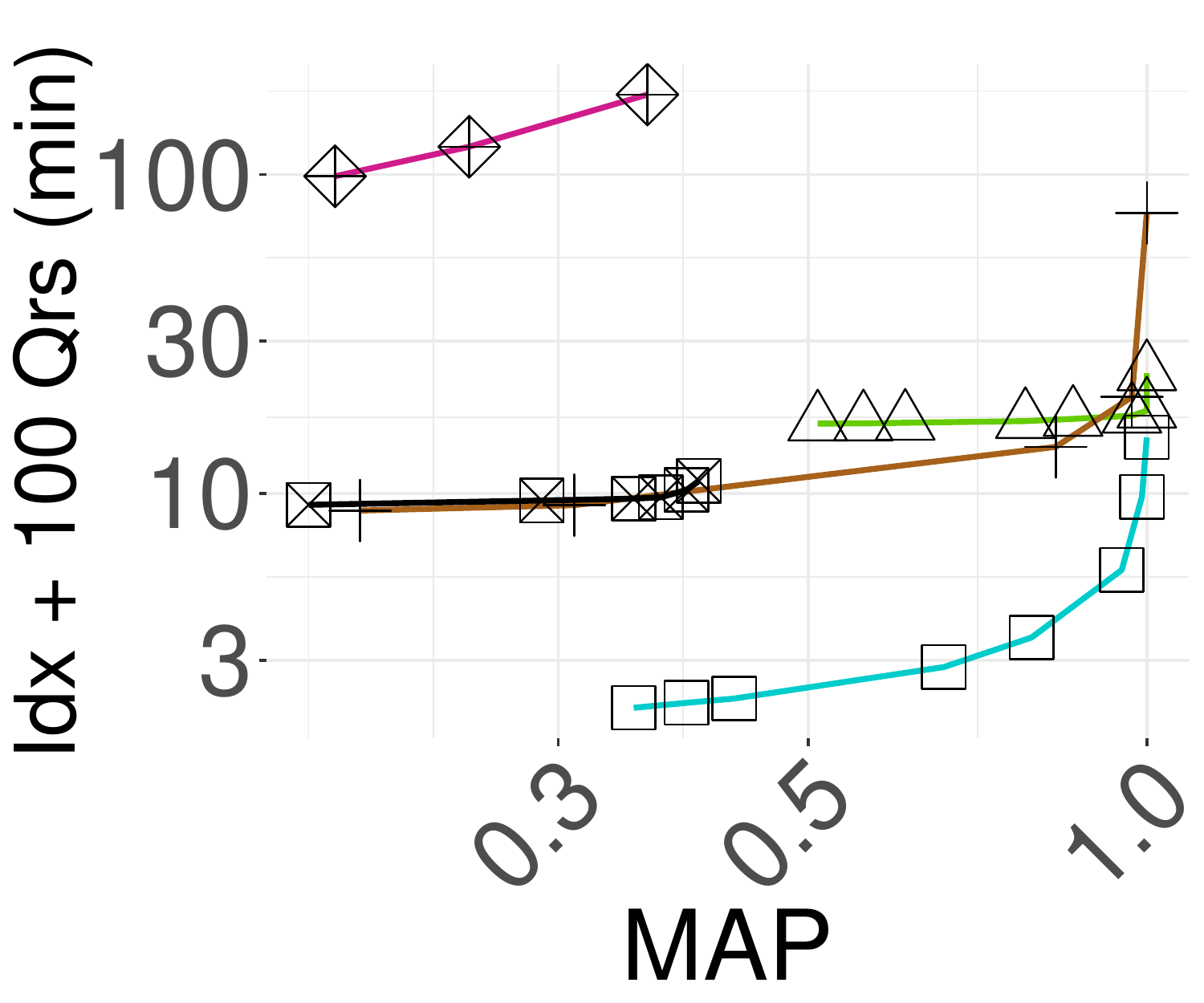}
		\scriptsize \caption{Rand25GB\\256 ($\bm{\delta\epsilon}$)} 
		\label{fig:approx:accuracy:efficiency:synthetic:25GB:256:hdd:de:100NN:100:nocache}
	\end{subfigure}
	\begin{subfigure}{0.16\textwidth}
		\centering
		\includegraphics[width=\textwidth]{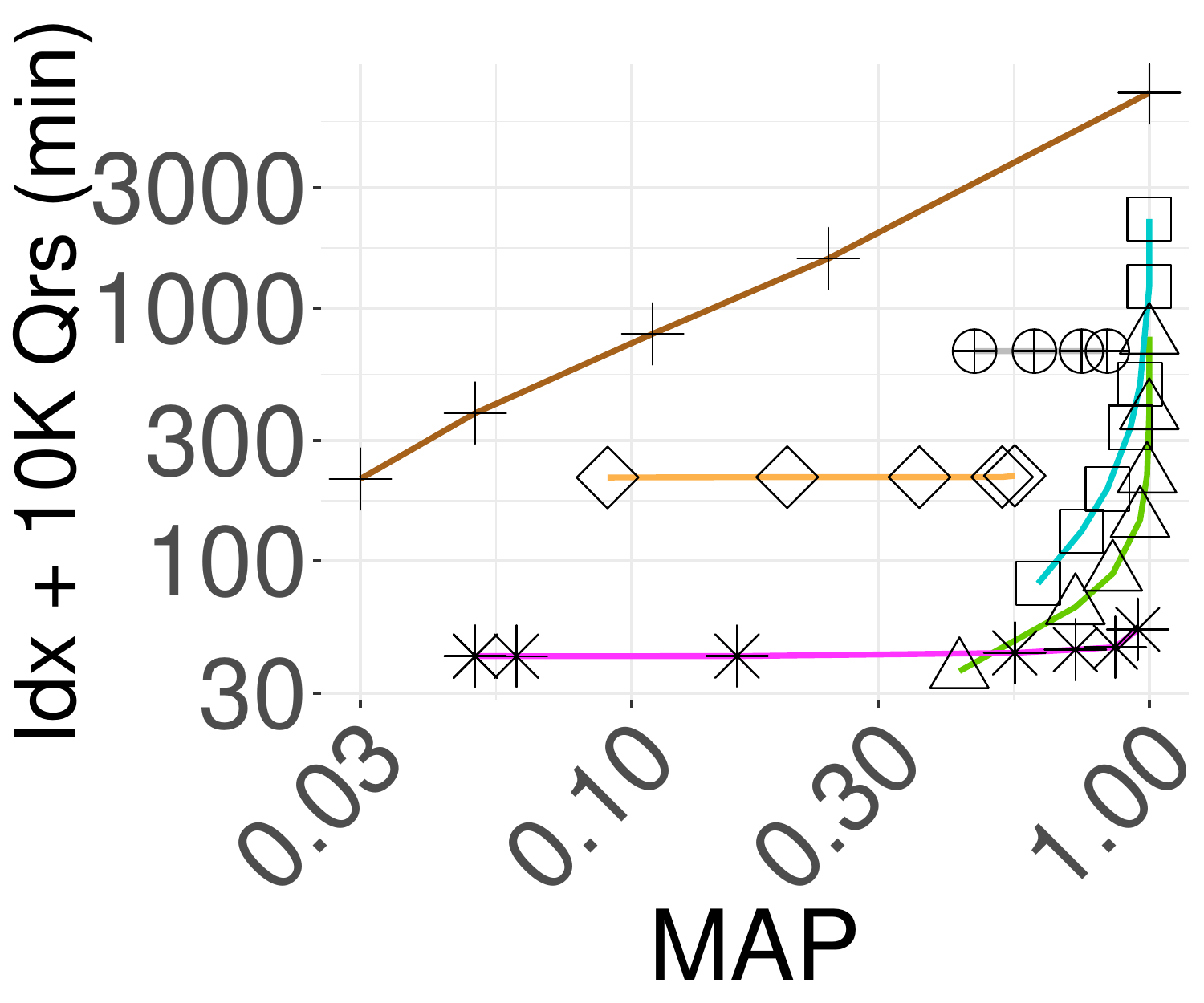}
		\scriptsize \caption{Rand25GB\\256 (ng)} 
		\label{fig:approx:accuracy:efficiency:synthetic:25GB:256:hdd:ng:100NN:10K:nocache}
	\end{subfigure}
	\begin{subfigure}{0.16\textwidth}
		\centering
		\includegraphics[width=\textwidth]{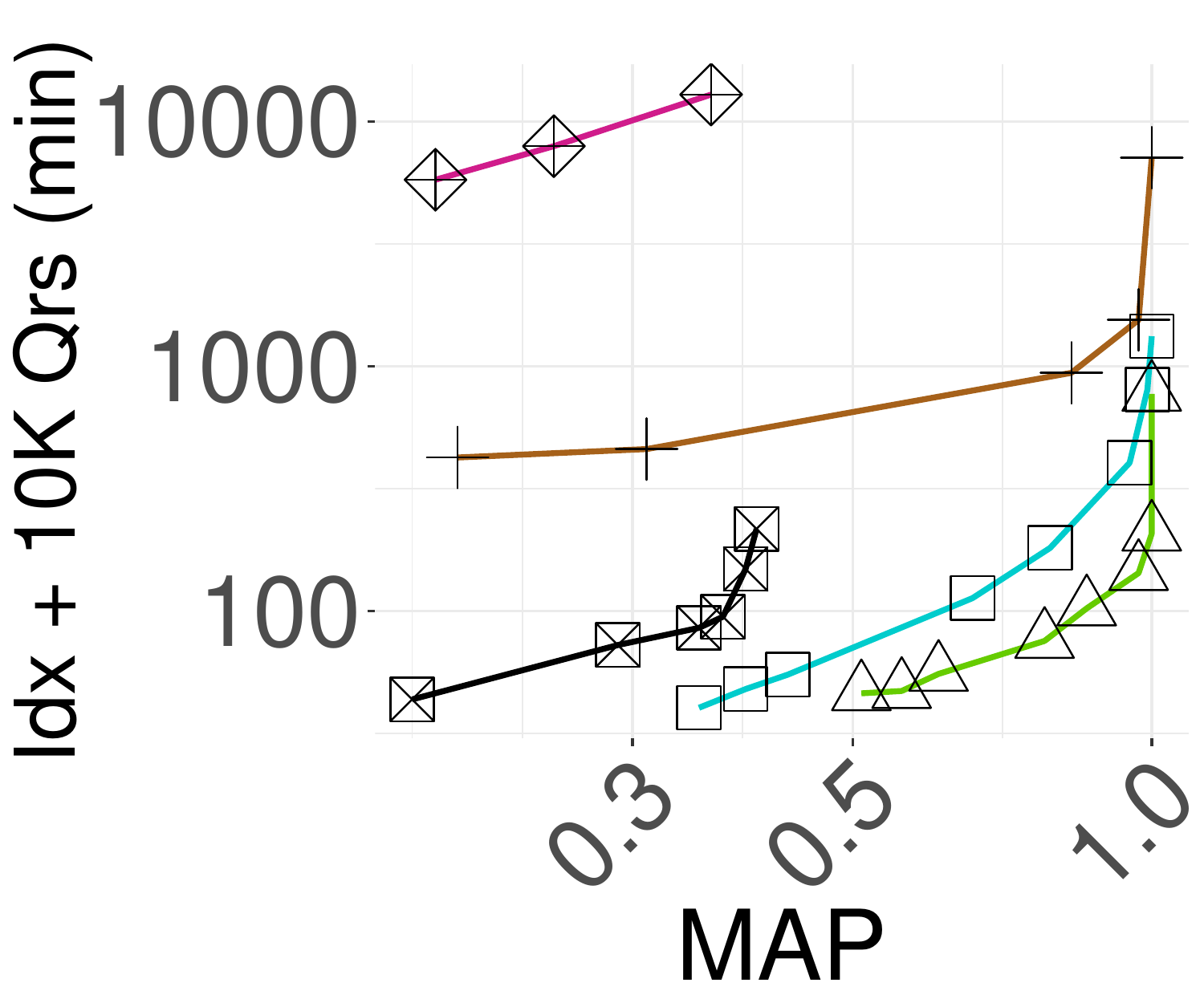}
		\scriptsize \caption{Rand25GB\\256 ($\bm{\delta\epsilon}$)} 
		\label{fig:approx:accuracy:efficiency:synthetic:25GB:256:hdd:de:100NN:10K:nocache}
	\end{subfigure}
	\begin{subfigure}{0.16\textwidth}
		\centering
		\includegraphics[width=\textwidth]{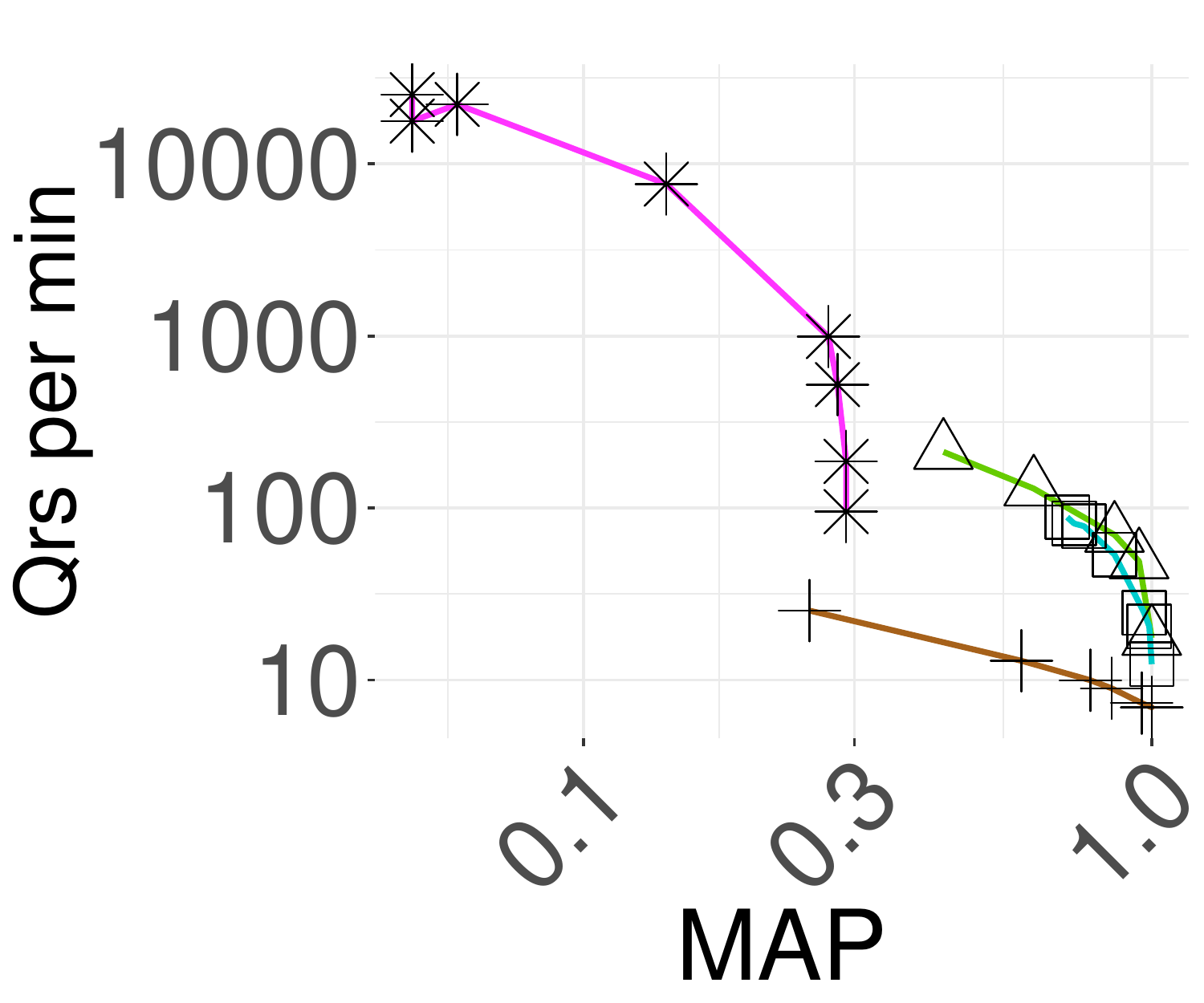}
		\scriptsize \caption{Rand25GB\\16384 (ng)} 
		\label{fig:approx:accuracy:qefficiency:synthetic:25GB:16384:ng:hdd:100NN:100:nocache}
	\end{subfigure}
	\begin{subfigure}{0.16\textwidth}
		\centering
		\includegraphics[width=\textwidth]{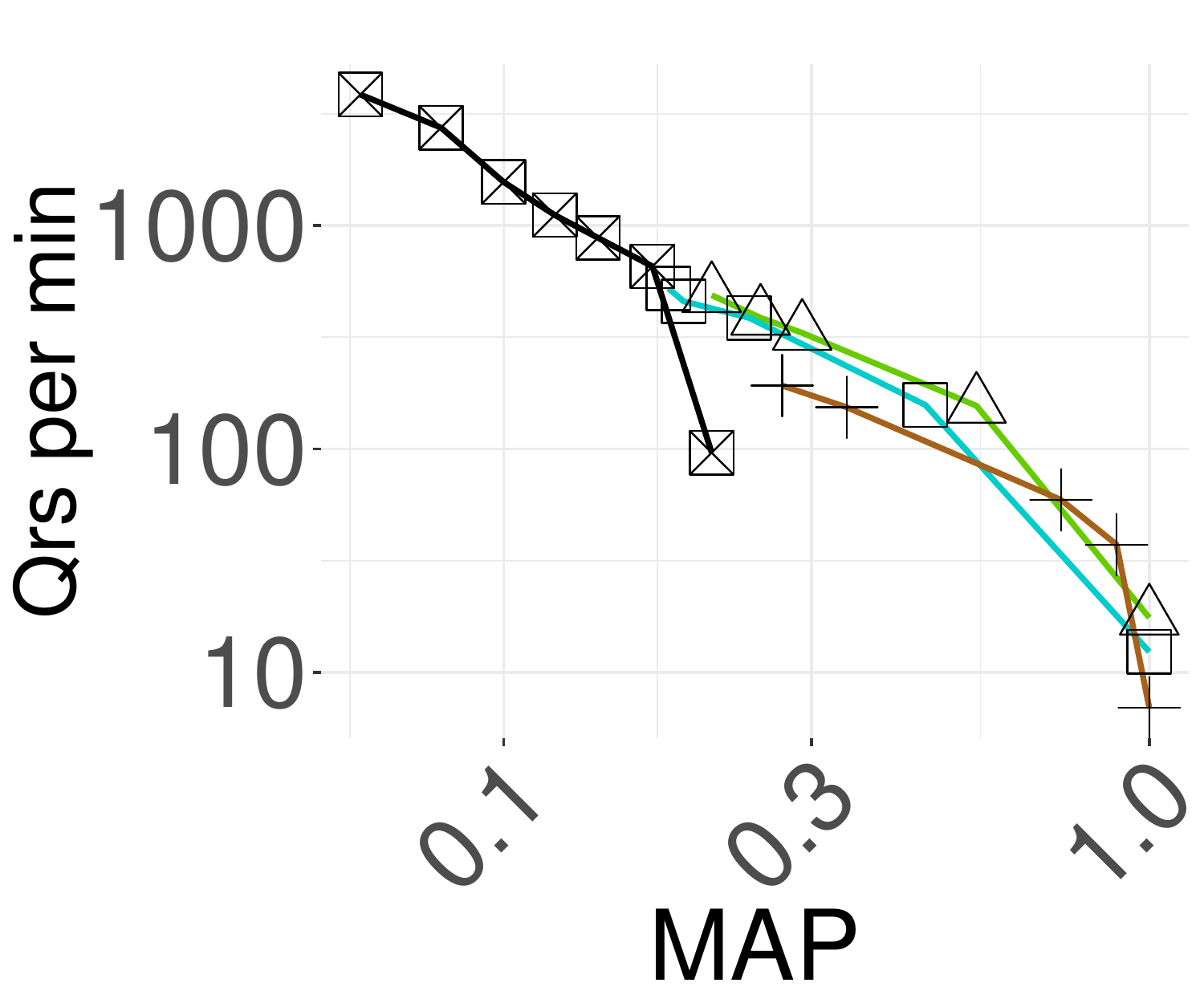}
		\scriptsize \caption{Rand25GB\\16384 ($\bm{\delta\epsilon}$)} 
		\label{fig:approx:accuracy:qefficiency:synthetic:25GB:16384:de:hdd:100NN:100:nocache}
	\end{subfigure}
	\begin{subfigure}{0.16\textwidth}
		\centering
		\includegraphics[width=\textwidth]{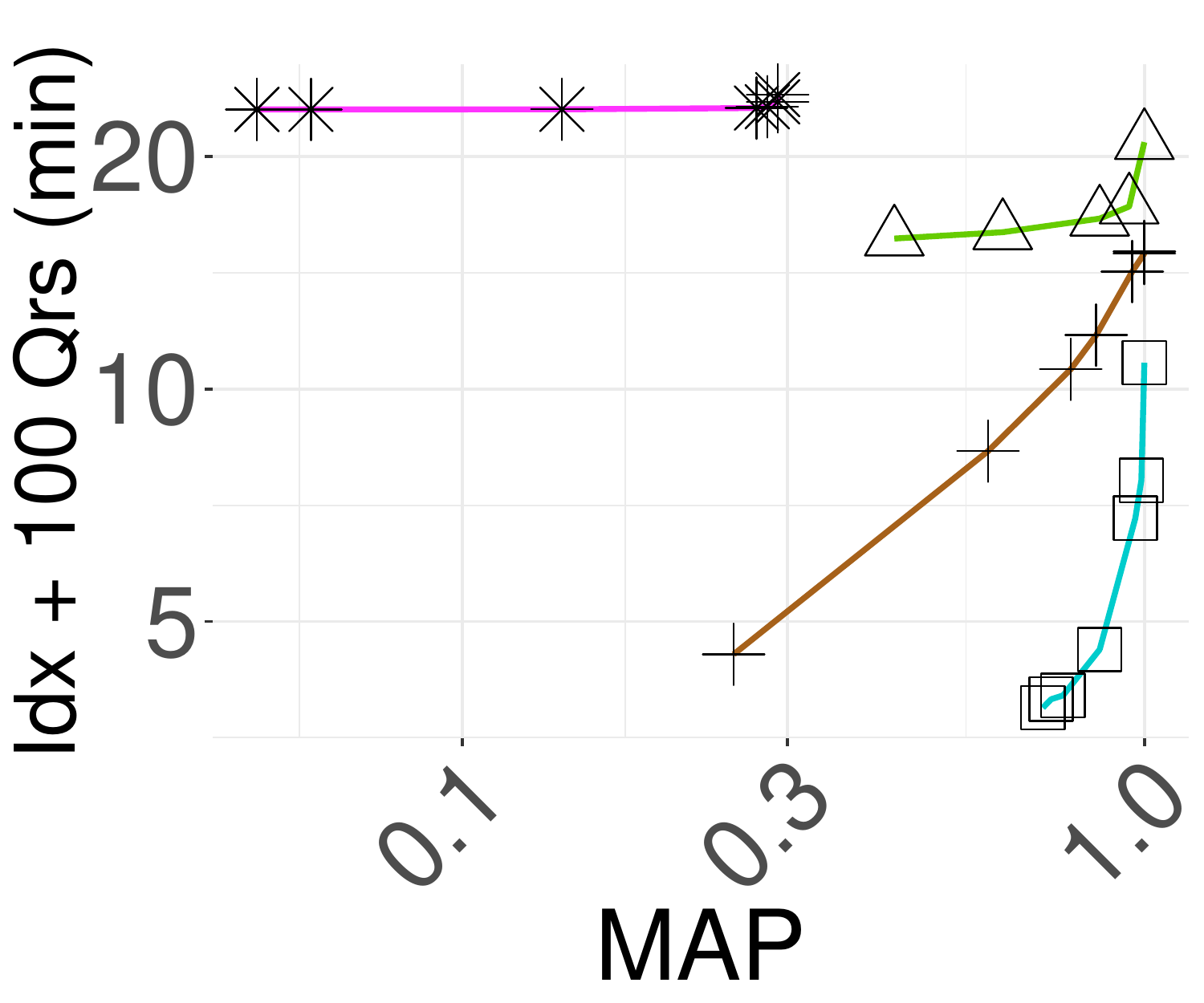}
		\scriptsize \caption{Rand25GB\\16384 (ng)} 
		\label{fig:approx:accuracy:efficiency:synthetic:25GB:16384:ng:hdd:100NN:100:nocache}
	\end{subfigure}
	\begin{subfigure}{0.16\textwidth}
		\centering
		\includegraphics[width=\textwidth]{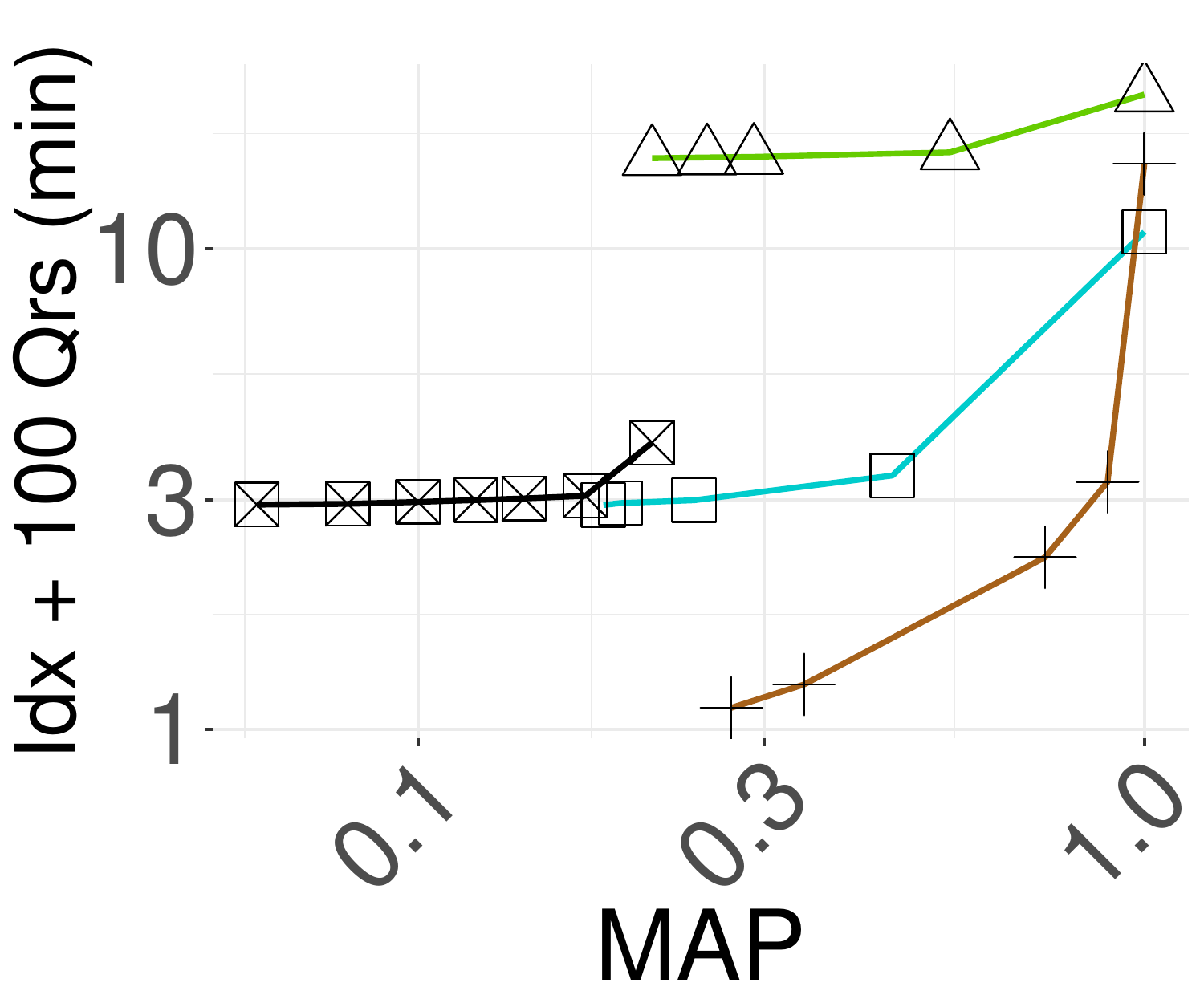}
		\scriptsize \caption{Rand25GB\\16384 ($\bm{\delta\epsilon}$)} 
		\label{fig:approx:accuracy:efficiency:synthetic:25GB:16384:de:hdd:100NN:100:nocache}
	\end{subfigure}
	\begin{subfigure}{0.16\textwidth}
		\centering
		\includegraphics[width=\textwidth]{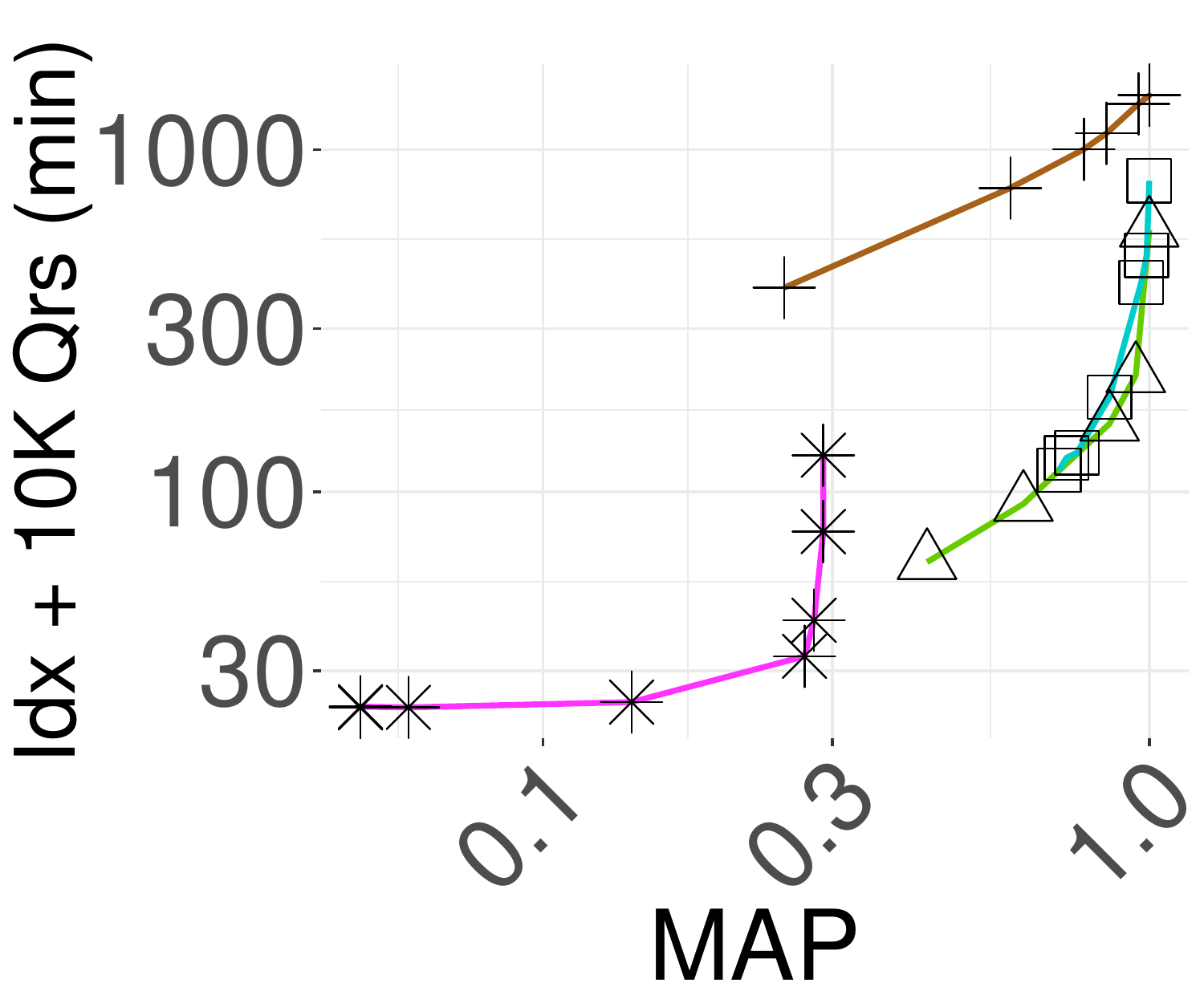}
		\scriptsize \caption{Rand25GB\\16384 (ng)} 
		\label{fig:approx:accuracy:efficiency:synthetic:25GB:16384:ng:hdd:100NN:10K:nocache}
	\end{subfigure}
	\begin{subfigure}{0.16\textwidth}
		\centering
		\includegraphics[width=\textwidth]{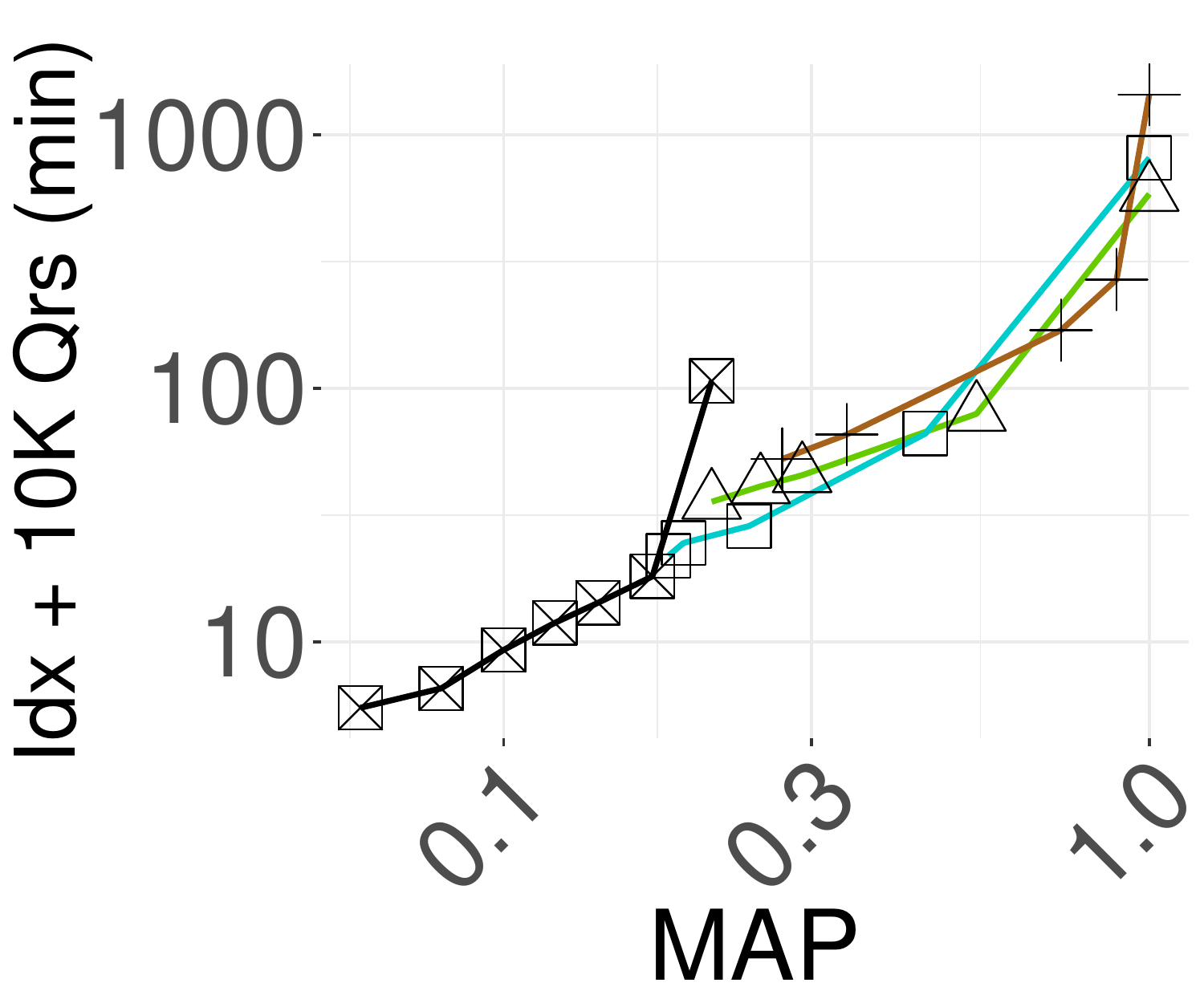}
		\scriptsize \caption{Rand25GB\\16384($\bm{\delta\epsilon}$)} 
		\label{fig:approx:accuracy:efficiency:synthetic:25GB:16384:de:hdd:100NN:10K:nocache}
	\end{subfigure}
	\begin{subfigure}{0.16\textwidth}
		\centering
		\includegraphics[width=\textwidth]{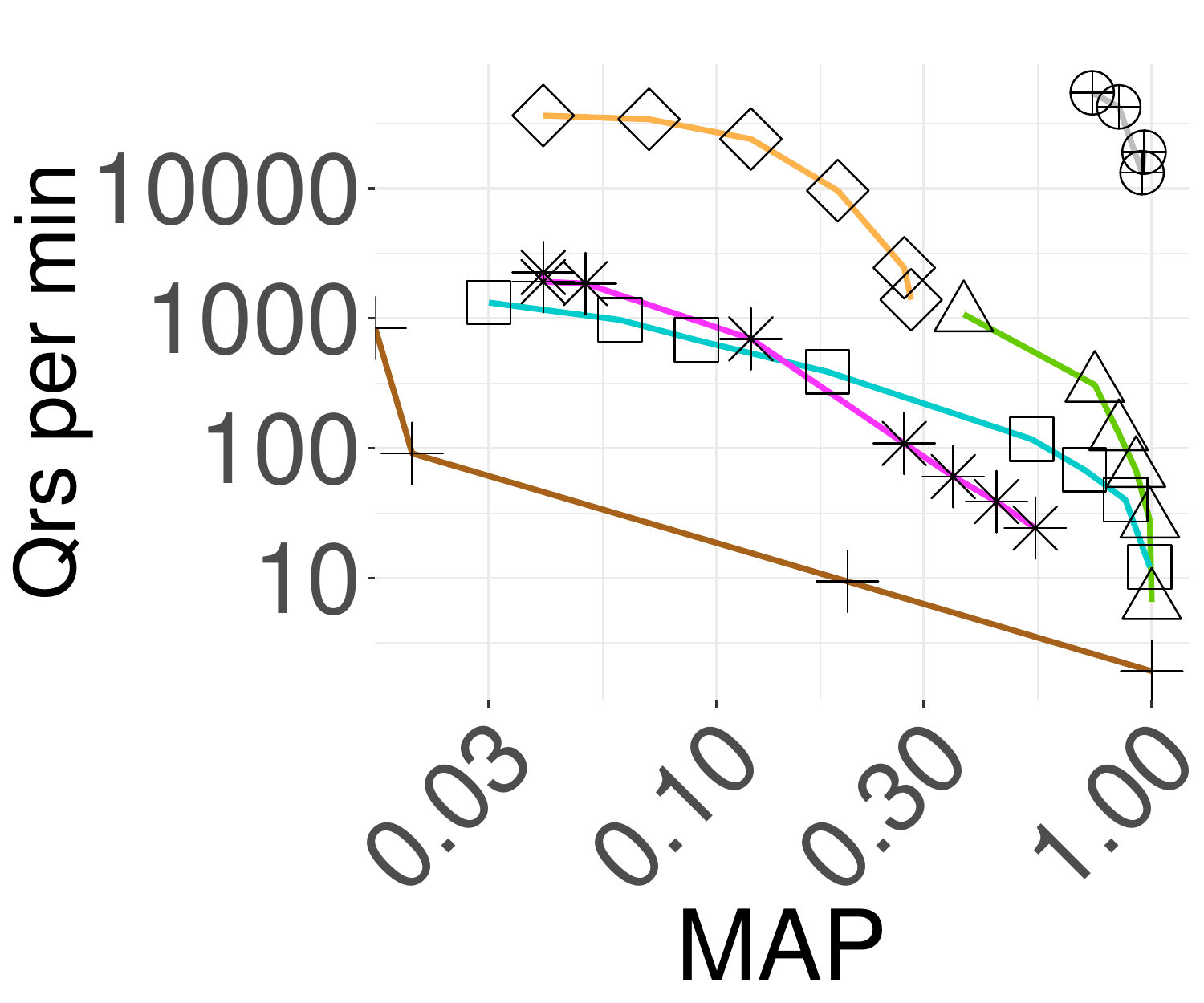}
		\scriptsize \caption{Sift25GB(ng)} 
		\label{fig:approx:accuracy:qefficiency:sift:25GB:ng:hdd:100NN:100:nocache}
	\end{subfigure}
	\begin{subfigure}{0.16\textwidth}
		\centering
		\includegraphics[width=\textwidth]{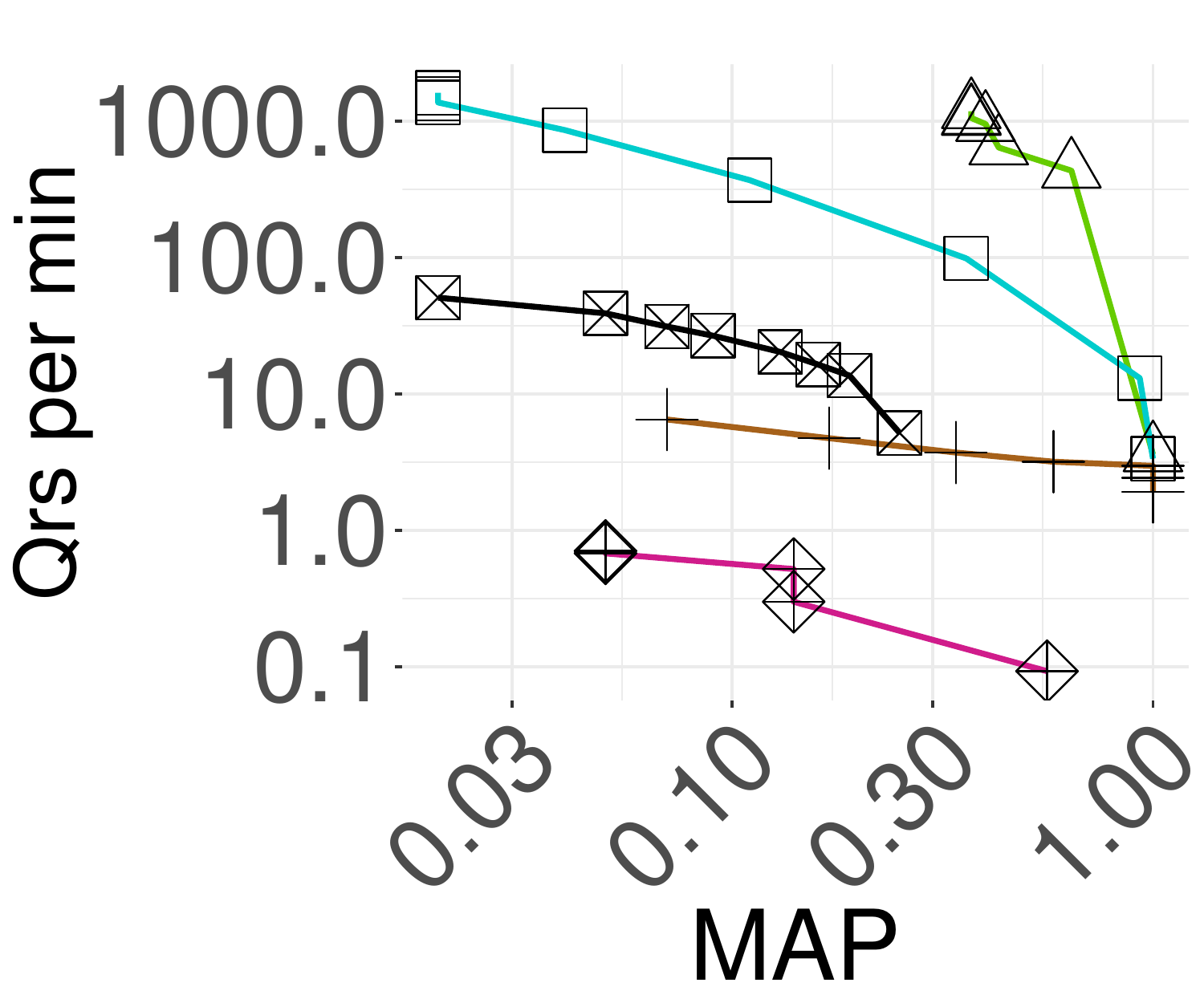}
		\scriptsize \caption{Sift25GB($\bm{\delta\epsilon}$)} 
		\label{fig:approx:accuracy:qefficiency:sift:25GB:de:hdd:100NN:100:nocache}
	\end{subfigure}
	\begin{subfigure}{0.16\textwidth}
		\centering
		\includegraphics[width=\textwidth]{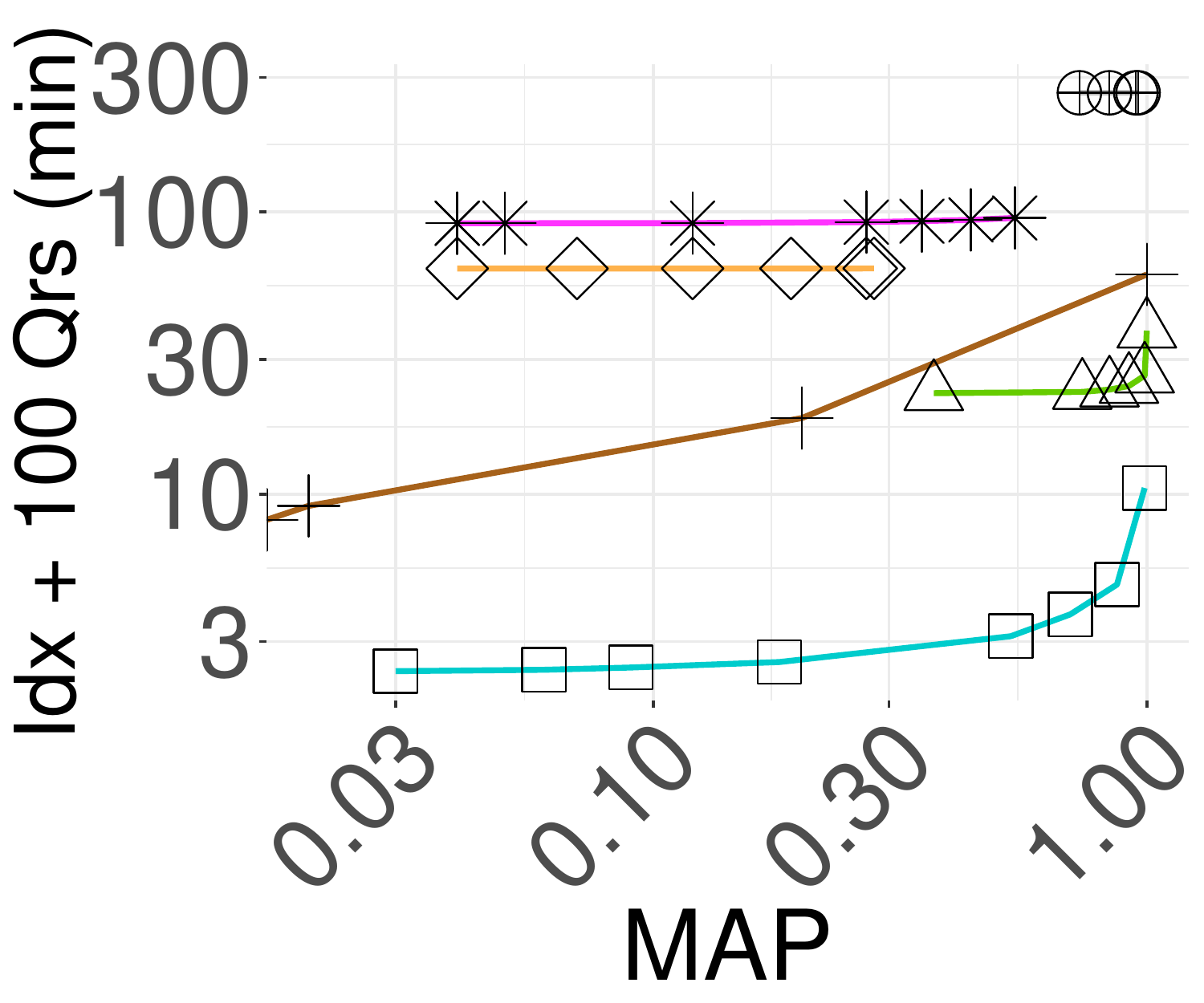}
		\scriptsize \caption{Sift25GB(ng)} 
		\label{fig:approx:accuracy:efficiency:sift:25GB:ng:hdd:100NN:100:nocache}
	\end{subfigure}
	\begin{subfigure}{0.16\textwidth}
		\centering
		\includegraphics[width=\textwidth]{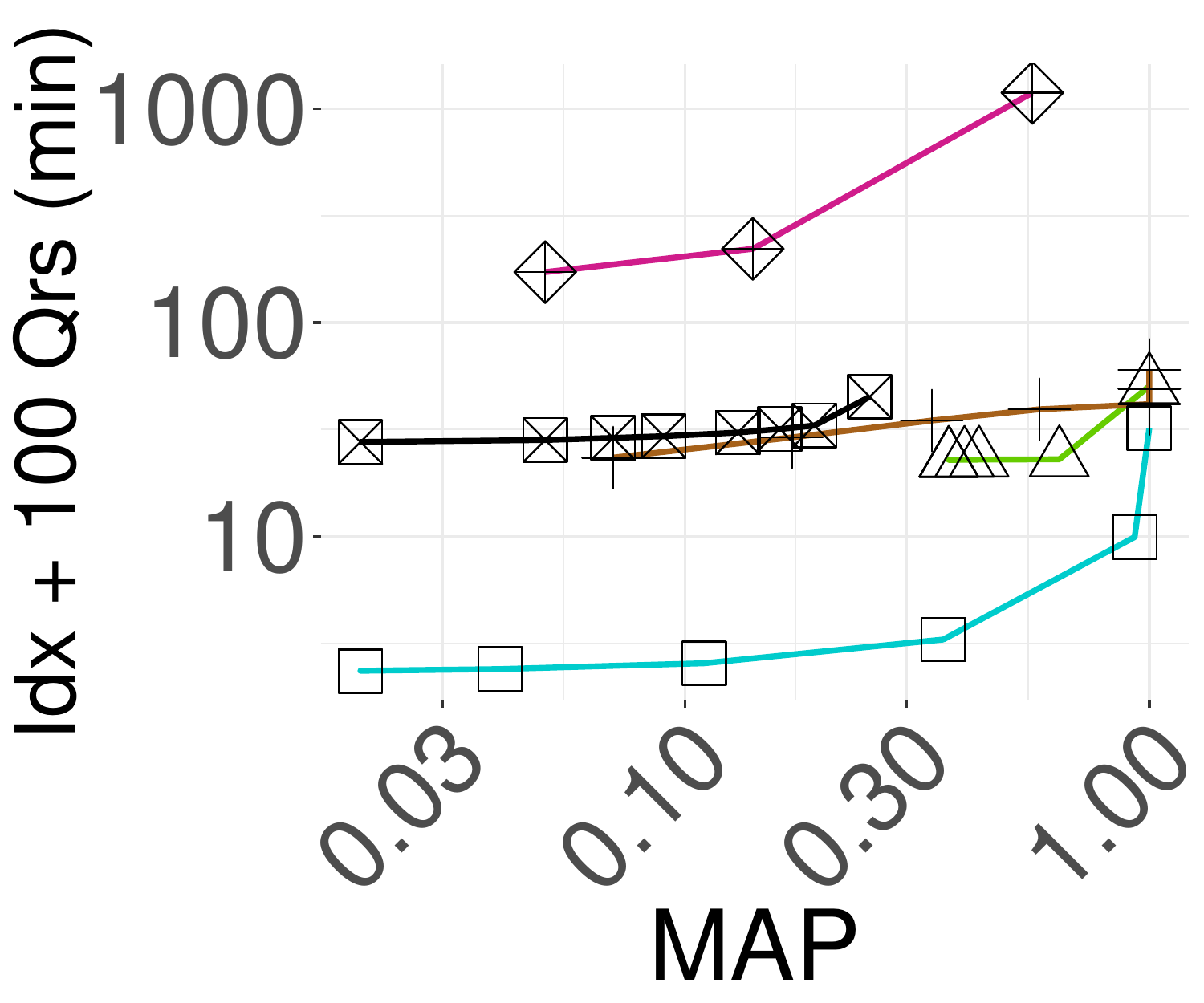}
		\scriptsize \caption{Sift25GB($\bm{\delta\epsilon}$)} 
		\label{fig:approx:accuracy:efficiency:sift:25GB:de:hdd:100NN:100:nocache}
	\end{subfigure}
	\begin{subfigure}{0.16\textwidth}
		\centering
		\includegraphics[width=\textwidth]{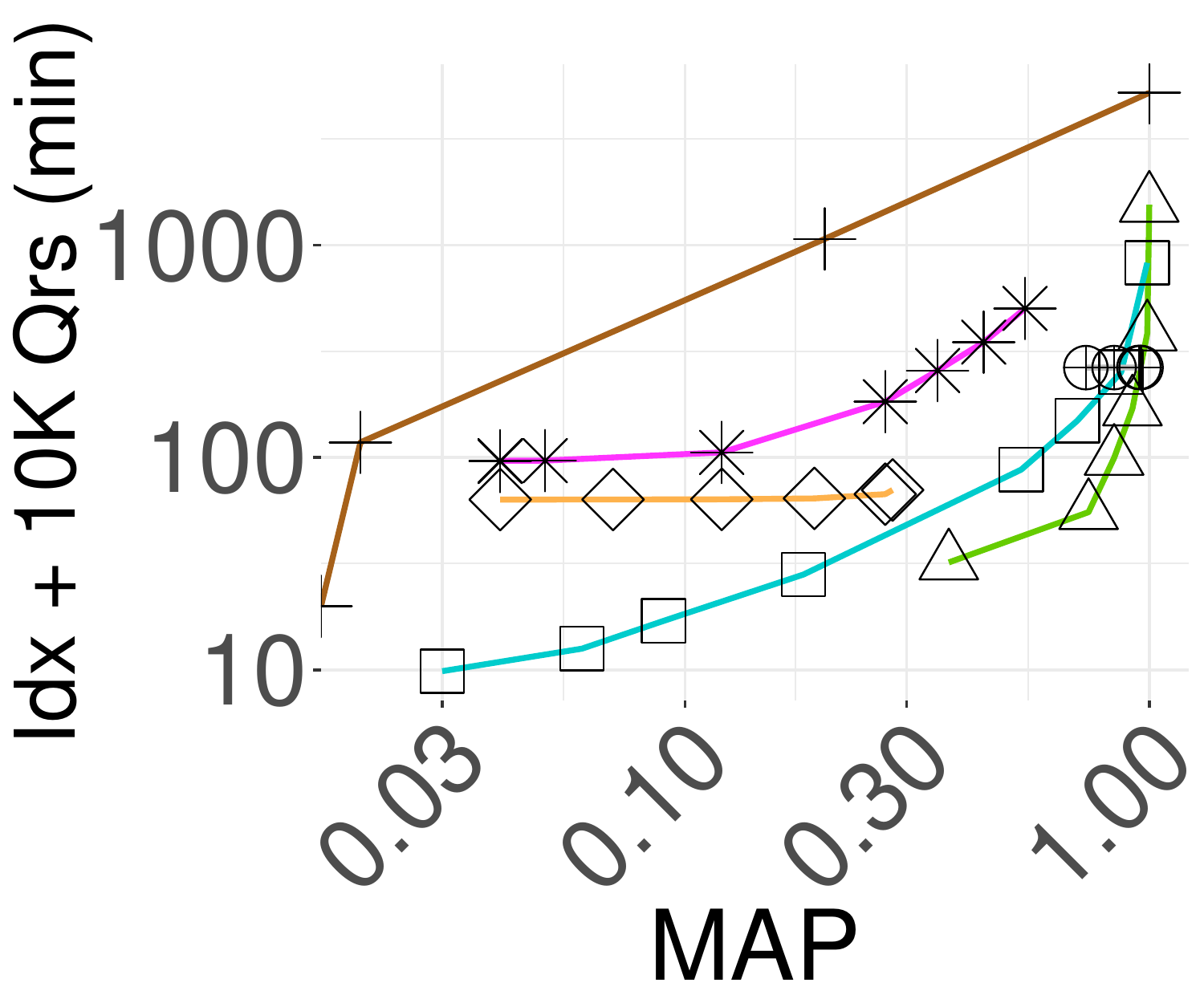}
		\scriptsize \caption{Sift25GB(ng)} 
		\label{fig:approx:accuracy:efficiency:sift:25GB:ng:hdd:100NN:10K:nocache}
	\end{subfigure}
	\begin{subfigure}{0.16\textwidth}
		\centering
		\includegraphics[width=\textwidth]{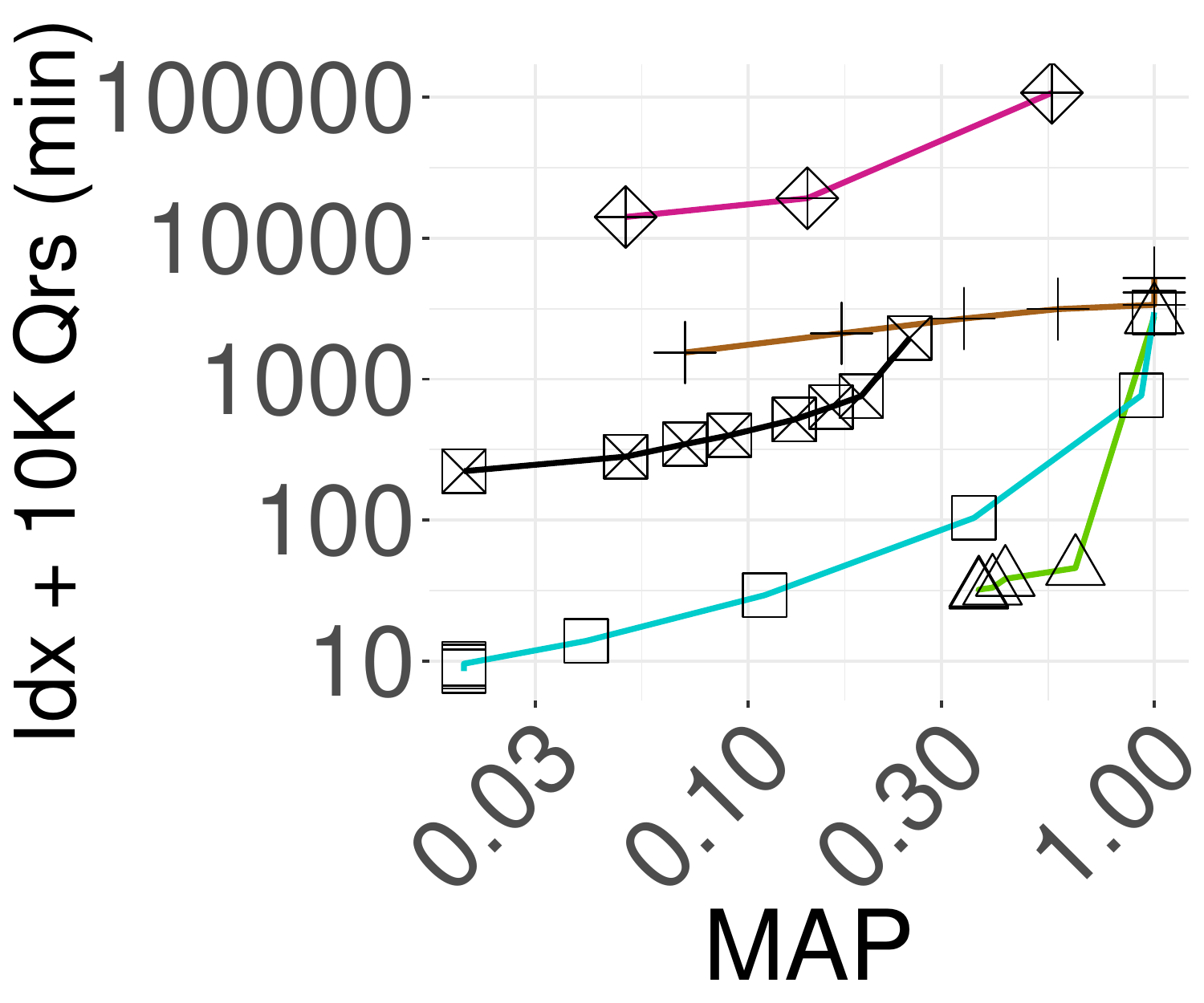}
		\scriptsize \caption{Sift25GB($\bm{\delta\epsilon}$)} 
		\label{fig:approx:accuracy:efficiency:sift:25GB:de:hdd:100NN:10K:nocache}
	\end{subfigure}
	\begin{subfigure}{0.16\textwidth}
		\centering
		\includegraphics[width=\textwidth]{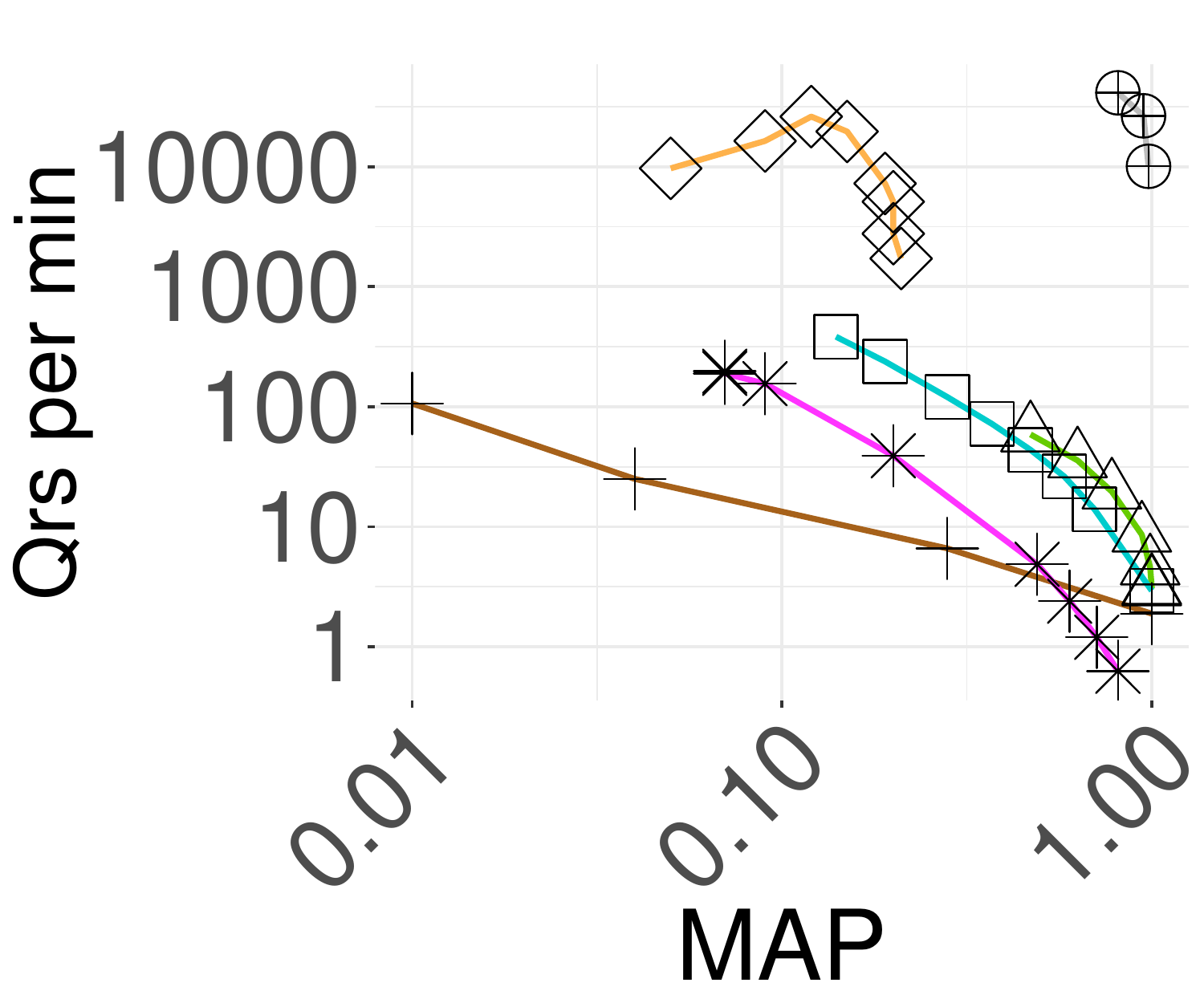}
		\scriptsize \caption{Deep25GB(ng)} 
		\label{fig:approx:accuracy:qefficiency:deep:25GB:96:hdd:ng:100NN:100:nocache}
	\end{subfigure}
	\begin{subfigure}{0.16\textwidth}
		\centering
		\includegraphics[width=\textwidth]{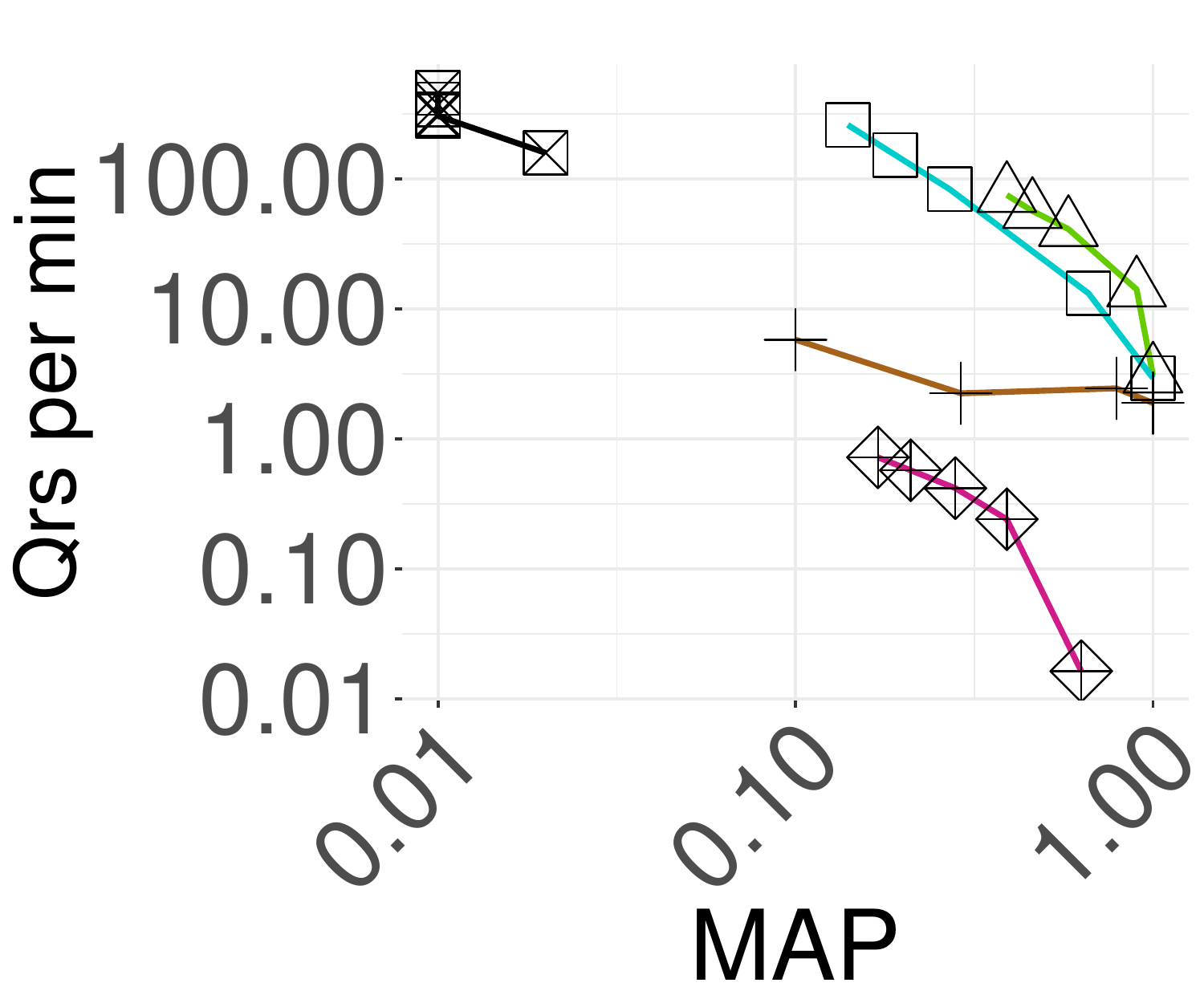}
		\scriptsize \caption{Deep25GB($\bm{\delta\epsilon}$)} 
		\label{fig:approx:accuracy:qefficiency:deep:25GB:96:hdd:de:100NN:100:nocache}
	\end{subfigure}
	\begin{subfigure}{0.16\textwidth}
		\centering
		\includegraphics[width=\textwidth]{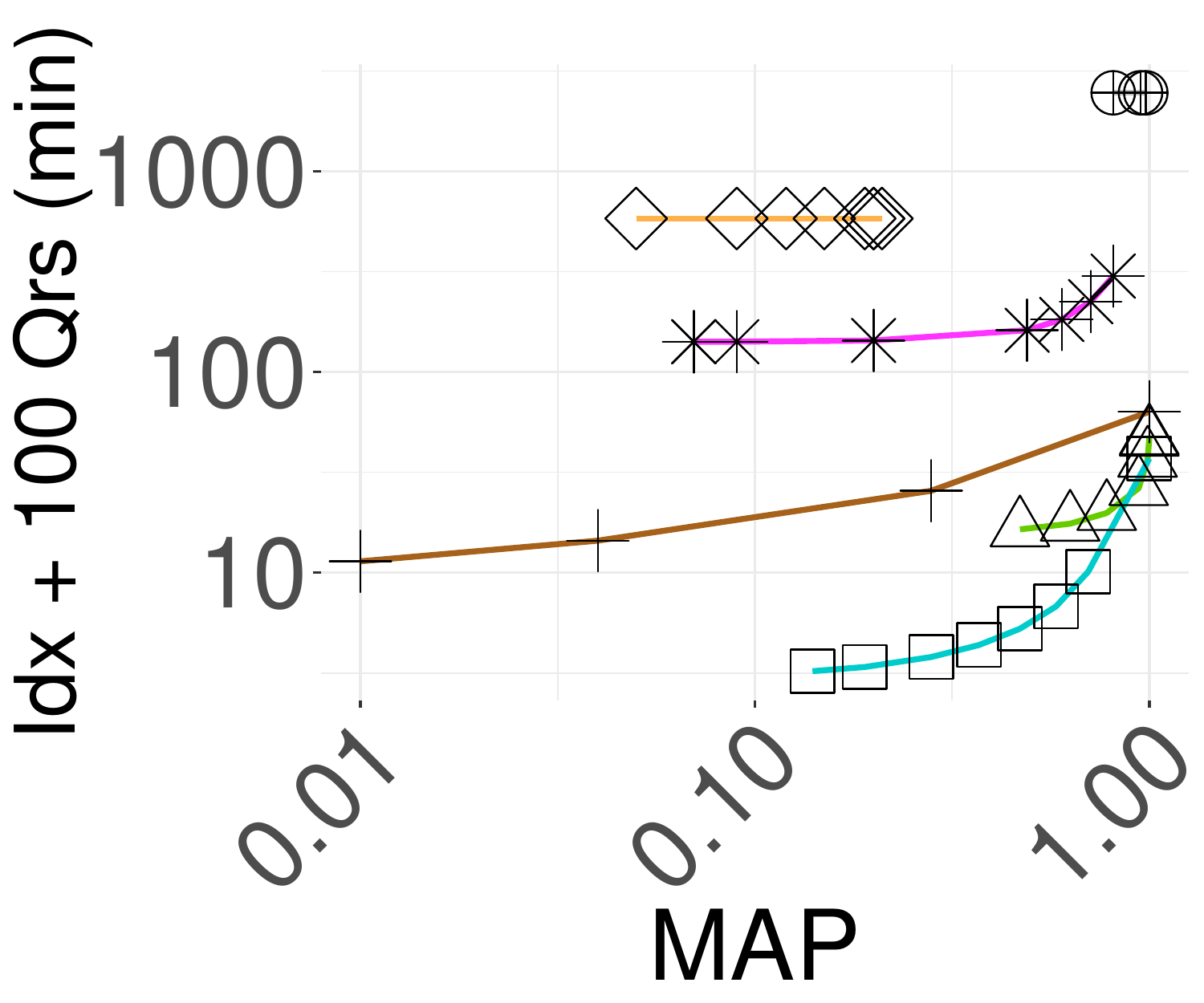}
		\scriptsize \caption{Deep25GB(ng)} 
		\label{fig:approx:accuracy:efficiency:deep:25GB:96:hdd:ng:100NN:100:nocache}
	\end{subfigure}
	\begin{subfigure}{0.16\textwidth}
		\centering
		\includegraphics[width=\textwidth]{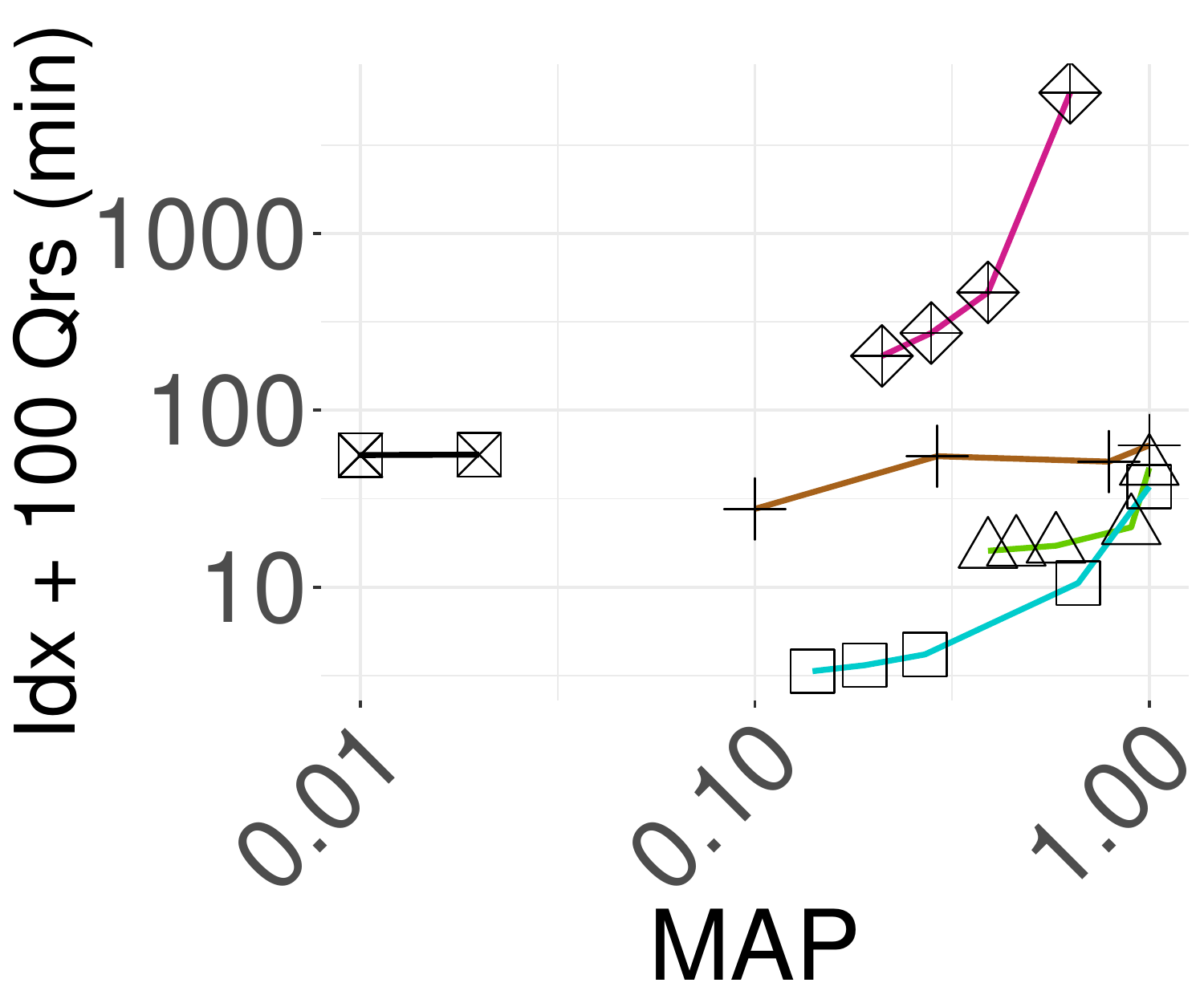}
		\scriptsize \caption{Deep25GB($\bm{\delta\epsilon}$)} 
		\label{fig:approx:accuracy:efficiency:deep:25GB:96:hdd:de:100NN:100:nocache}
	\end{subfigure}
	\begin{subfigure}{0.165\textwidth}
		\centering
		\includegraphics[width=\textwidth]{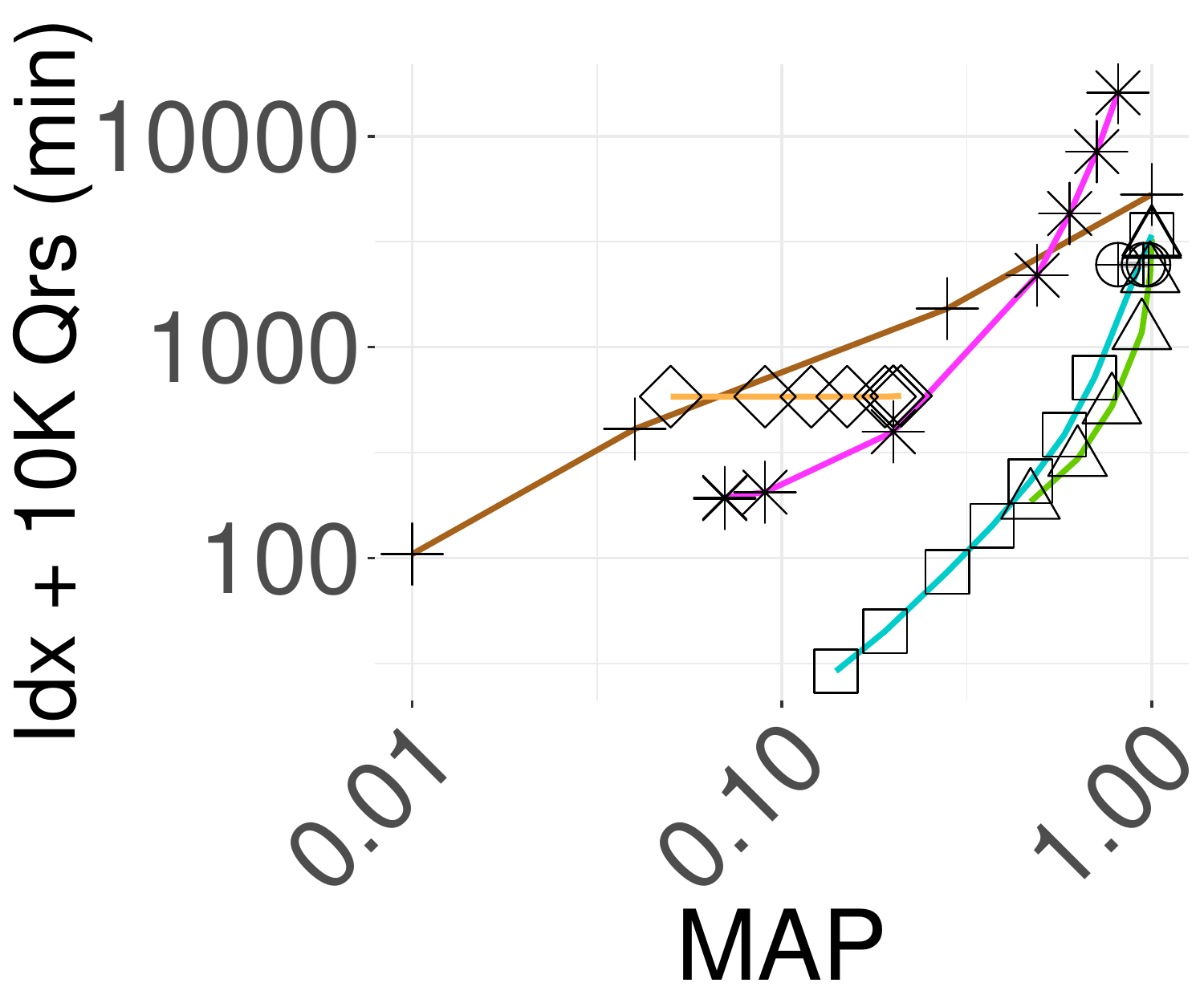}
		\scriptsize\caption{Deep25GB(ng)} 
		\label{fig:approx:accuracy:efficiency:deep:25GB:96:hdd:ng:100NN:10K:nocache}
	\end{subfigure}
	\begin{subfigure}{0.16\textwidth}
		\centering
		\includegraphics[width=\textwidth]{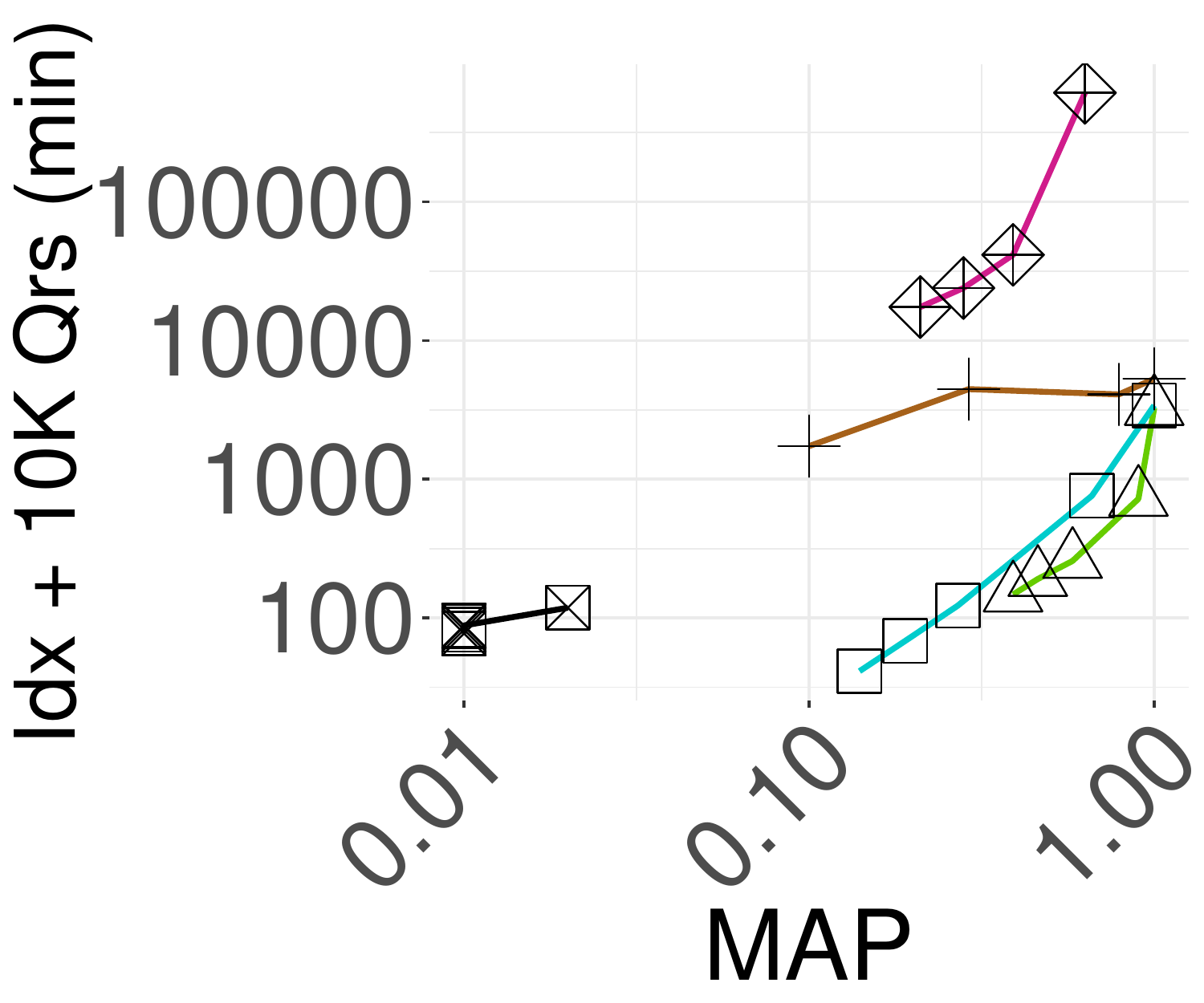}
		\scriptsize \caption{Deep25GB($\bm{\delta\epsilon}$)} 
		\label{fig:approx:accuracy:efficiency:deep:25GB:96:hdd:de:100NN:10K:nocache}
	\end{subfigure}	
	\vspace*{-0.2cm}
	\caption{{\color{black} Efficiency vs. accuracy in memory (100NN queries)}}	
	\vspace*{-0.2cm}
	\label{fig:approx:accuracy:efficiency:synthetic:25GB:inmemory:hdd}
\end{figure*}

\noindent\textbf{Long Series}. 
In this experiment, we use dataset sizes of 25GB, and query length of 16384.
For $ng$-approximate search, we report the results only for iSAX2+, DSTree and VA+file. We ran several experiments with IMI and HNSW building the indexes using different parameters, but obtained a MAP of 0 for IMI for all index configurations we tried, and ran into a segmentation fault during query answering with HNSW. 
DSTree outperforms both iSAX2+ and VA+file in terms of throughput and combined total cost for the larger workload (Figures~\ref{fig:approx:accuracy:qefficiency:synthetic:25GB:16384:ng:hdd:100NN:100:nocache} and~\ref{fig:approx:accuracy:efficiency:synthetic:25GB:16384:ng:hdd:100NN:10K:nocache}), whereas iSAX2+ wins for the smaller workload when the combined total cost is considered (Figure~\ref{fig:approx:accuracy:efficiency:synthetic:25GB:16384:ng:hdd:100NN:100:nocache}). {\color{black}We note also that the performance of FLANN deteriorates with the increased dimensionality.}

For $\delta$-$\epsilon$-approximate queries, Figure~\ref{fig:approx:accuracy:qefficiency:synthetic:25GB:16384:de:hdd:100NN:100:nocache} shows that DSTree and VA+file outperform all other methods for large MAP values, while DSTree and iSAX2+ have higher throughput for small MAP values. 
Note that the SRS accuracy decreases when compared to series of length 256, with the best MAP value now being 0.25. 
This is due to the increased information loss, as for both series lengths the number of dimensions in the projected space is 16. 
When index building time is considered, VA+file wins for the small workload (Figure~\ref{fig:approx:accuracy:efficiency:synthetic:25GB:16384:de:hdd:100NN:100:nocache}), and iSAX2+ and DSTree win for the large one (Figure~\ref{fig:approx:accuracy:efficiency:synthetic:25GB:16384:de:hdd:100NN:10K:nocache}). {\color{black}We do not report numbers for QALSH because the algorithm ran into a segmentation fault for series of length 16384.}

\noindent\textbf{Real Data}. 
We ran the same set of experiments with real datasets. 
For $ng$-approximate queries, HNSW outperforms the query performance of other methods by a large margin (Figures~\ref{fig:approx:accuracy:qefficiency:sift:25GB:ng:hdd:100NN:100:nocache} and~\ref{fig:approx:accuracy:qefficiency:deep:25GB:96:hdd:ng:100NN:100:nocache}). When indexing time is considered, HNSW loses its edge due to its high indexing cost to iSAX2+ when the query workload consists of 100 queries (Figures~\ref{fig:approx:accuracy:efficiency:sift:25GB:ng:hdd:100NN:100:nocache} and~\ref{fig:approx:accuracy:efficiency:deep:25GB:96:hdd:ng:100NN:100:nocache}) and to DSTree for the 10K workload (Figures~\ref{fig:approx:accuracy:efficiency:sift:25GB:ng:hdd:100NN:10K:nocache} and ~\ref{fig:approx:accuracy:efficiency:deep:25GB:96:hdd:ng:100NN:10K:nocache}). 
HNSW does not achieve a MAP of 1, while DSTree and ISAX2+ both do,  yet at a high cost. 

DSTree clearly wins on Sift25GB and Deep25GB among $\delta$-$\epsilon$-approximate methods (Figures~\ref{fig:approx:accuracy:qefficiency:sift:25GB:de:hdd:100NN:100:nocache}, ~\ref{fig:approx:accuracy:qefficiency:deep:25GB:96:hdd:de:100NN:100:nocache}, ~\ref{fig:approx:accuracy:efficiency:sift:25GB:de:hdd:100NN:10K:nocache}, and ~\ref{fig:approx:accuracy:efficiency:deep:25GB:96:hdd:de:100NN:10K:nocache}), except for the scenario of indexing plus answering 100 queries, where iSAX2+ has the least combined cost (Figures~\ref{fig:approx:accuracy:efficiency:sift:25GB:de:hdd:100NN:100:nocache} and~\ref{fig:approx:accuracy:efficiency:deep:25GB:96:hdd:de:100NN:100:nocache}). 
This is because DSTree's query answering is very fast, but its indexing cost is high, so it is amortized only with a large query workload (Figures~\ref{fig:approx:accuracy:efficiency:sift:25GB:de:hdd:100NN:10K:nocache} and~\ref{fig:approx:accuracy:efficiency:deep:25GB:96:hdd:de:100NN:10K:nocache}). 
We observe a similar trend for both Sift25GB and Deep25GB, except the degradation of the performance of SRS, which achieves a very low accuracy of 0.01 on Deep25GB, despite using the most restrictive parameters ($\delta$ = 0.99 and $\epsilon$ = 0). 

\noindent\textbf{Comparison of Accuracy Measures.} In the approximate similarity search literature, the most commonly used accuracy measures are approximation error and recall. The approximation error evaluates how far the approximate neighbors are from the true neighbors, whereas recall assesses how many true neighbors are returned. 
In our study, we refer to the recall and approximation error of a workload as Avg\_Recall and MRE respectively. 
In addition, we use a third measure called MAP because it takes into account the order of the returned candidates and thus is more sensitive than recall. Figures~\ref{fig:approx:map:recall:sift:25GB:128:ng:hdd} and~\ref{fig:approx:map:mre:sift:25GB:128:ng:hdd} compare all three measures for the popular real dataset Sift25GB (we use the 25GB subset to include in-memory methods as well). 
We observe that for any given workload, the Avg\_Recall is equal to MAP for all methods, except for IMI. 
This is because IMI returns the short-listed candidates based on distance calculations on the compressed vectors, while the other methods further refine the candidates by sorting them based on the Euclidean distance of the query to the raw data. 
Figure~\ref{fig:approx:map:mre:sift:25GB:128:ng:hdd} illustrates the relationship between MAP and MRE. 
Note that the value of the approximation error is not always indicative of the actual accuracy. 
For instance, an MRE of about 0.5 for iSAX2 sounds acceptable (some popular LSH methods only work with $\epsilon>=3$~\cite{journal/tods/tao2010,conf/sigmod/gan2012}), yet it corresponds to a very low accuracy of 0.03 as measured by MAP (Figures~\ref{fig:approx:map:mre:sift:25GB:128:ng:hdd}). 
Note that MAP can be more useful in practice, since it takes into account the actual ranks of the true neighbors returned, whereas MRE is evaluated only on the distances between the query and its neighbors.

\subsubsection{\textbf{Query Answering Efficiency and Accuracy: on-Disk Datasets}}
\label{ssec:query_efficiency_disk}

{\color{black}We now report results (Figure~\ref{fig:approx:accuracy:efficiency:synthetic:250GB:ondisk:hdd}) for on-disk experiments, excluding the in-memory only HNSW, {QALSH and FLANN}}. 

\noindent\textbf{Synthetic Data}. DSTree and iSAX2+ outperform by far the rest of the techniques on both $ng$-approximate and $\delta$-$\epsilon$-approximate queries. iSAX2+ is particularly competitive when the total cost is considered with the smaller workload (Figures~\ref{fig:approx:accuracy:efficiency:synthetic:250GB:256:ng:hdd:100NN:100:nocache} and~\ref{fig:approx:accuracy:efficiency:synthetic:250GB:256:de:hdd:100NN:100:nocache}). The querying performance of SRS degraded on-disk due to severe swapping issues (Figure~\ref{fig:approx:accuracy:qefficiency:synthetic:250GB:256:de:hdd:10NN:100:nocache}), therefore we do not include this method in further disk-based experiments. Although IMI is much faster than both iSAX2+ and DSTree on $ng$-approximate search, its accuracy is extremely low. In fact, the best MAP accuracy achieved by IMI plummets to 0.05, whereas DSTree and iSAX2+ have much higher MAP values (Figure~\ref{fig:approx:accuracy:qefficiency:synthetic:250GB:256:ng:hdd:100NN:100:nocache}).

\noindent\textbf{Real Data}. DSTree outperforms all methods 
on both Sift250GB and Deep250GB. The only exception is iSAX2+ having an edge when the combined indexing and search costs are considered for the smaller workload (Figures~\ref{fig:approx:accuracy:efficiency:sift:250GB:ng:hdd:100NN:100:nocache},~\ref{fig:approx:accuracy:efficiency:sift:250GB:de:hdd:100NN:100:nocache},~\ref{fig:approx:accuracy:efficiency:deep:250GB:96:hdd:ng:100NN:100:nocache} and~\ref{fig:approx:accuracy:efficiency:deep:250GB:96:hdd:de:100NN:100:nocache}) and being equally competitive on $ng$-approximate query answering (Figures~\ref{fig:approx:accuracy:qefficiency:sift:250GB:ng:hdd:100NN:100:nocache},~\ref{fig:approx:accuracy:qefficiency:sift:250GB:de:hdd:100NN:100:nocache}).

\begin{figure*}[tb]
	\captionsetup{justification=centering}
	\captionsetup[subfigure]{justification=centering}
	\begin{subfigure}{\textwidth}
		\centering
		\includegraphics[scale=0.18]{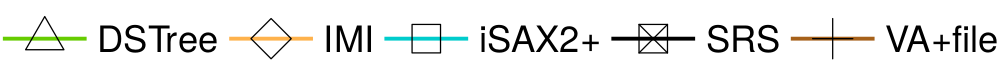}\\
	\end{subfigure}	
	\begin{subfigure}{0.16\textwidth}
		\centering
		\includegraphics[width=\textwidth]{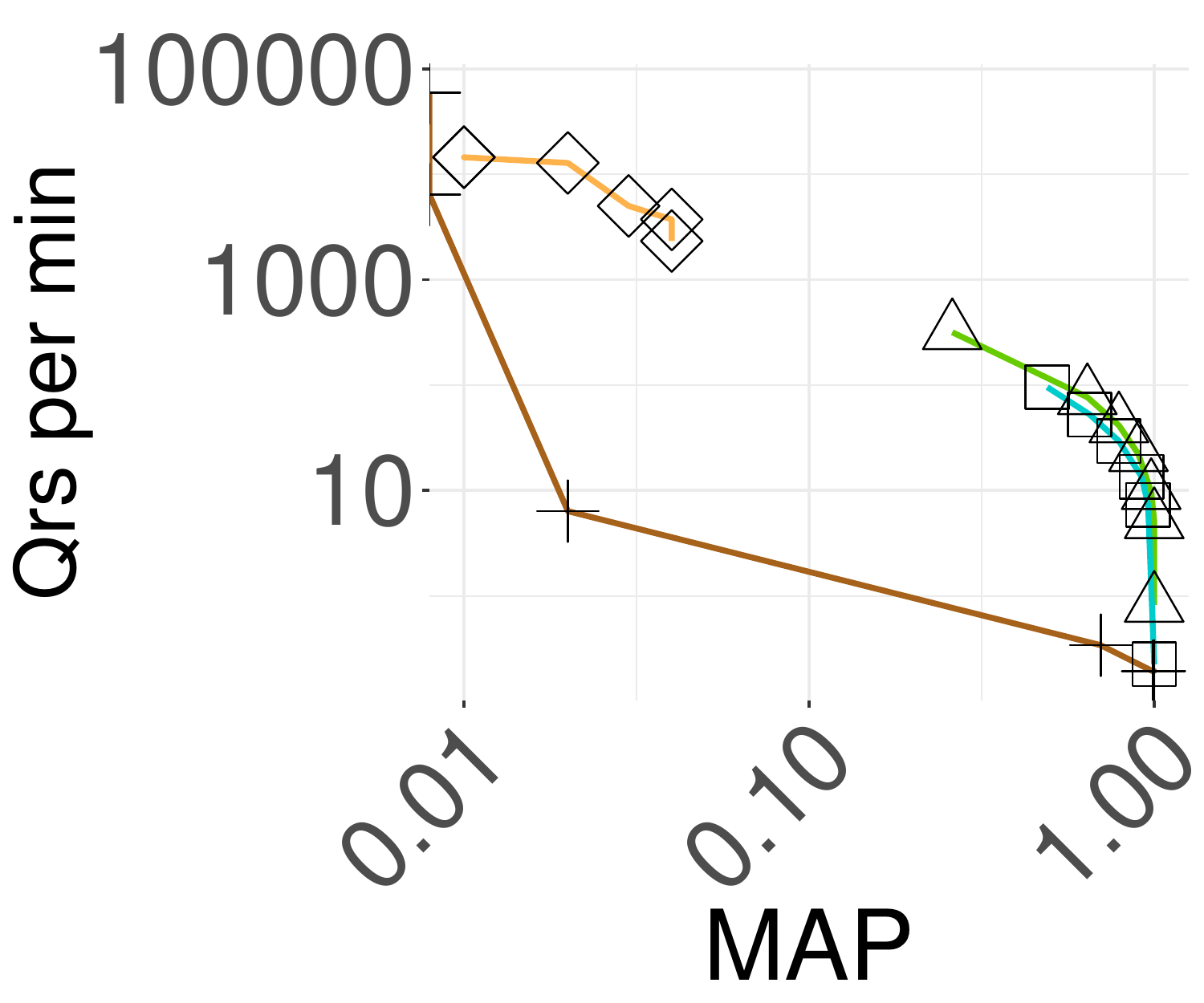}
		\scriptsize \caption{Rand250GB(ng)} 
		\label{fig:approx:accuracy:qefficiency:synthetic:250GB:256:ng:hdd:100NN:100:nocache}
	\end{subfigure}
	\begin{subfigure}{0.16\textwidth}
		\centering
		\includegraphics[width=\textwidth]{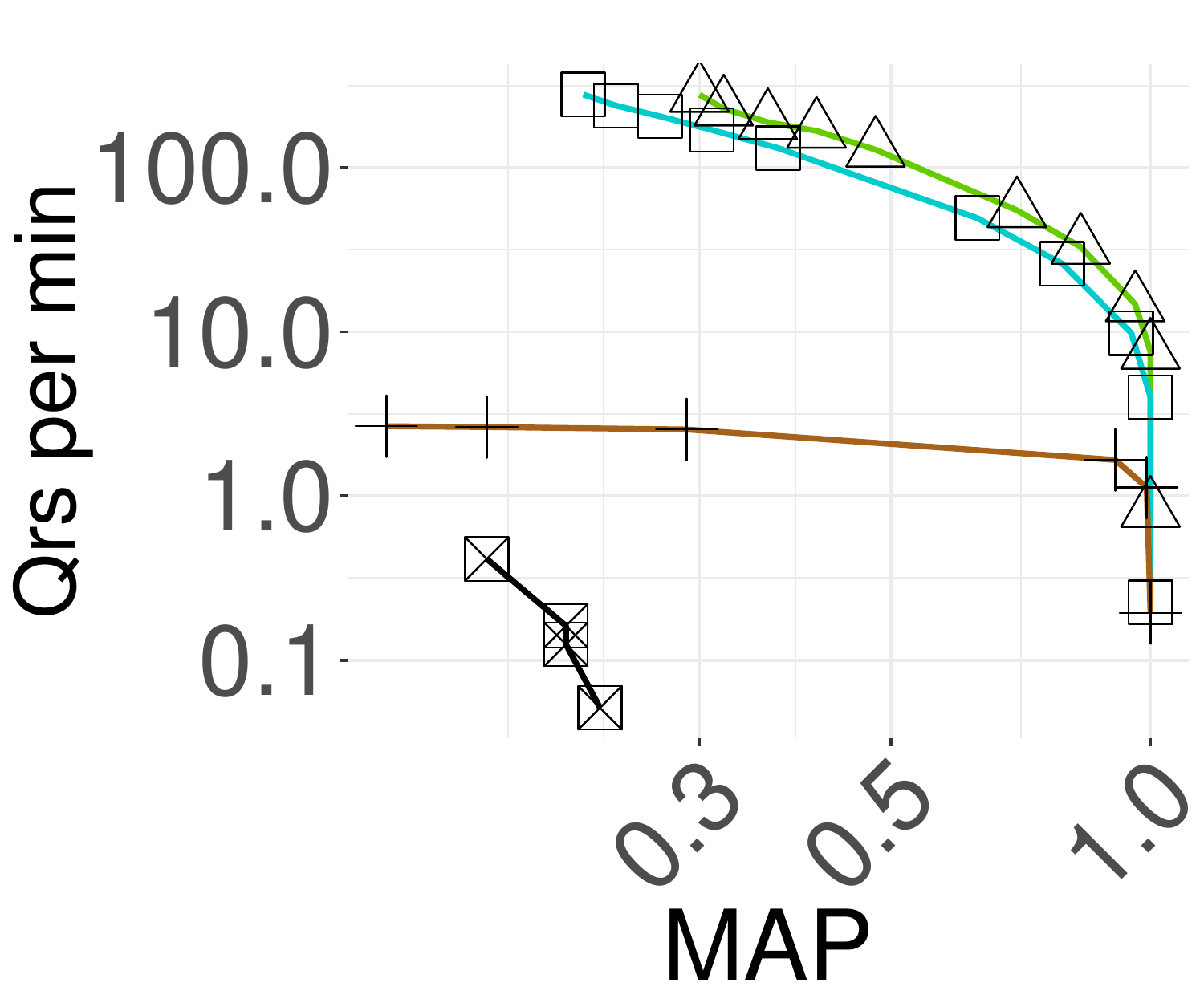}
		\scriptsize \caption{Rand250GB($\bm{\delta\epsilon}$)} 
		\label{fig:approx:accuracy:qefficiency:synthetic:250GB:256:de:hdd:10NN:100:nocache}
	\end{subfigure}
	\begin{subfigure}{0.16\textwidth}
		\centering
		\includegraphics[width=\textwidth]{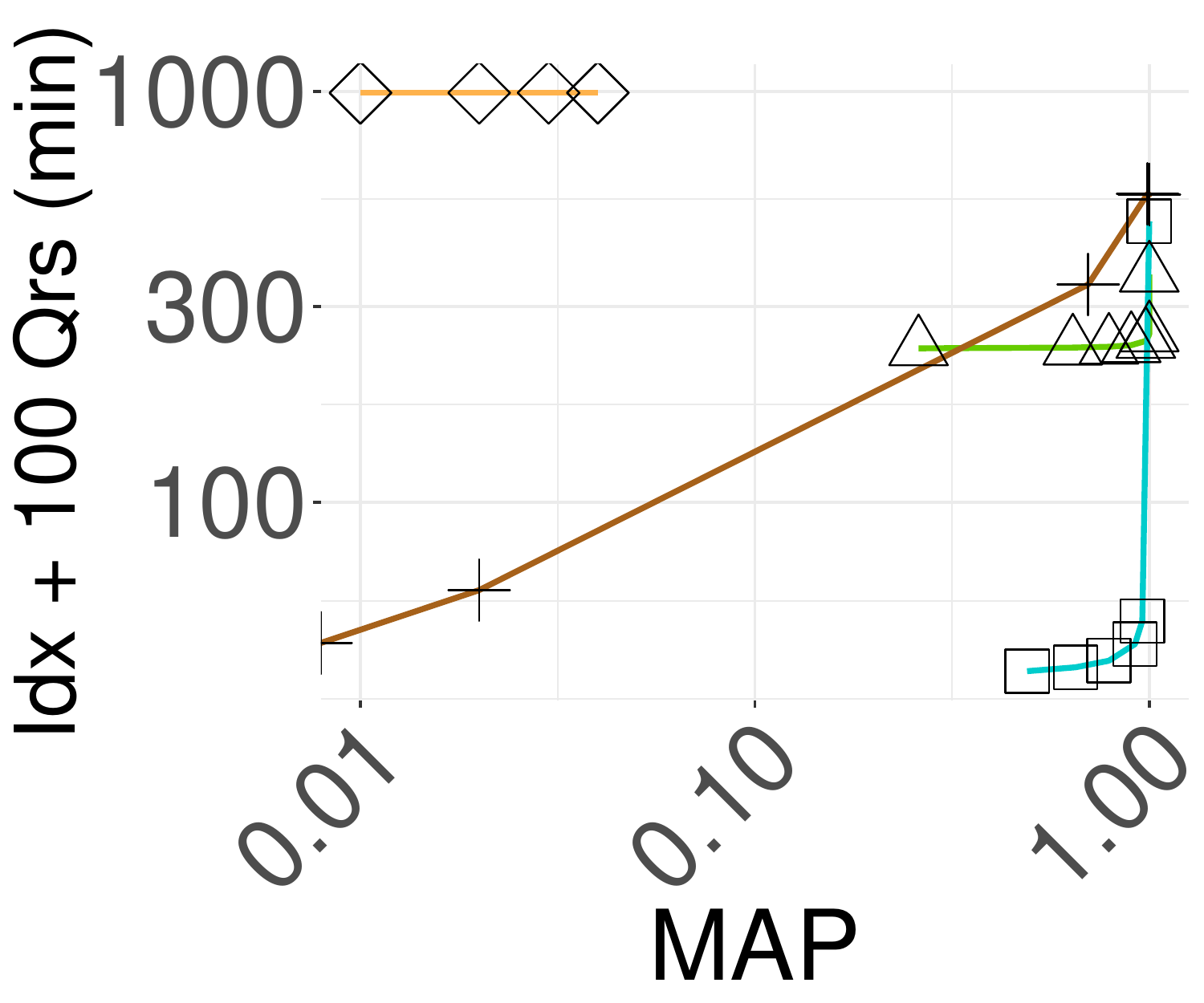}
		\scriptsize \caption{Rand250GB(ng)} 
		\label{fig:approx:accuracy:efficiency:synthetic:250GB:256:ng:hdd:100NN:100:nocache}
	\end{subfigure}
	\begin{subfigure}{0.16\textwidth}
		\centering
		\includegraphics[width=\textwidth]{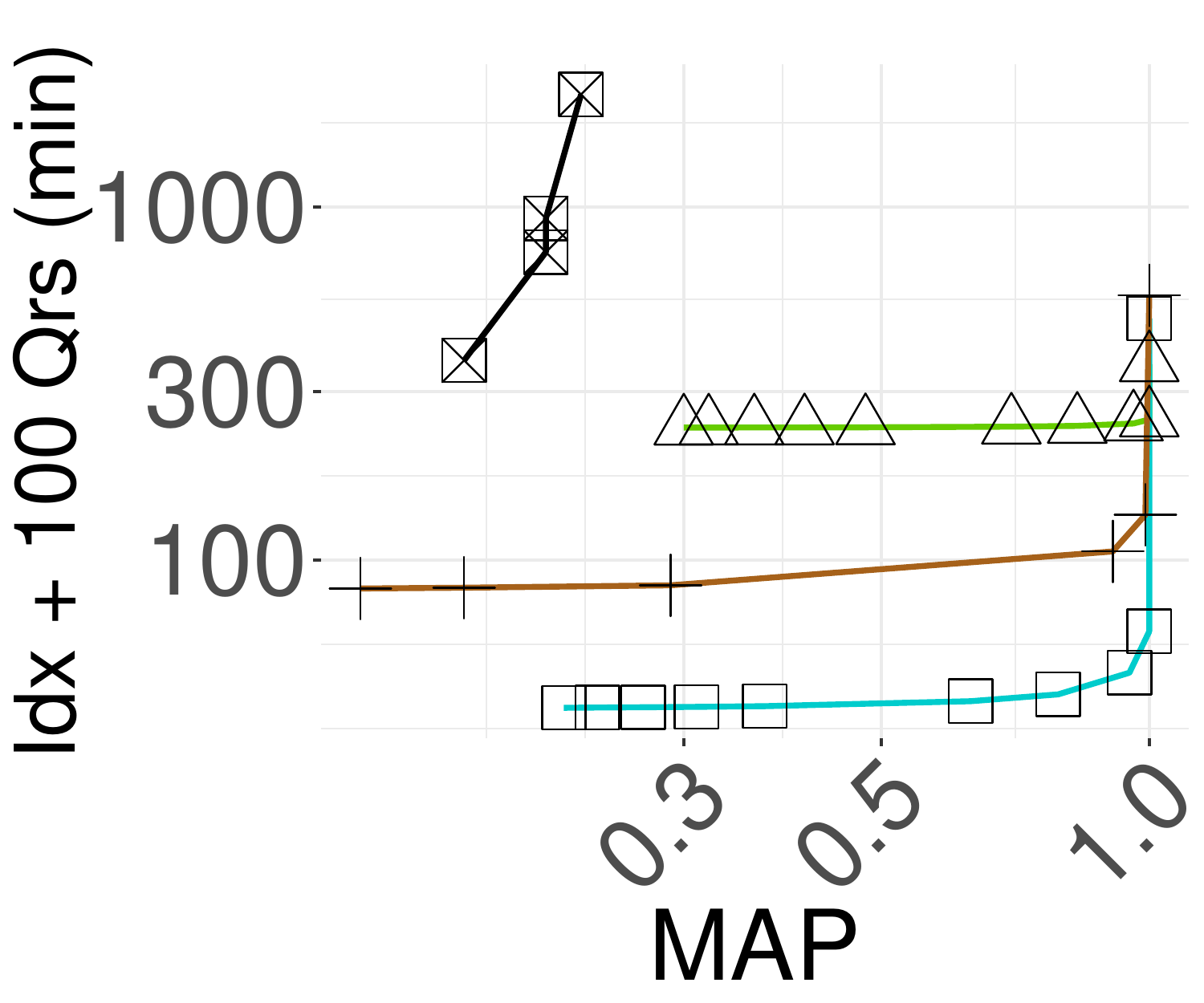}
		\scriptsize \caption{Rand250GB($\bm{\delta\epsilon}$)} 
		\label{fig:approx:accuracy:efficiency:synthetic:250GB:256:de:hdd:100NN:100:nocache}
	\end{subfigure}
	\begin{subfigure}{0.16\textwidth}
		\centering
		\includegraphics[width=\textwidth]{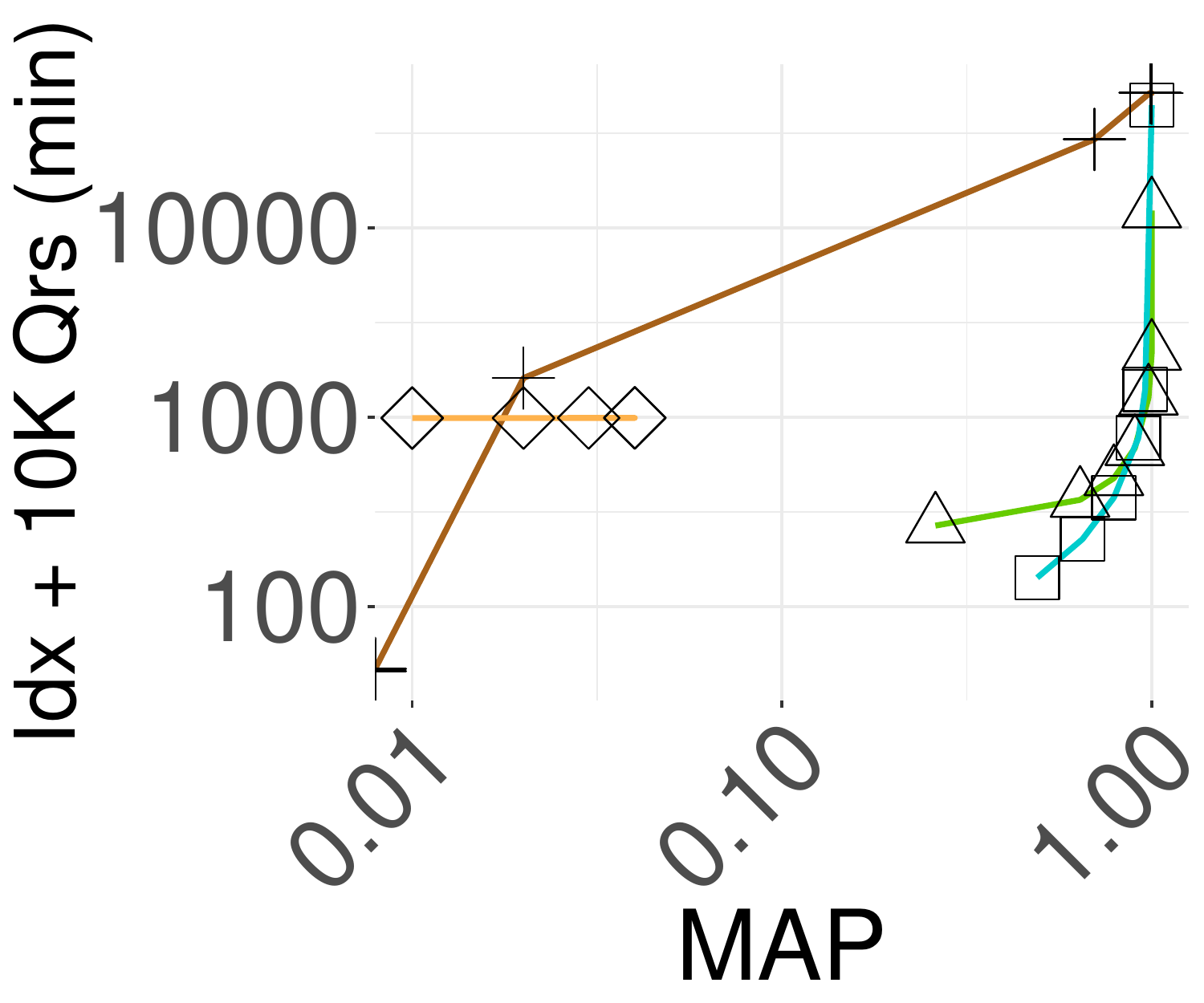}
		\scriptsize \caption{Rand250GB(ng)} 
		\label{fig:approx:accuracy:efficiency:synthetic:250GB:256:ng:hdd:10NN:10K:nocache}
	\end{subfigure}
	\begin{subfigure}{0.16\textwidth}
		\centering
		\includegraphics[width=\textwidth]{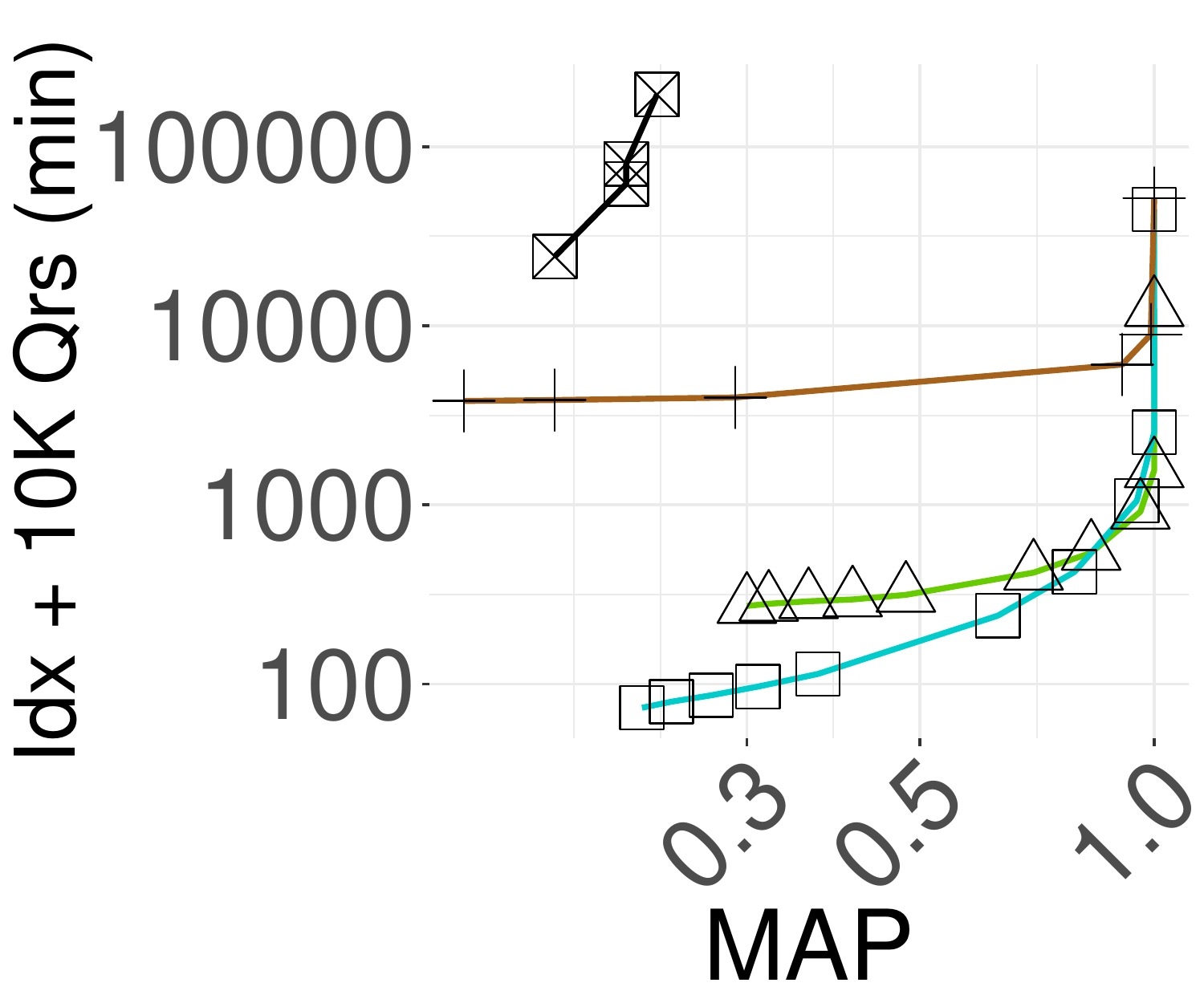}
		\scriptsize \caption{Rand250GB($\bm{\delta\epsilon}$)} 
		\label{fig:approx:accuracy:efficiency:synthetic:250GB:256:de:hdd:100NN:10K:nocache}
	\end{subfigure}
	\begin{subfigure}{0.16\textwidth}
		\centering
		\includegraphics[width=\textwidth]{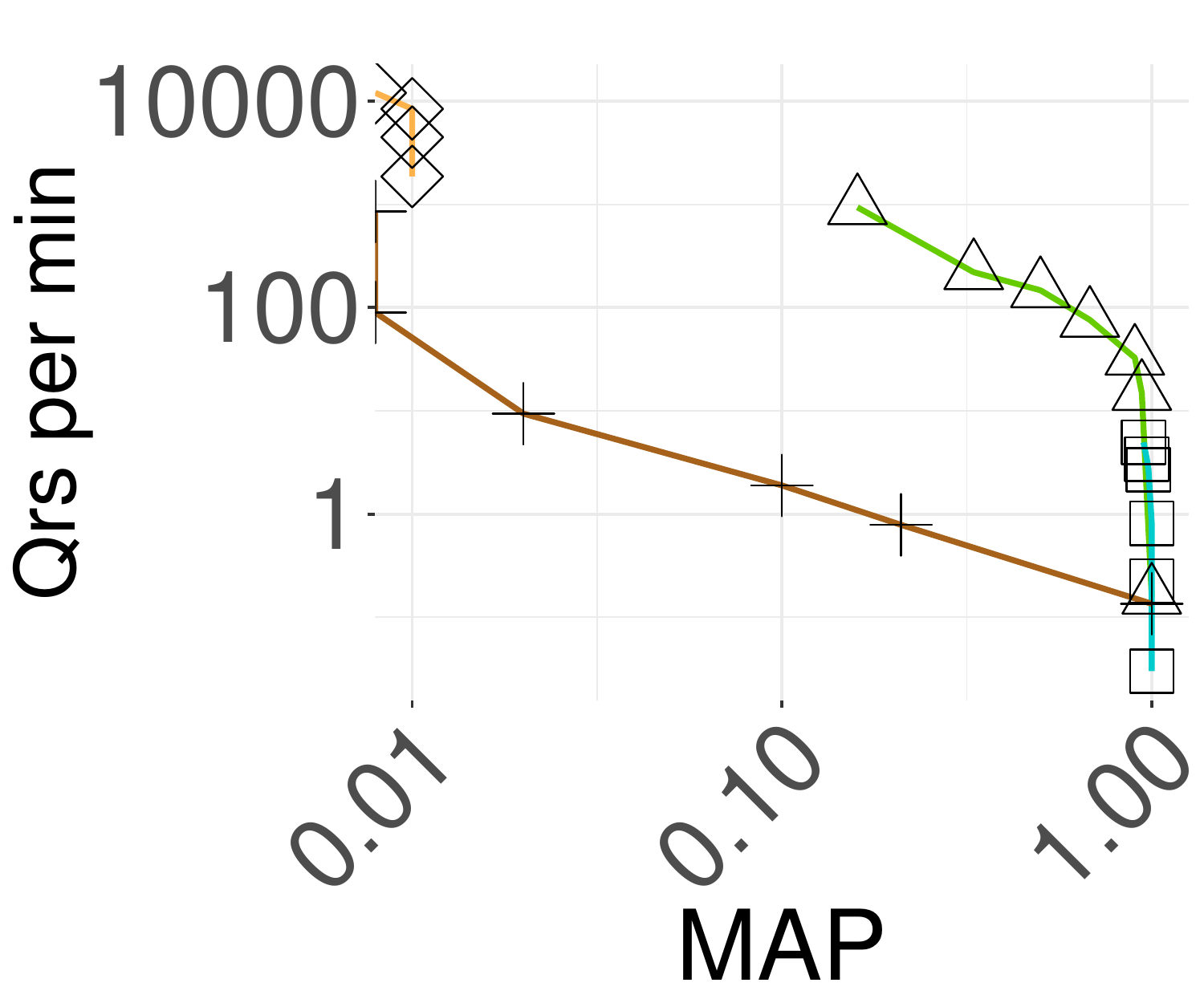}
		\scriptsize \caption{Sift250GB(ng)} 
		\label{fig:approx:accuracy:qefficiency:sift:250GB:ng:hdd:100NN:100:nocache}
	\end{subfigure}
	\begin{subfigure}{0.16\textwidth}
		\centering
		\includegraphics[width=\textwidth]{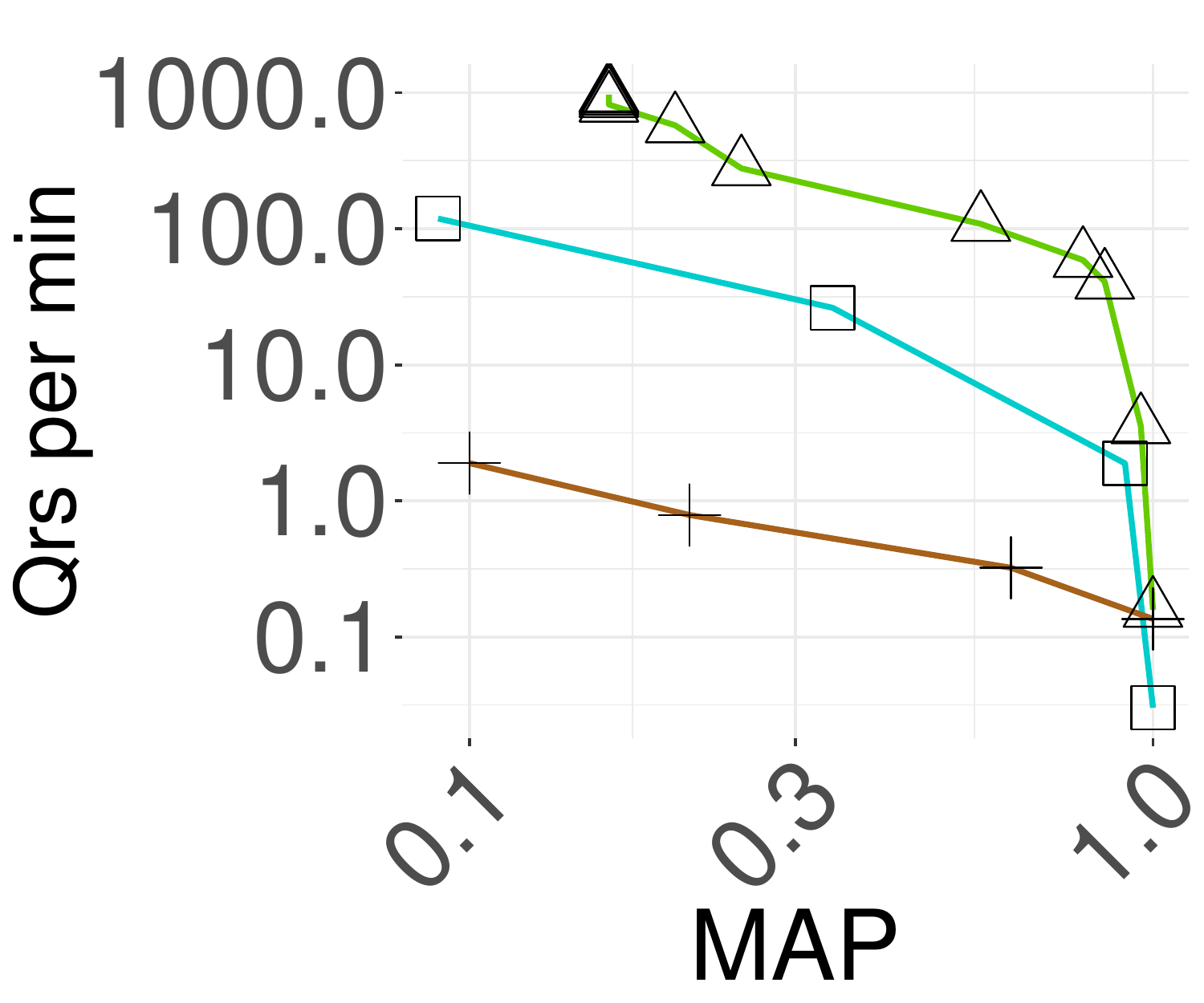}
		\scriptsize \caption{Sift250GB($\bm{\delta\epsilon}$)} 
		\label{fig:approx:accuracy:qefficiency:sift:250GB:de:hdd:100NN:100:nocache}
	\end{subfigure}
	\begin{subfigure}{0.16\textwidth}
		\centering
		\includegraphics[width=\textwidth]{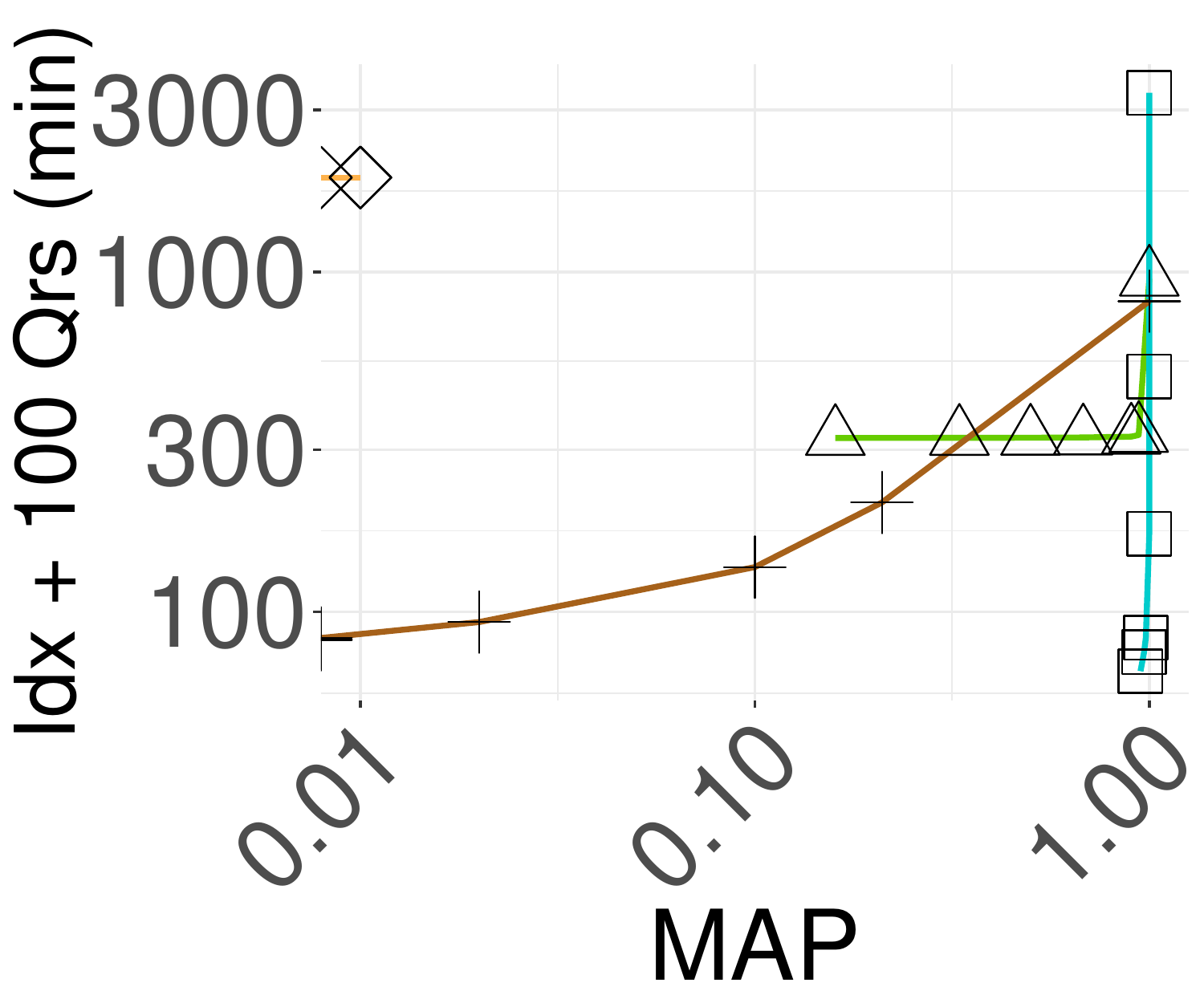}
		\scriptsize \caption{Sift250GB(ng)} 
		\label{fig:approx:accuracy:efficiency:sift:250GB:ng:hdd:100NN:100:nocache}
	\end{subfigure}
	\begin{subfigure}{0.16\textwidth}
		\centering
		\includegraphics[width=\textwidth]{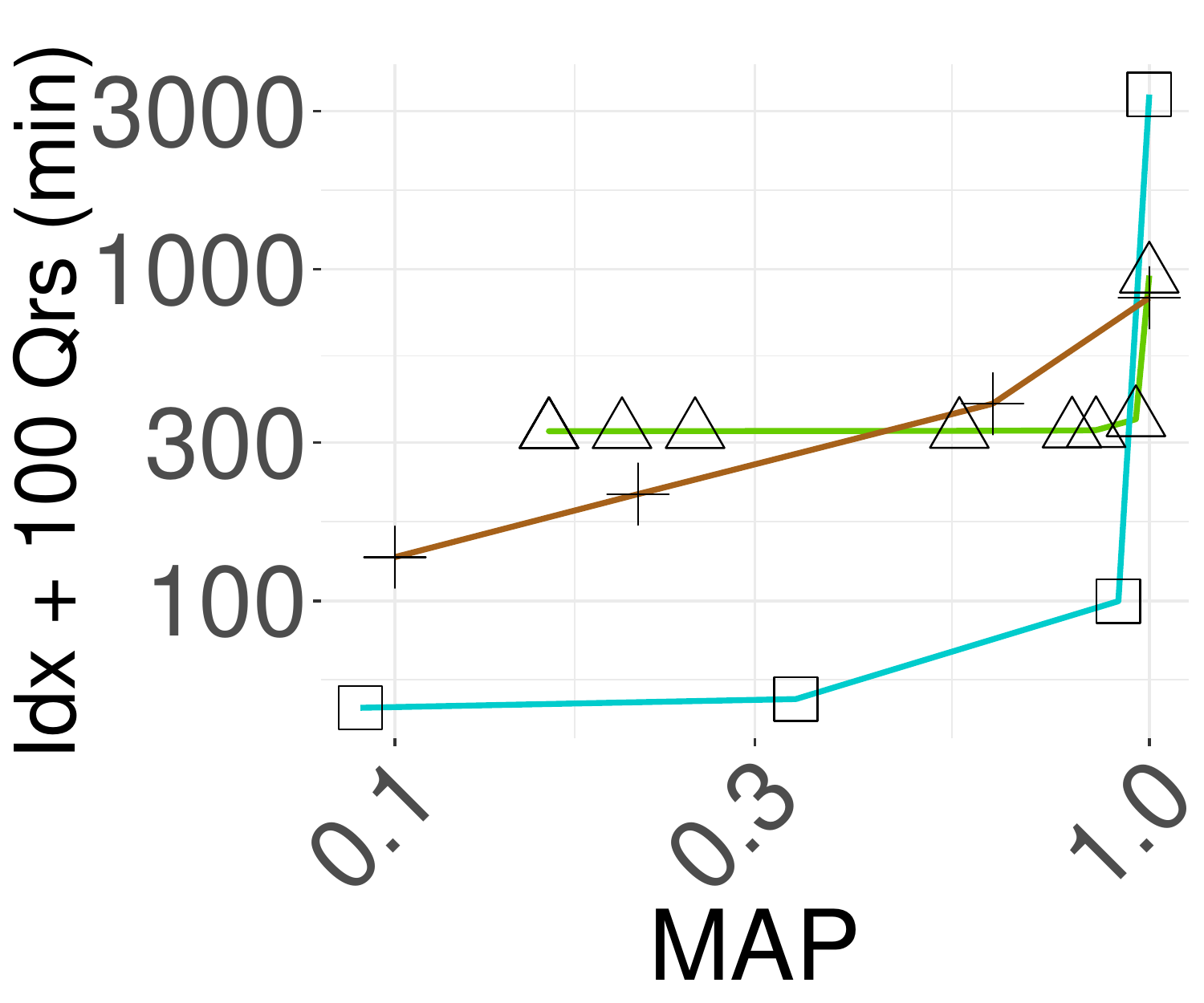}
		\scriptsize \caption{Sift250GB($\bm{\delta\epsilon}$)} 
		\label{fig:approx:accuracy:efficiency:sift:250GB:de:hdd:100NN:100:nocache}
	\end{subfigure}
	\begin{subfigure}{0.16\textwidth}
		\centering
		\includegraphics[width=\textwidth]{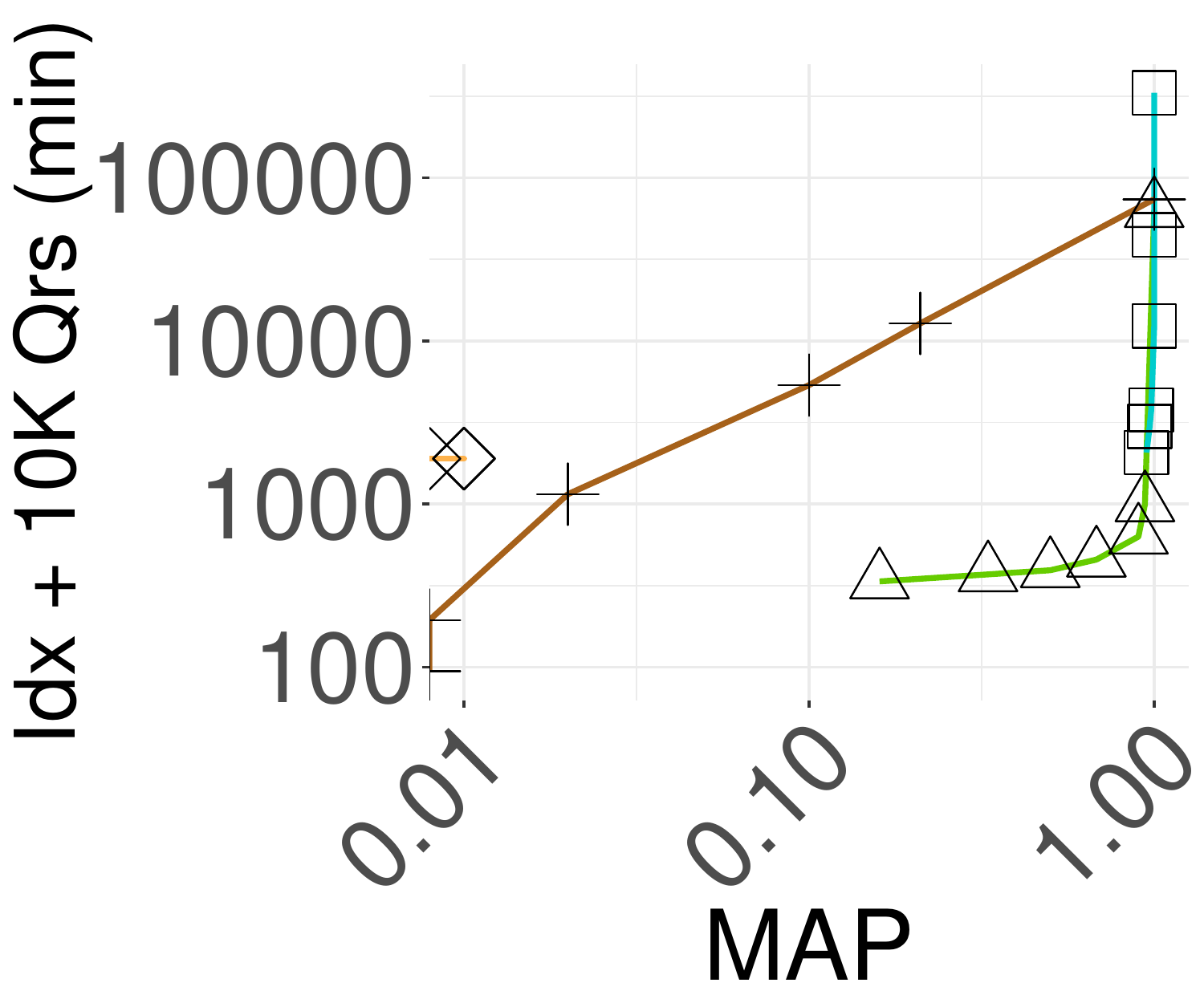}
		\scriptsize \caption{Sift250GB(ng)} 
		\label{fig:approx:accuracy:efficiency:sift:250GB:ng:hdd:100NN:10K:nocache}
	\end{subfigure}
	\begin{subfigure}{0.16\textwidth}
		\centering
		\includegraphics[width=\textwidth]{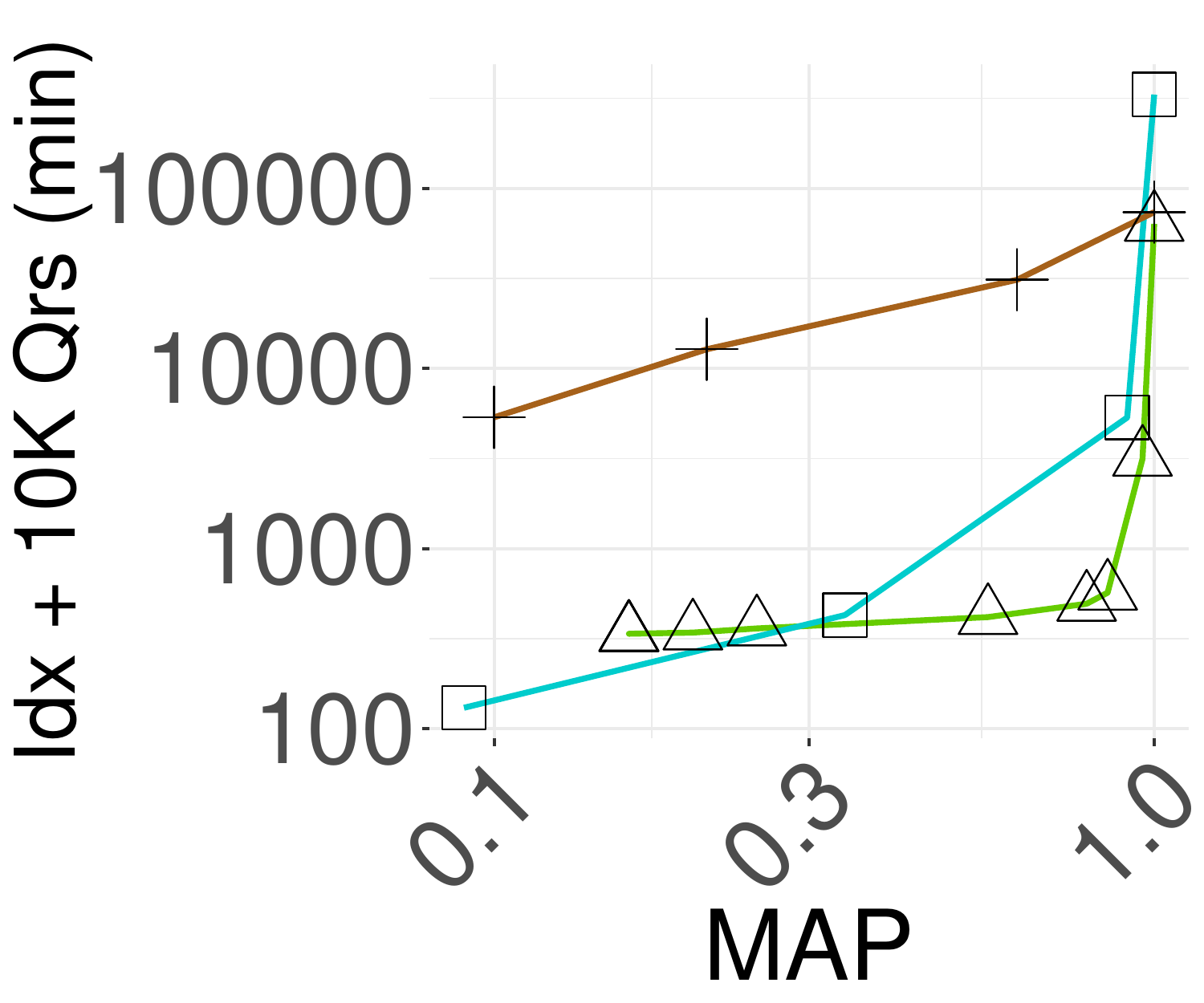}
		\scriptsize \caption{Sift250GB($\bm{\delta\epsilon}$)} 
		\label{fig:approx:accuracy:efficiency:sift:250GB:de:hdd:100NN:10K:nocache}
	\end{subfigure}	
	\begin{subfigure}{0.16\textwidth}
		\centering
		\includegraphics[width=\textwidth]{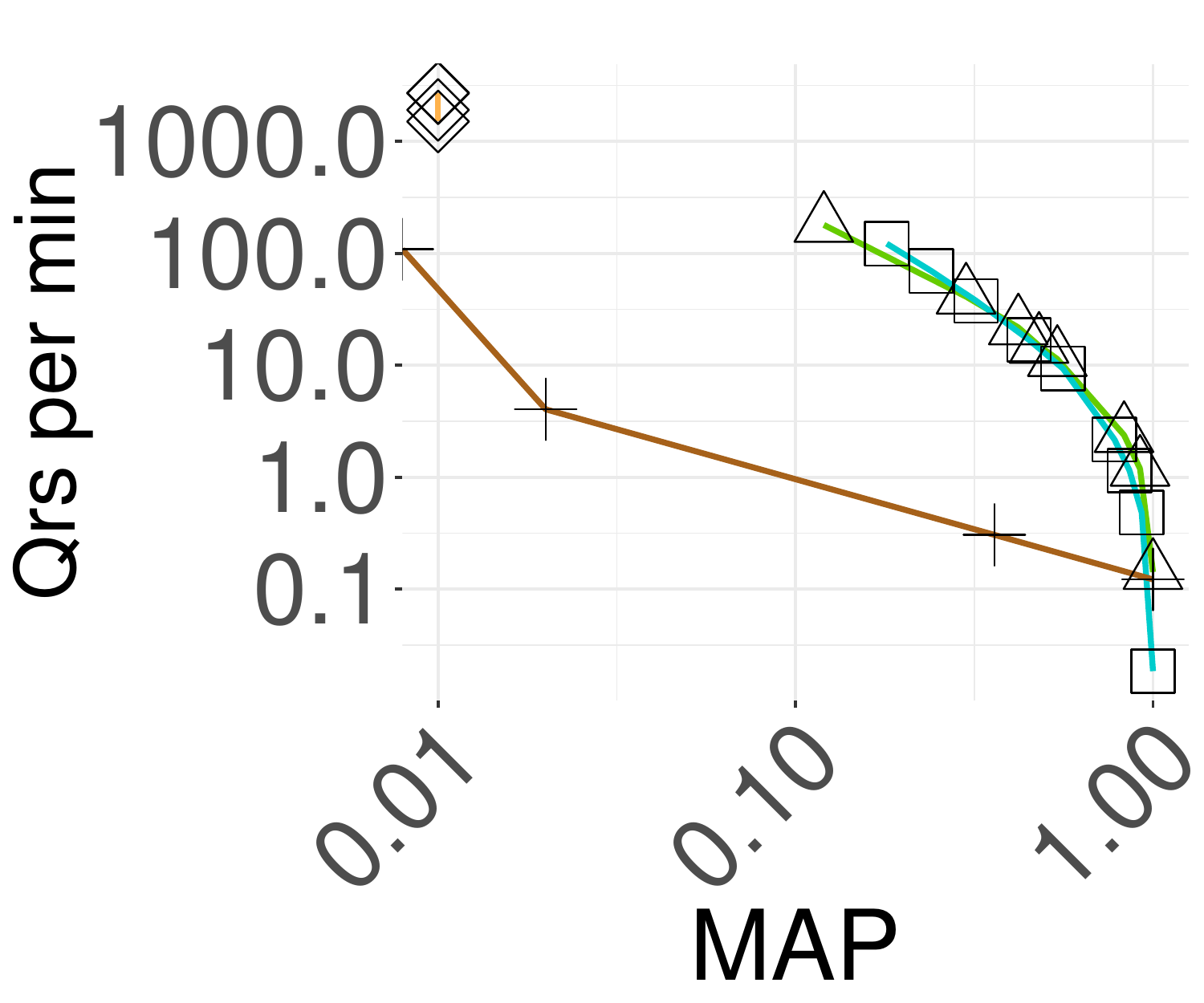}
		\scriptsize \caption{Deep250GB(ng)} 
		\label{fig:approx:accuracy:qefficiency:deep:250GB:96:hdd:ng:100NN:100:nocache}
	\end{subfigure}
	\begin{subfigure}{0.16\textwidth}
		\centering
		\includegraphics[width=\textwidth]{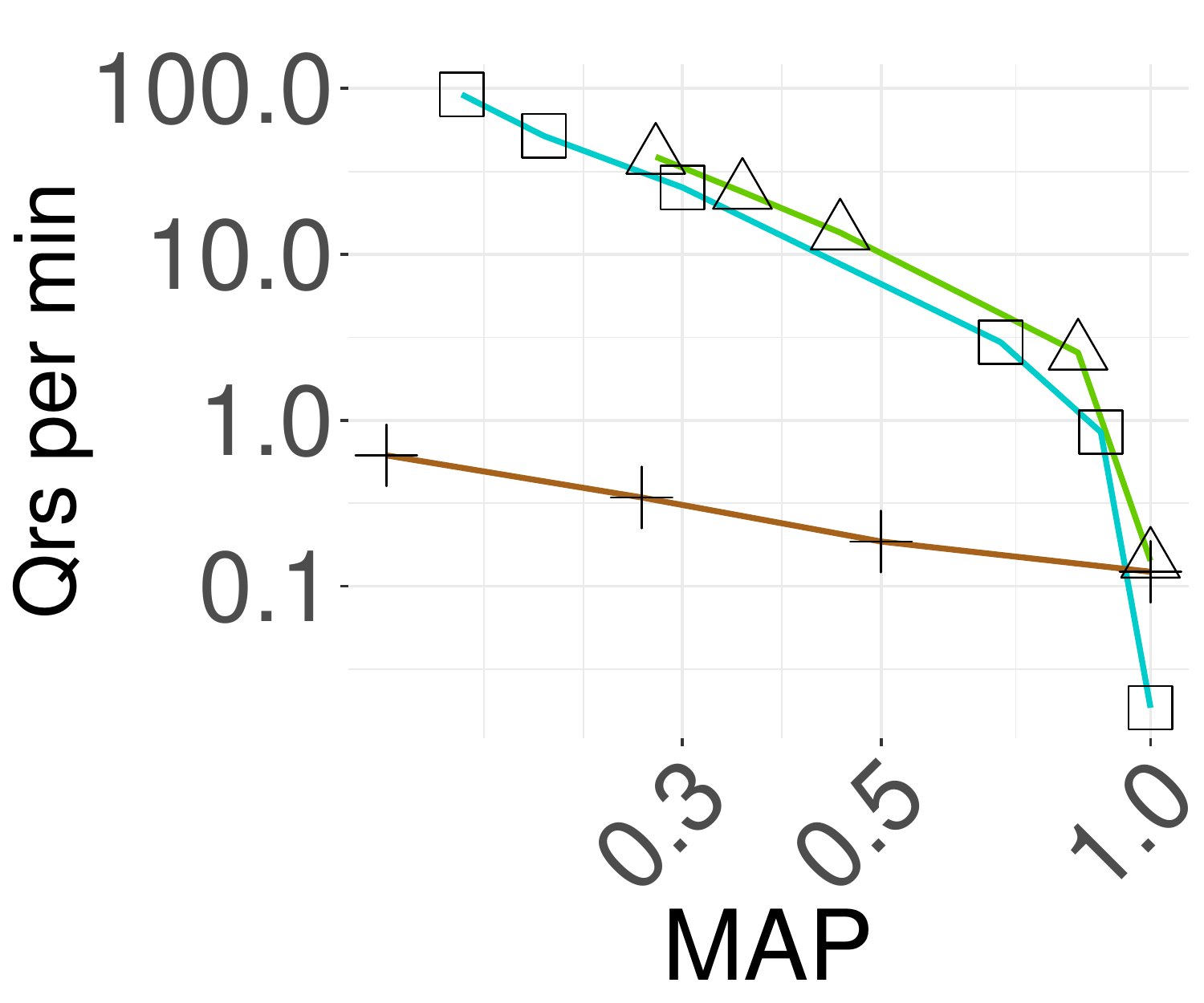}
		\scriptsize \caption{Deep250GB($\bm{\delta\epsilon}$)} 
		\label{fig:approx:accuracy:qefficiency:deep:250GB:96:hdd:de:100NN:100:nocache}
	\end{subfigure}
	\begin{subfigure}{0.16\textwidth}
		\centering
		\includegraphics[width=\textwidth]{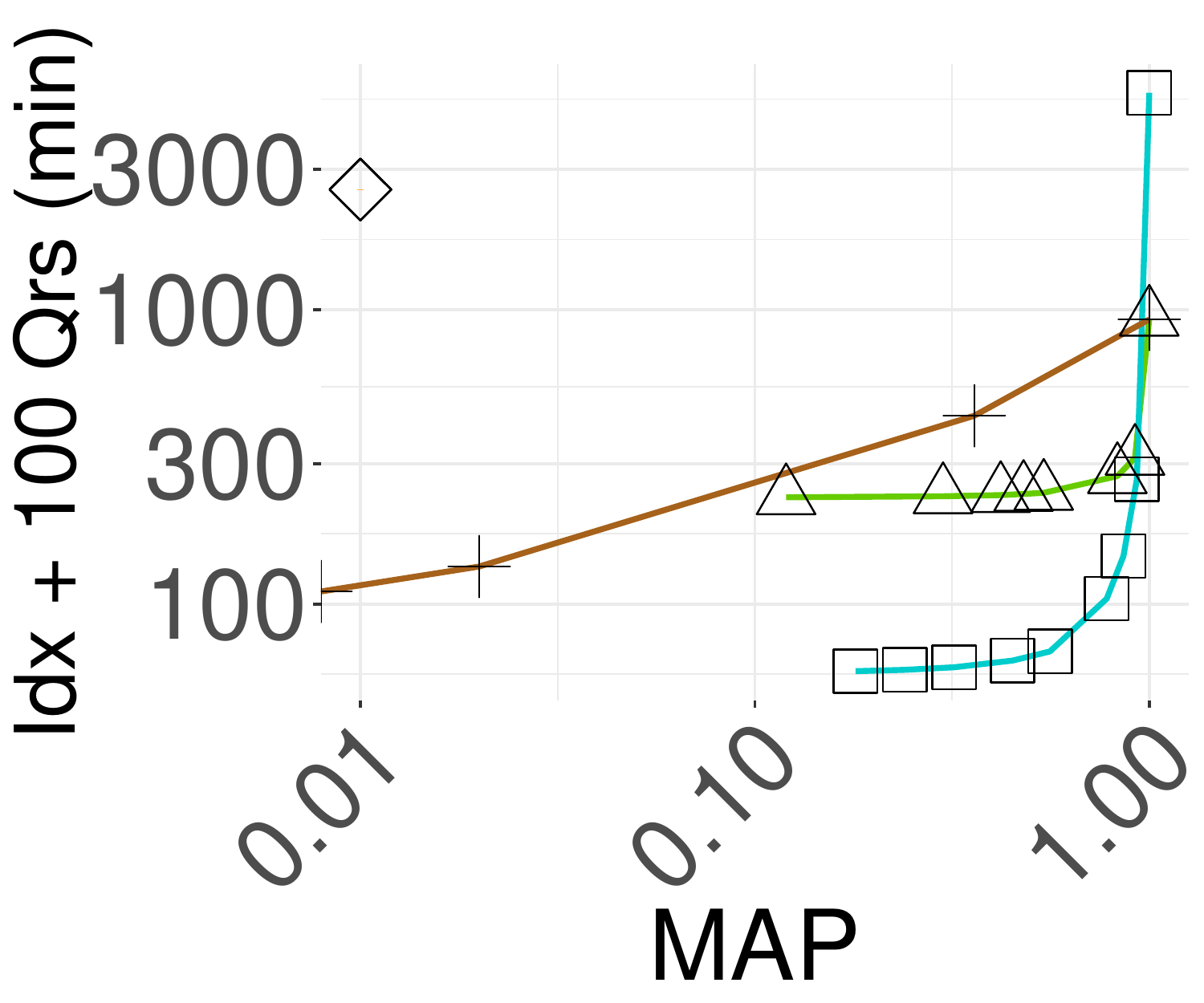}
		\scriptsize \caption{Deep250GB(ng)} 
		\label{fig:approx:accuracy:efficiency:deep:250GB:96:hdd:ng:100NN:100:nocache}
	\end{subfigure}
	\begin{subfigure}{0.16\textwidth}
		\centering
		\includegraphics[width=\textwidth]{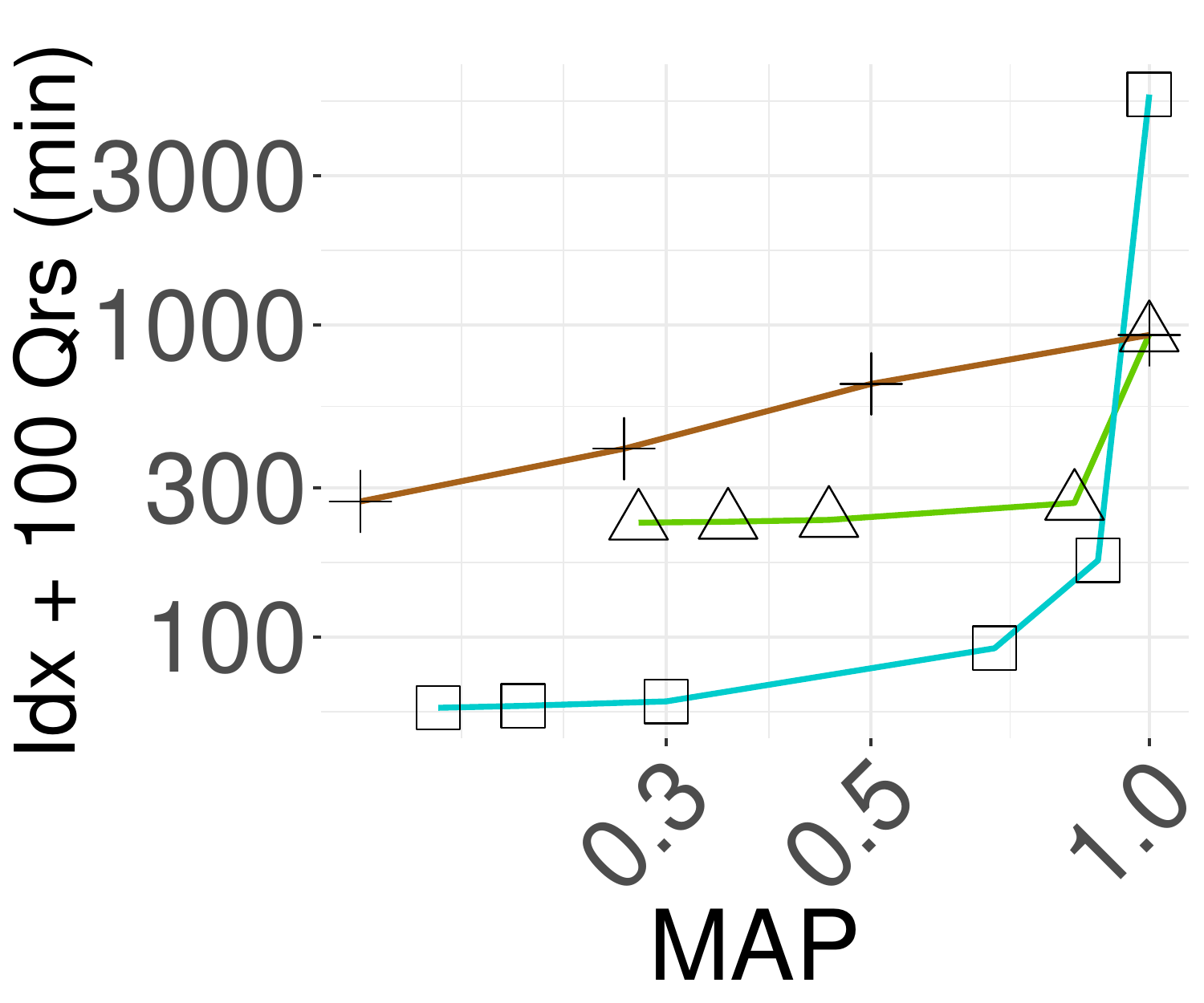}
		\scriptsize \caption{Deep250GB($\bm{\delta\epsilon}$)} 
		\label{fig:approx:accuracy:efficiency:deep:250GB:96:hdd:de:100NN:100:nocache}
	\end{subfigure}
	\begin{subfigure}{0.16\textwidth}
		\centering
		\includegraphics[width=\textwidth]{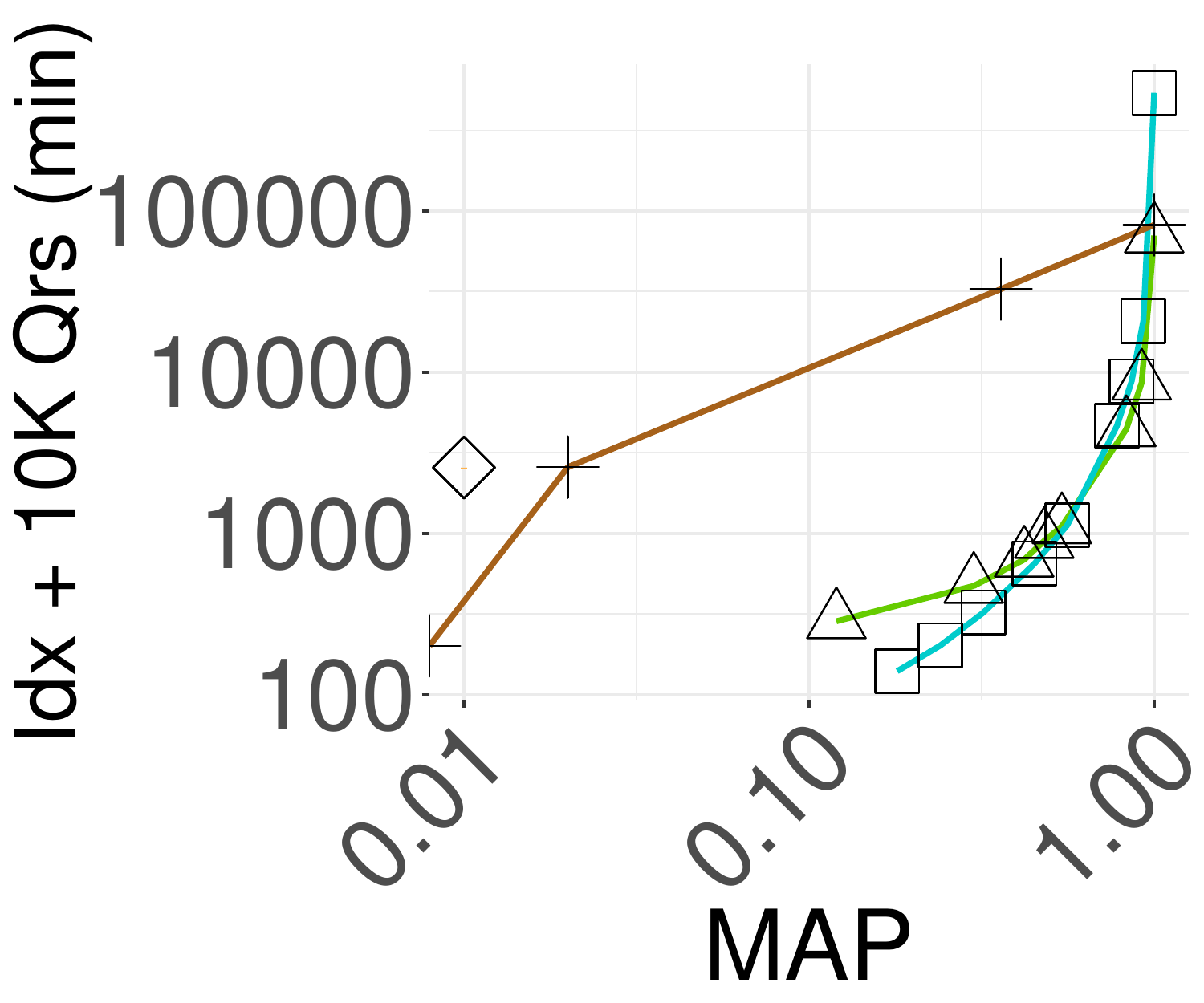}
		\scriptsize \caption{Deep250GB(ng)} 
		\label{fig:approx:accuracy:efficiency:deep:250GB:96:hdd:ng:100NN:10K:nocache}
	\end{subfigure}
	\begin{subfigure}{0.16\textwidth}
		\centering
		\includegraphics[width=\textwidth]{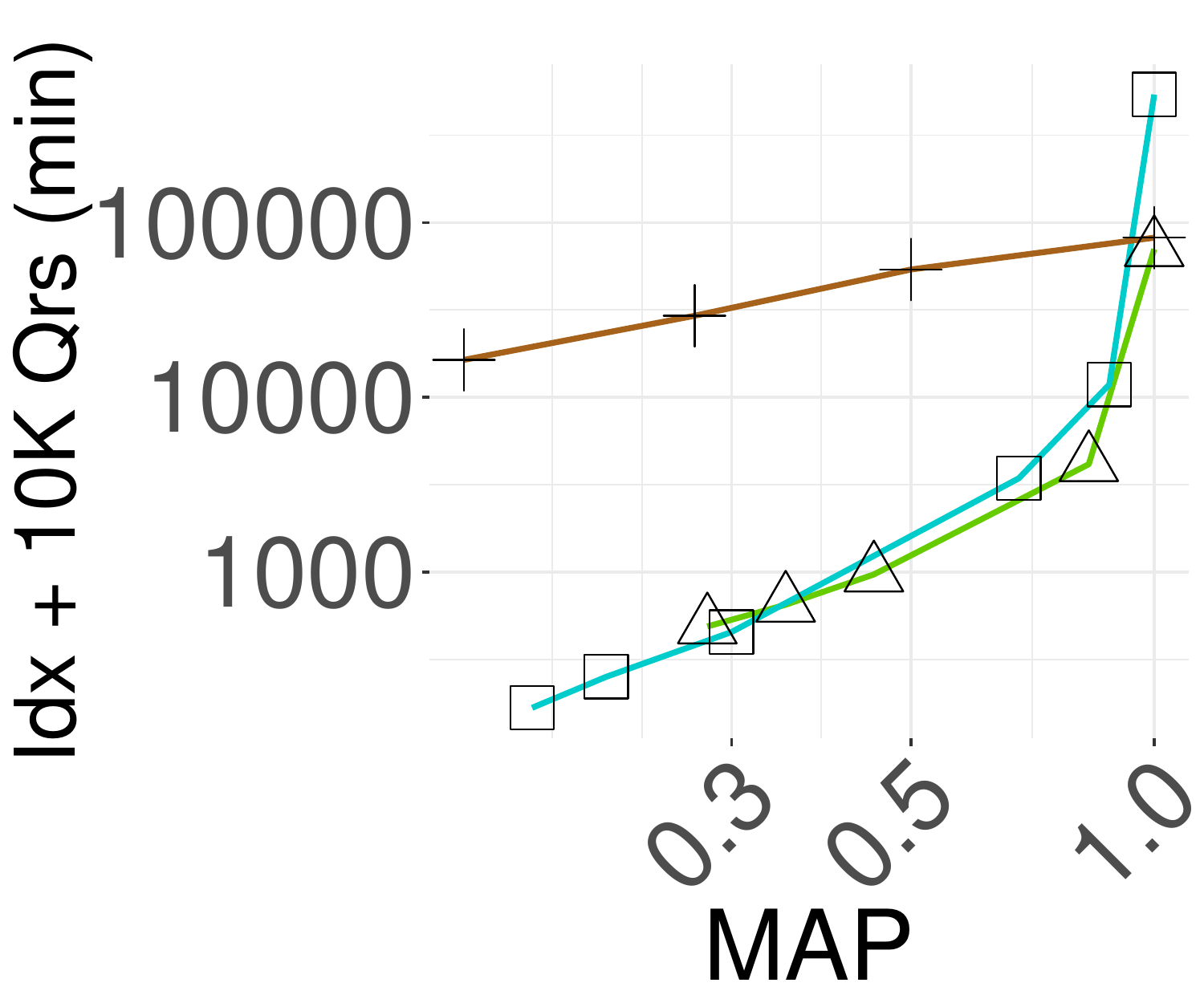}
		\scriptsize \caption{Deep250GB($\bm{\delta\epsilon}$)} 
		\label{fig:approx:accuracy:efficiency:deep:250GB:96:hdd:de:100NN:10K:nocache}
	\end{subfigure}
	\caption{{\color{black} Efficiency vs. accuracy on disk (100NN queries)}}	
	\vspace*{-0.2cm}
	\label{fig:approx:accuracy:efficiency:synthetic:250GB:ondisk:hdd}
\end{figure*}

{\color{black} \noindent\textbf{Best Performing Methods.}} The earlier results show that VA+file is 
outperformed by DSTree and iSAX2+, and that SRS and IMI have very low accuracy on the large datasets. 
We thus conduct further experiments considering only iSAX2+ and DSTree (recall that HNSW is an in-memory approach only): 
see Figures~\ref{fig:approx:accuracy:data:250GB:hdd:best},~\ref{fig:approx:efficiency:k:hdd} and~\ref{fig:approx:accuracy_efficiency:delta:epsilon:synthetic:250GB:hdd}.
In terms of query efficiency/accuracy tradeoff, DSTree outperforms iSAX2+ on all datasets, except for Sald100GB (Figure~\ref{fig:approx:accuracy:throughput:sald:100GB:128:hdd:ng:100NN:100:nocache:best}), and for low MAP values on Seismic100GB (Figure~\ref{fig:approx:accuracy:throughput:seismic:100GB:256:hdd:ng:100NN:100:nocache:best}).

\begin{figure*}[tb]
\begin{minipage}{\textwidth}
	\centering
\includegraphics[scale=0.18]{{full_epsilon_legend_25GB}}
\end{minipage}
\begin{minipage}{0.22\textwidth}
		\captionsetup{justification=centering}
		\captionsetup[subfigure]{justification=centering}
		\begin{subfigure}{\textwidth}
			\includegraphics[width=0.95\textwidth]{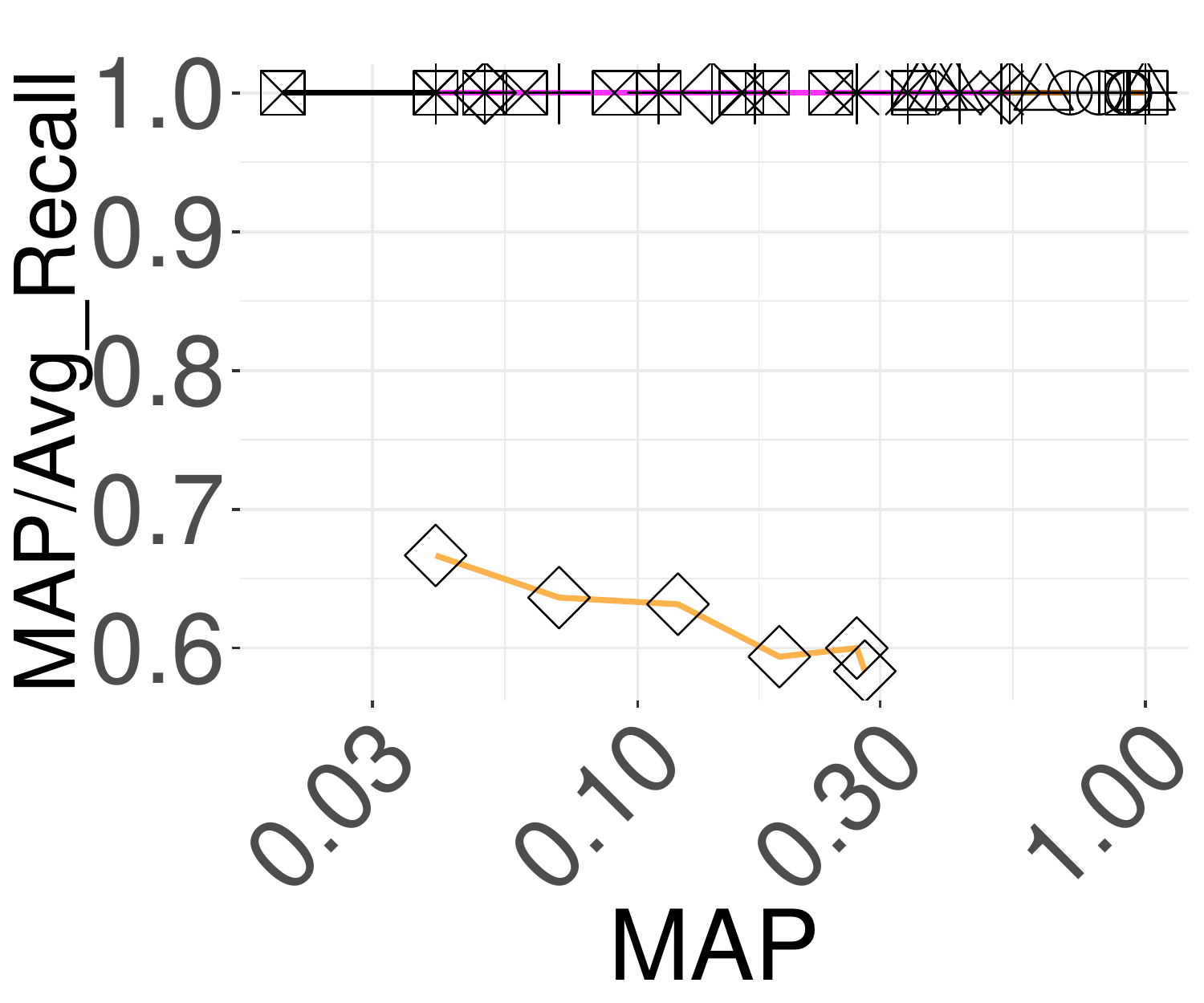}
			\caption{Recall vs. MAP}  
			\label{fig:approx:map:recall:sift:25GB:128:ng:hdd}
		\end{subfigure}
		\begin{subfigure}{\textwidth}
			\includegraphics[width=0.95\textwidth]{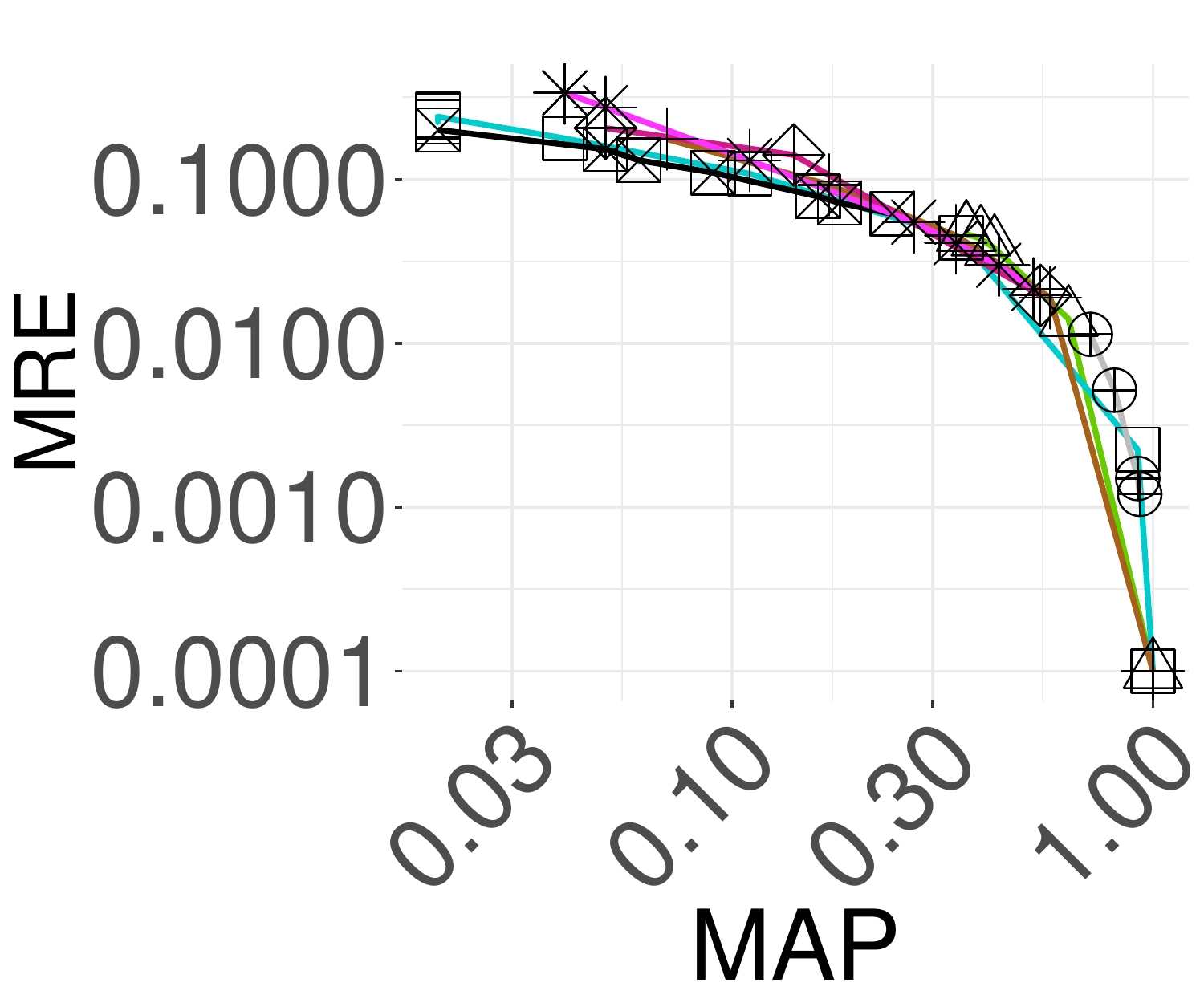}
			\caption{MRE vs. MAP}  
			\label{fig:approx:map:mre:sift:25GB:128:ng:hdd}
		\end{subfigure}
		\caption{{\color{black} Comparison of measures (Sift25GB)}}	
		\label{fig:approx:map:recall:hdd}
\end{minipage}
\begin{minipage}{0.04\textwidth}
\end{minipage}
\begin{minipage}{0.76\textwidth}
	\captionsetup{justification=centering}
	\captionsetup[subfigure]{justification=centering}
	\begin{subfigure}{0.18\textwidth}
		\centering
		\includegraphics[width=\textwidth]{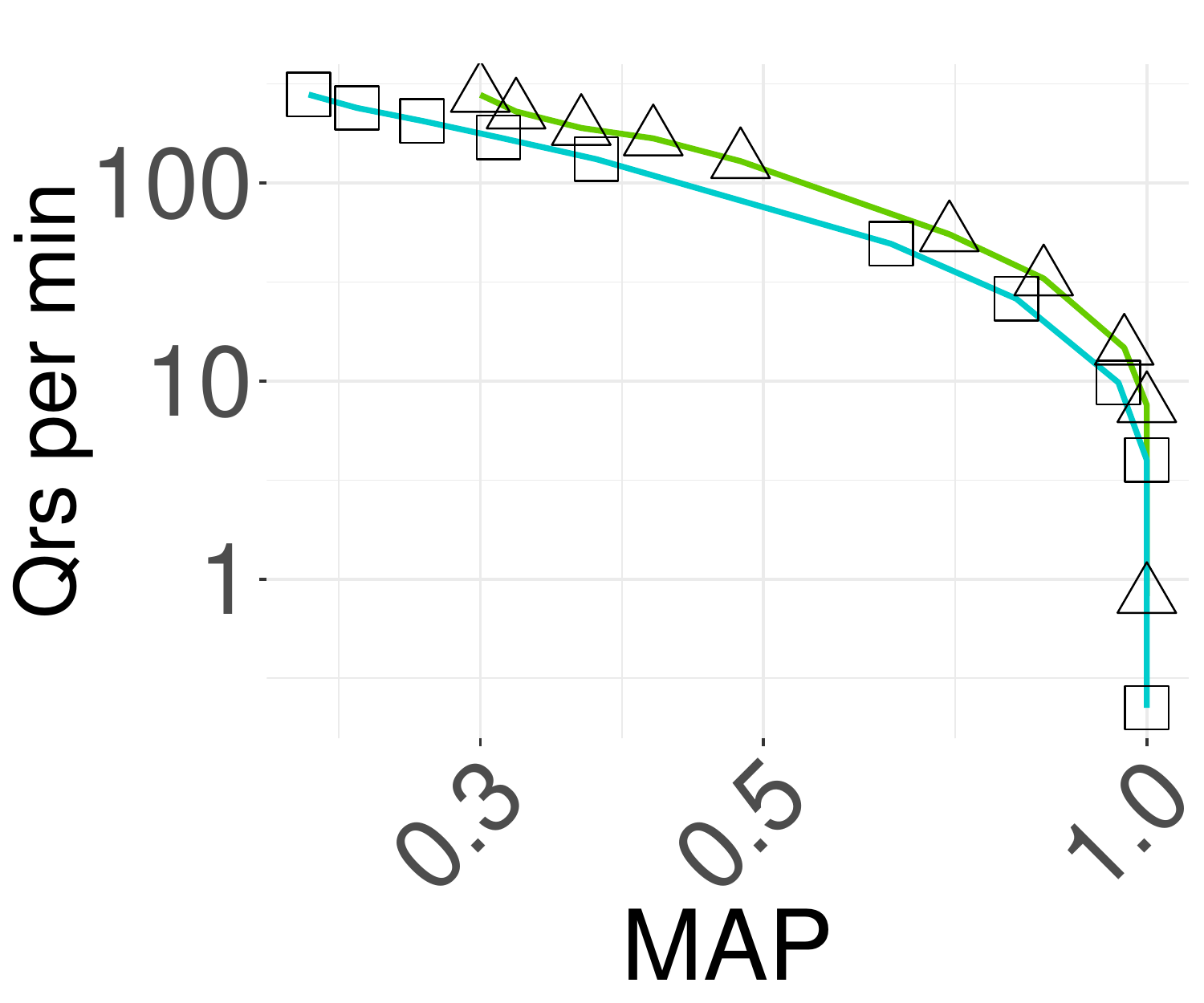}
		\caption{Rand250GB} 
		\label{fig:approx:accuracy:throughput:synthetic:250GB:256:hdd:de:100NN:100:nocache:best}
	\end{subfigure}
	\begin{subfigure}{0.18\textwidth}
		\centering
		\includegraphics[width=\textwidth]{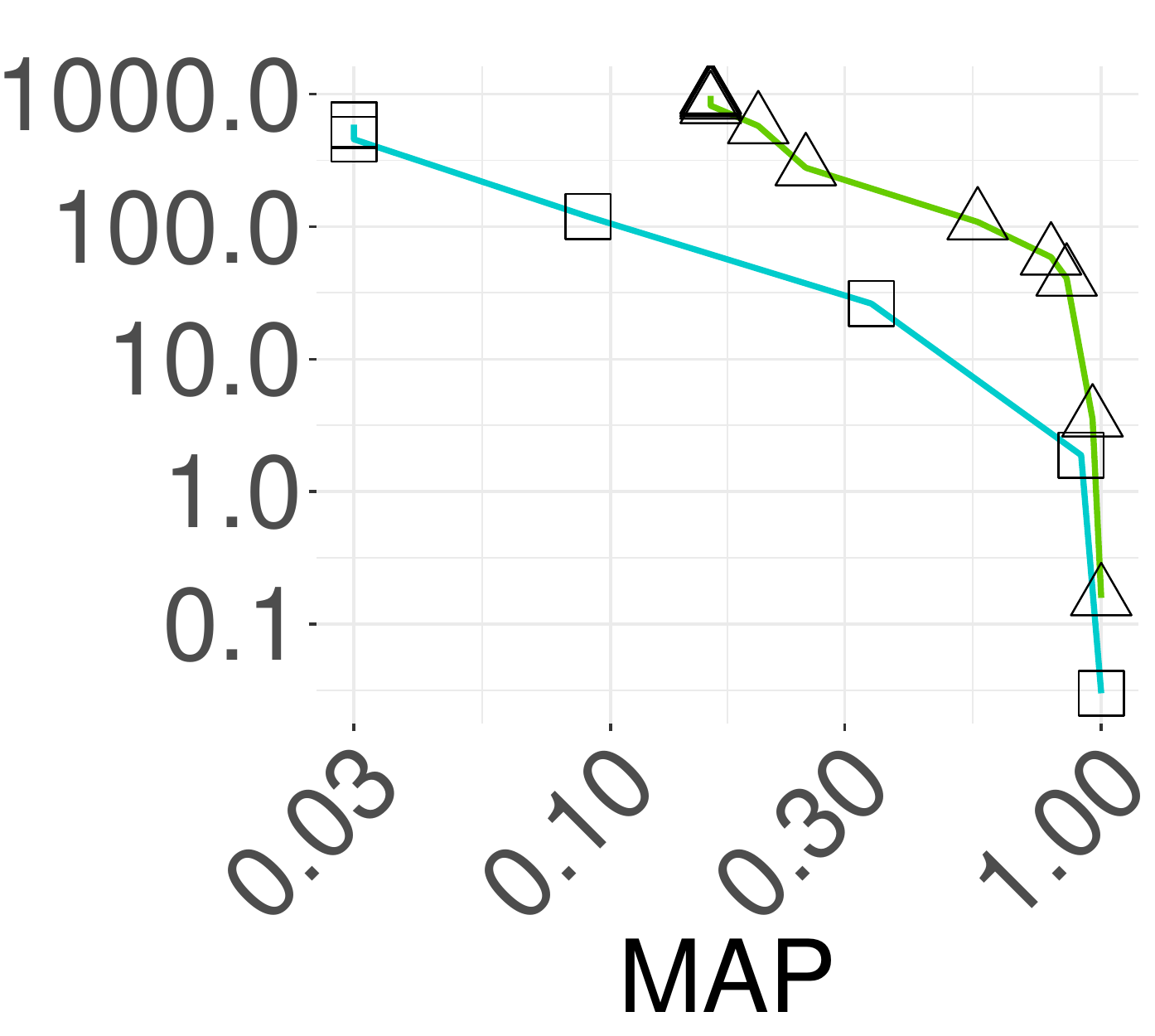}
		\caption{Sift250GB} 
		\label{fig:approx:accuracy:throughput:sift:250GB:128:hdd:ng:100NN:100:nocache:best}
	\end{subfigure}
	\begin{subfigure}{0.18\textwidth}
		\centering
		\includegraphics[width=\textwidth]{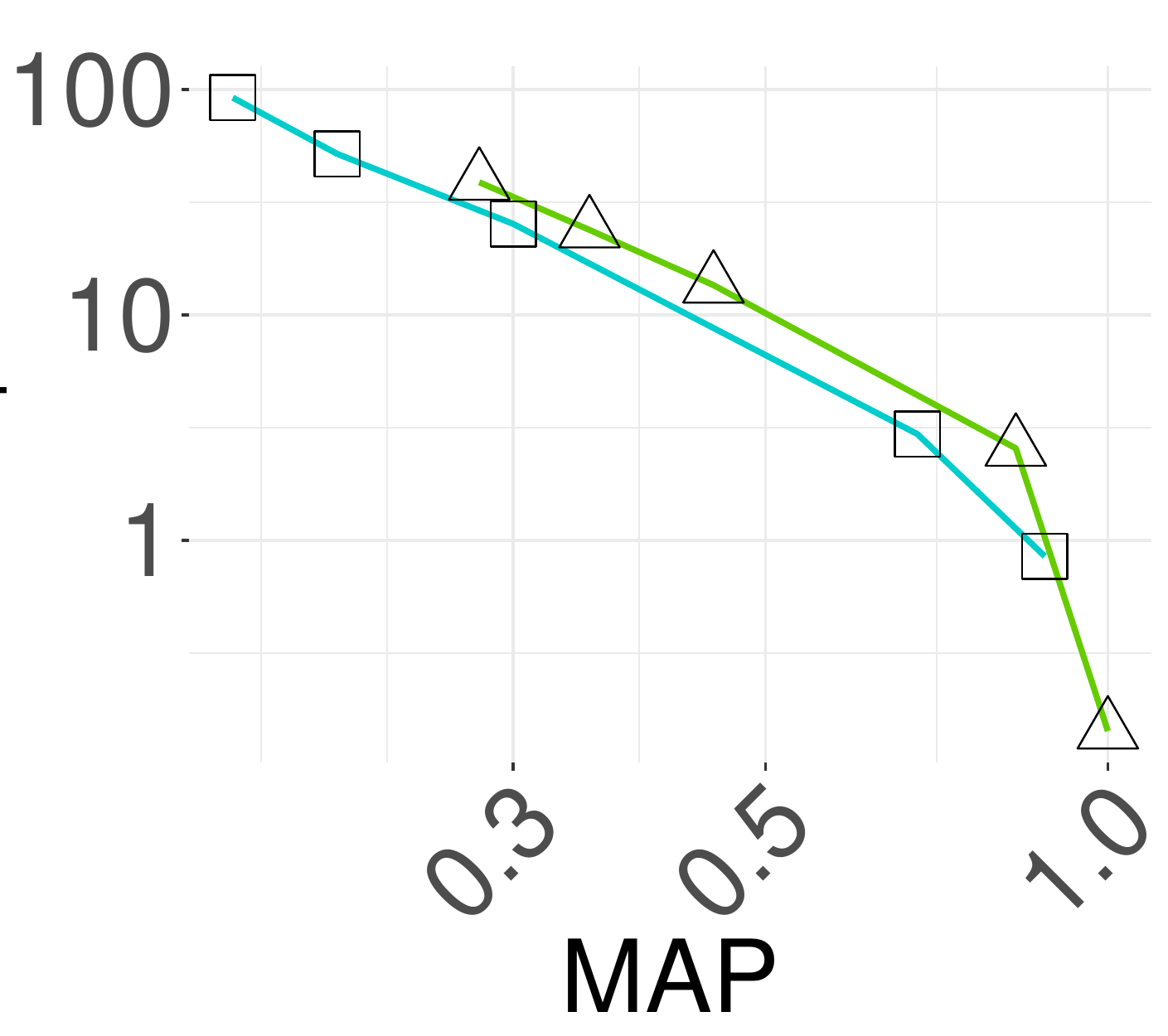}
		\caption{Deep250GB} 
		\label{fig:approx:accuracy:throughput:deep:250GB:96:hdd:ng:100NN:100:nocache:best}
	\end{subfigure}
	\begin{subfigure}{0.18\textwidth}
		\centering
		\includegraphics[width=\textwidth]{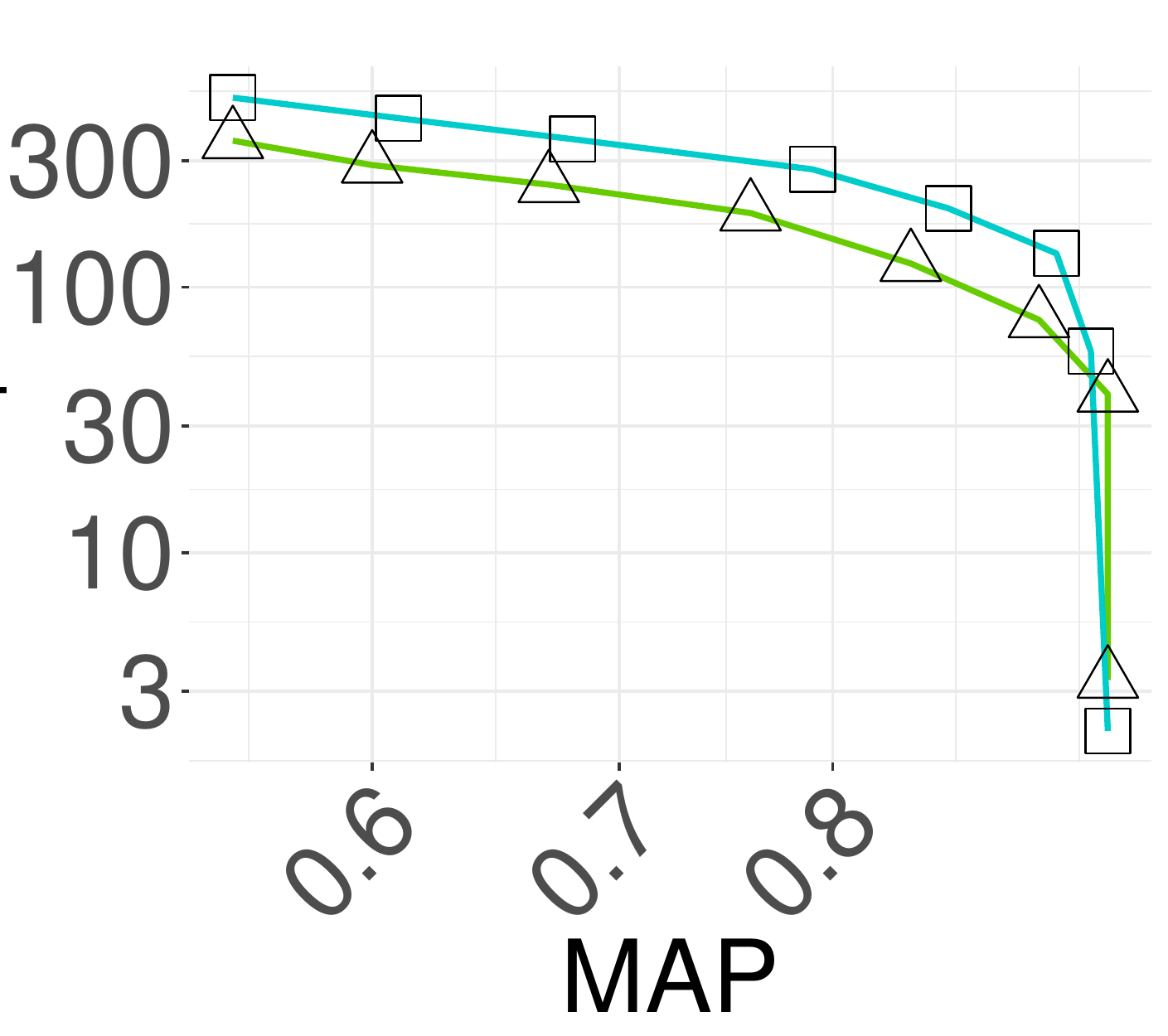}
		\caption{Sald100GB} 
		\label{fig:approx:accuracy:throughput:sald:100GB:128:hdd:ng:100NN:100:nocache:best}
	\end{subfigure}	
	\begin{subfigure}{0.18\textwidth}
		\centering
		\includegraphics[width=\textwidth]{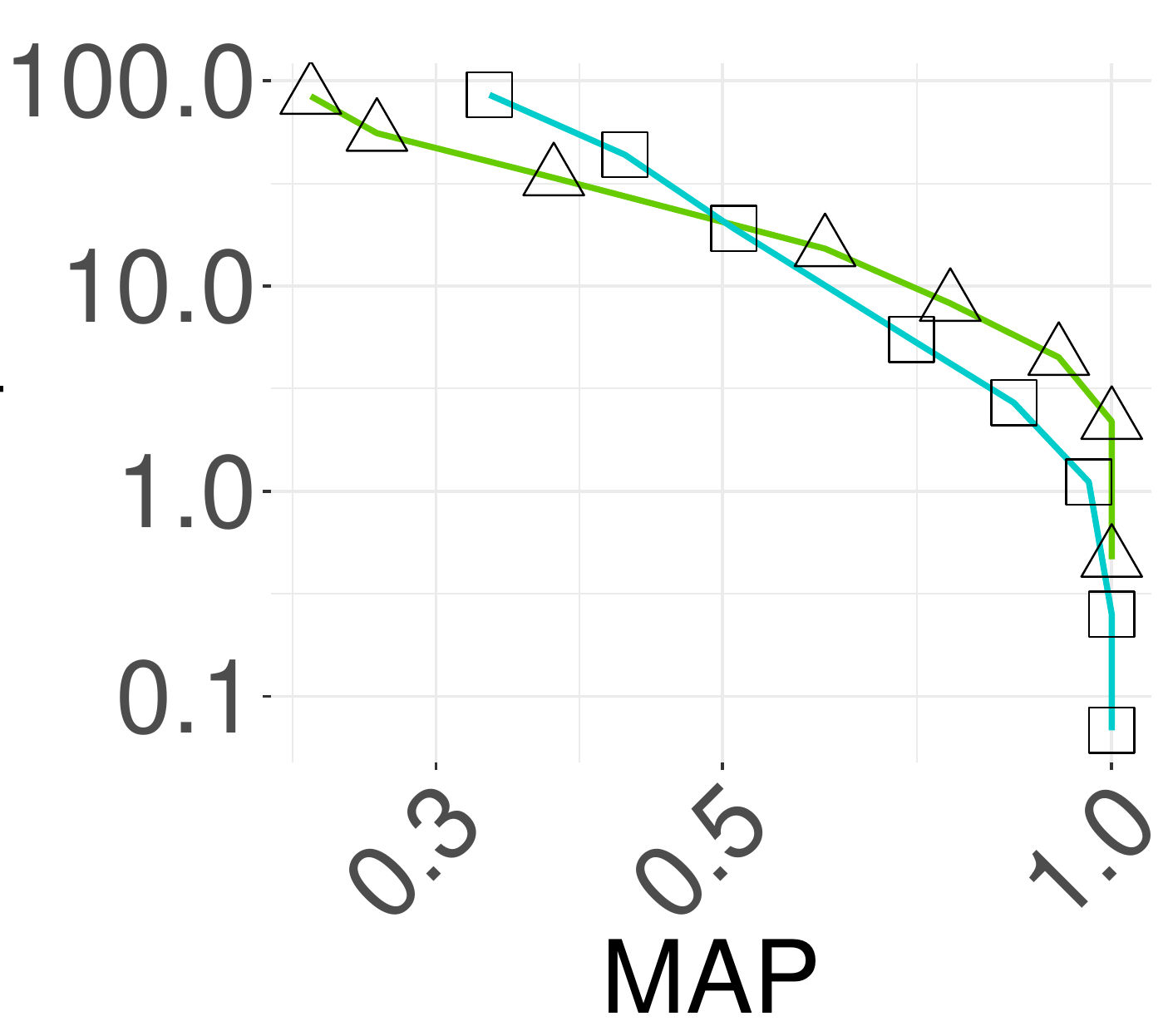}
		\caption{Seismic100GB} 
		\label{fig:approx:accuracy:throughput:seismic:100GB:256:hdd:ng:100NN:100:nocache:best}
	\end{subfigure}	

	\begin{subfigure}{0.18\textwidth}
		\centering
		\includegraphics[width=\textwidth]{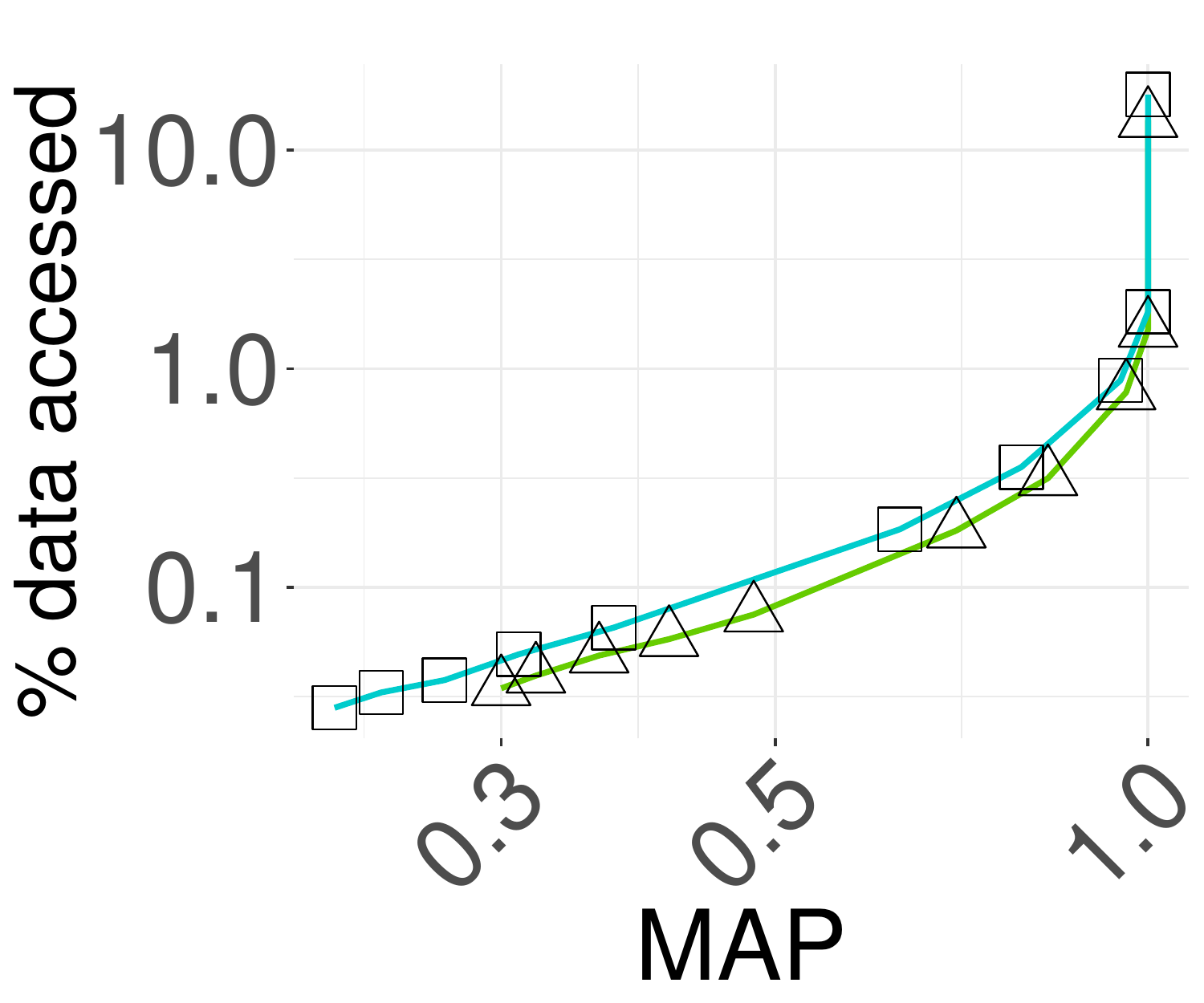}
		\caption{Rand250GB} 
		\label{fig:approx:accuracy:data:synthetic:250GB:256:hdd:de:100NN:100:nocache:best}
	\end{subfigure}
	\begin{subfigure}{0.18\textwidth}
		\centering
		\includegraphics[width=\textwidth]{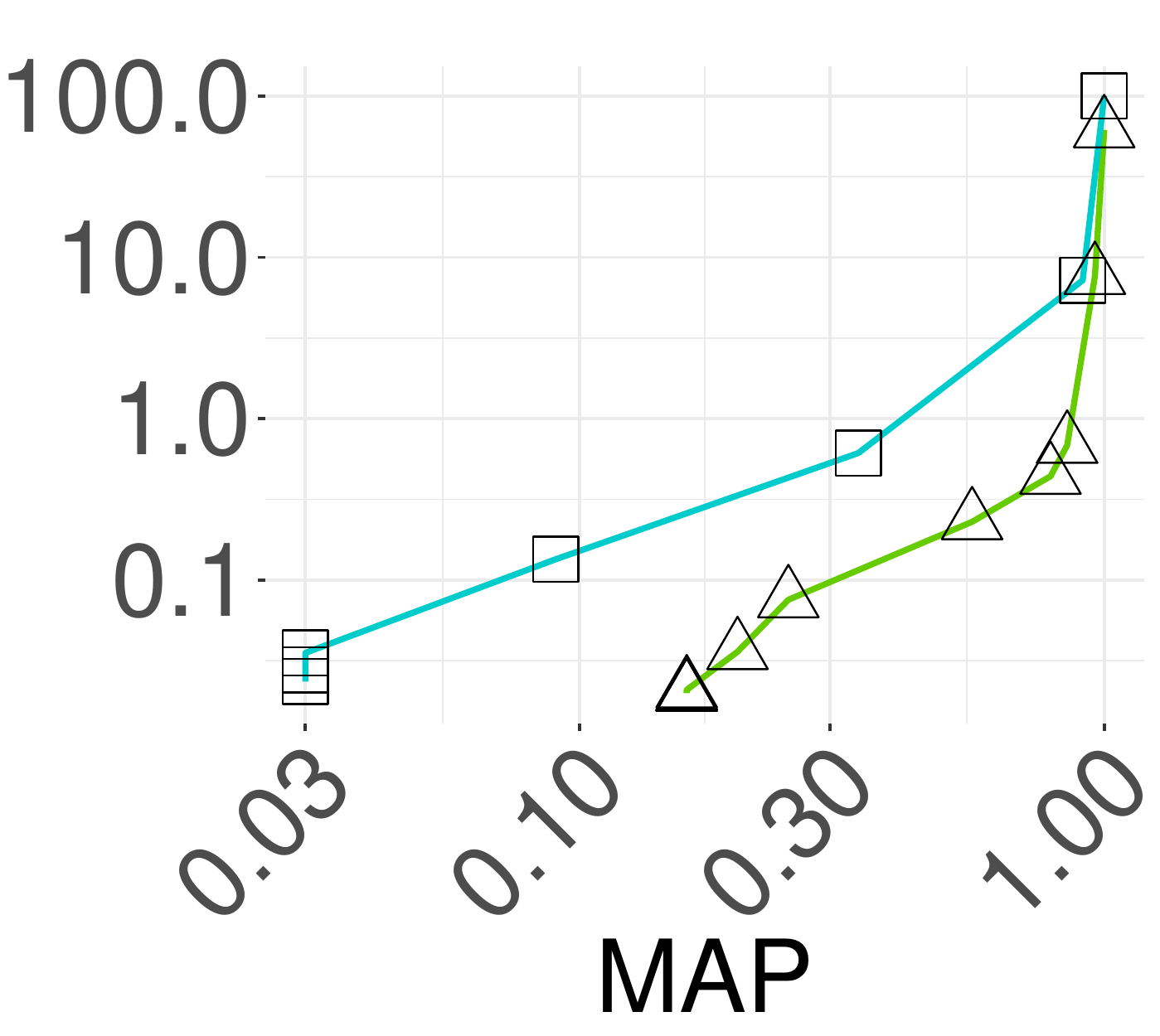}
		\caption{Sift250GB} 
		\label{fig:approx:accuracy:data:sift:250GB:128:hdd:de:100NN:100:nocache:best}
	\end{subfigure}
	\begin{subfigure}{0.18\textwidth}
		\centering
		\includegraphics[width=\textwidth]{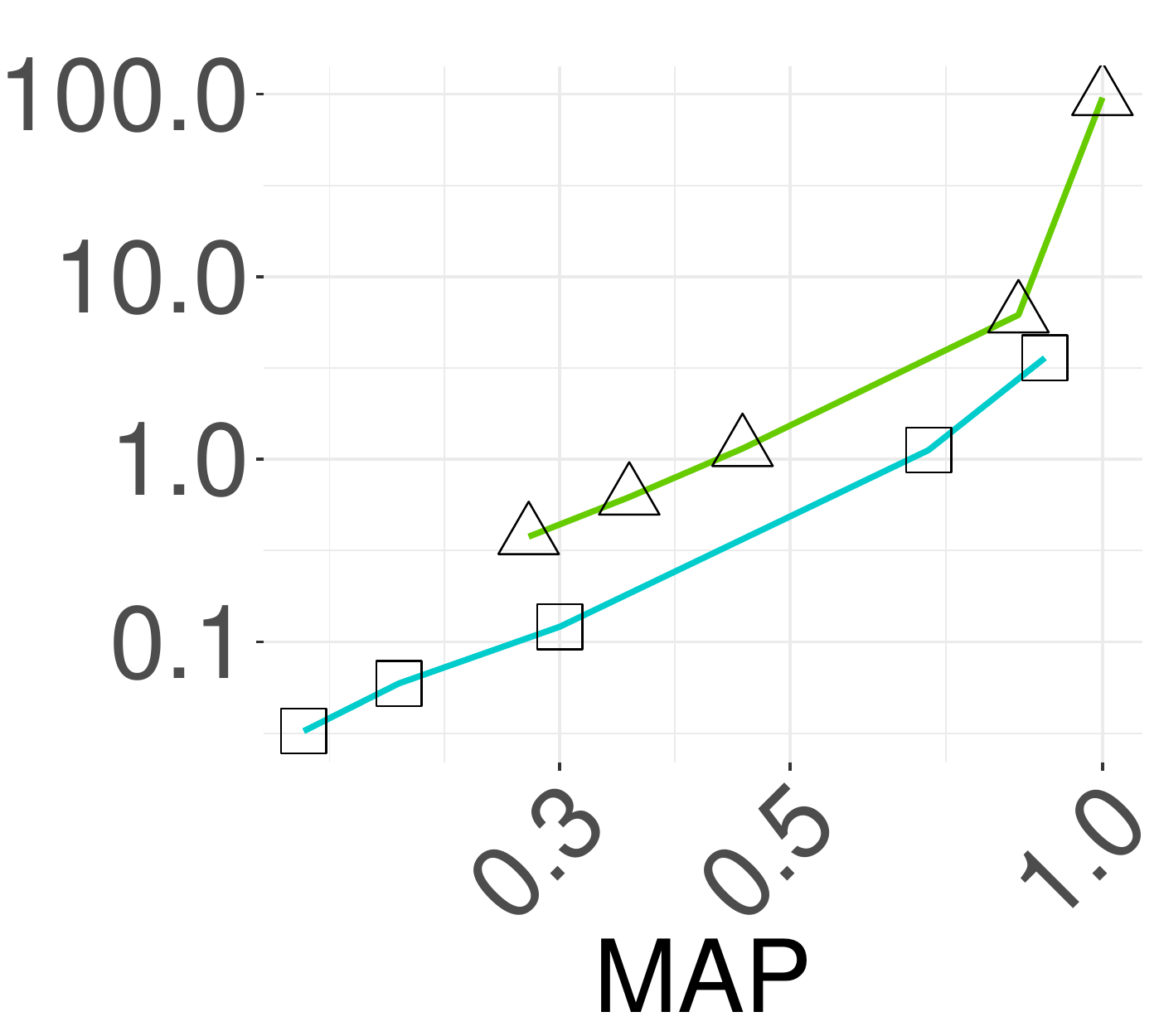}
		\caption{Deep250GB} 
		\label{fig:approx:accuracy:data:deep:250GB:96:hdd:de:100NN:100:nocache:best}
	\end{subfigure}
	\begin{subfigure}{0.18\textwidth}
		\centering
		\includegraphics[width=\textwidth]{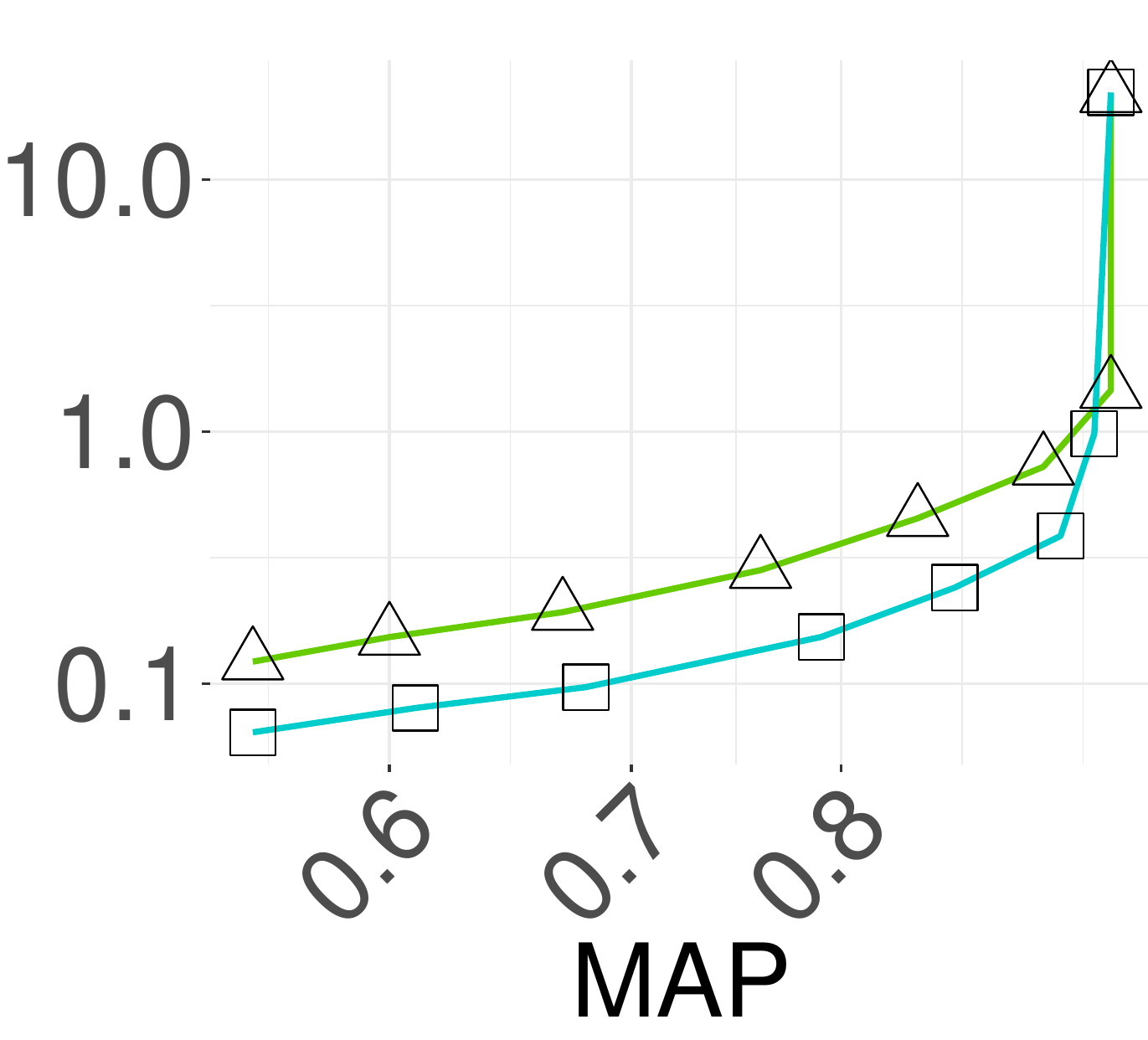}
		\caption{Sald100GB} 
		\label{fig:approx:accuracy:data:sald:100GB:128:hdd:de:100NN:100:nocache:best}
	\end{subfigure}
	\begin{subfigure}{0.18\textwidth}
		\centering
		\includegraphics[width=\textwidth]{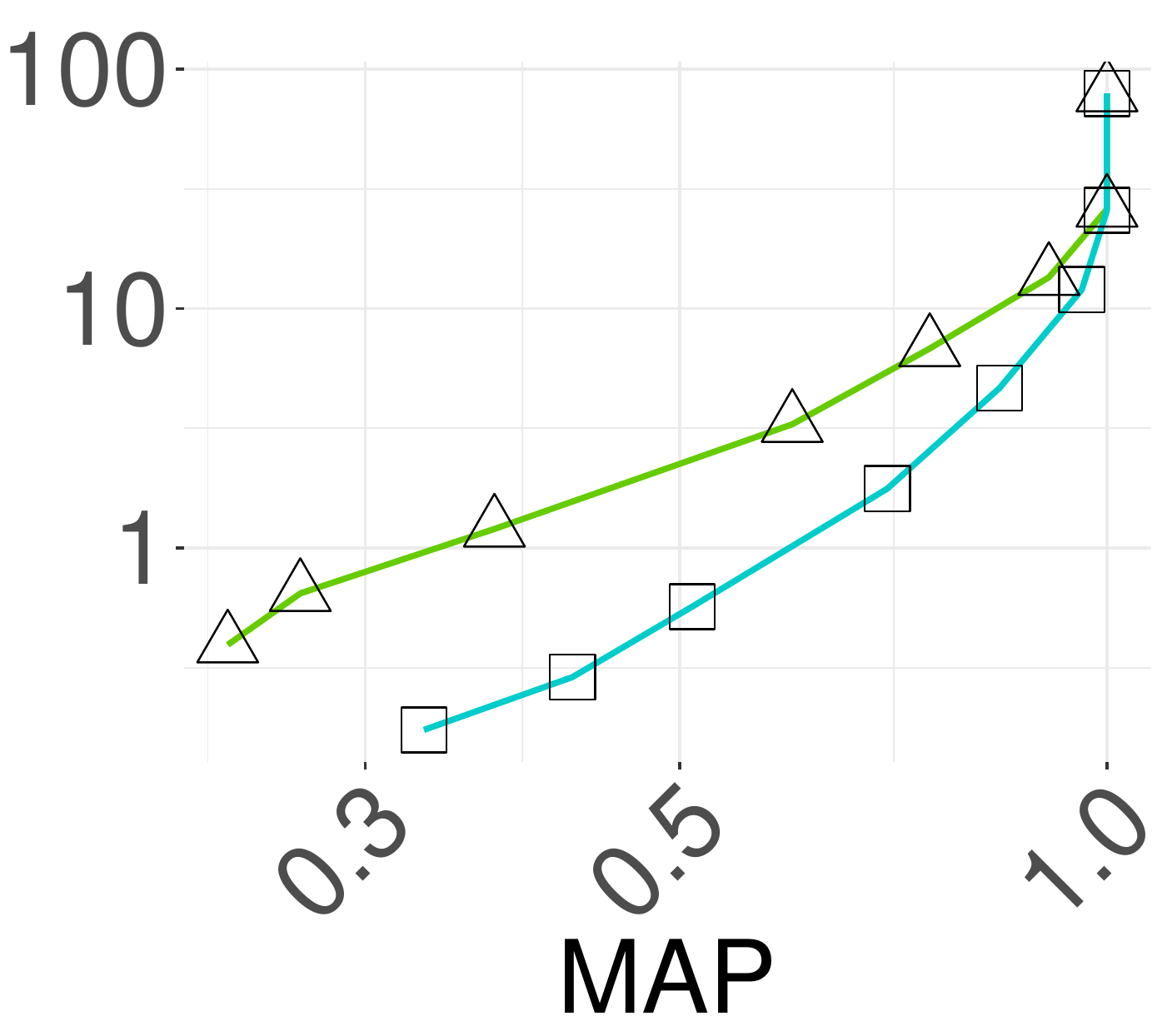}
		\caption{Seismic100GB} 
		\label{fig:approx:accuracy:data:seismic:100GB:256:hdd:de:100NN:100:nocache:best}
	\end{subfigure}

	\begin{subfigure}{0.18\textwidth}
		\centering
		\includegraphics[width=\textwidth]{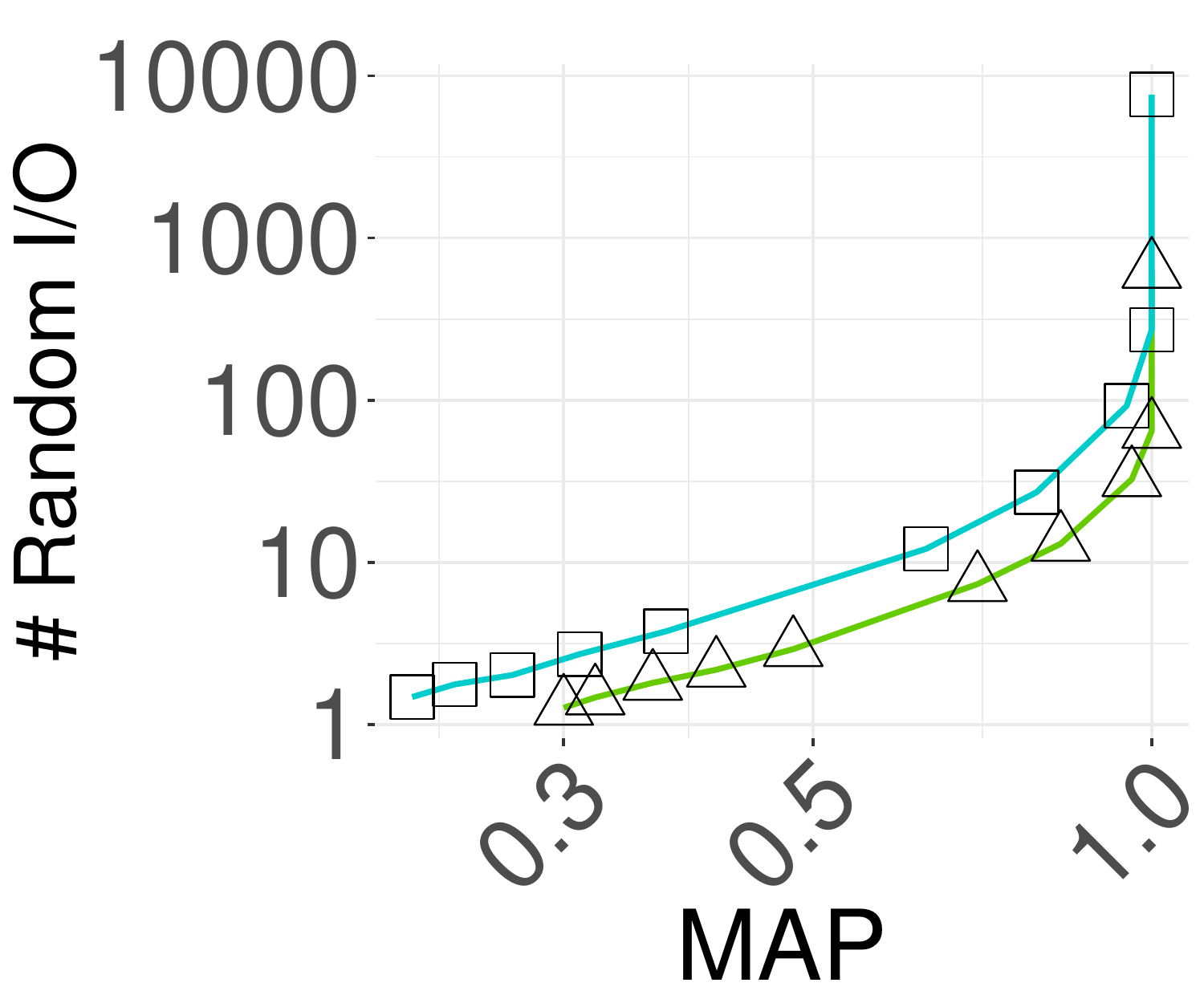}
		\caption{Rand250GB} 
		\label{fig:approx:accuracy:random:synthetic:250GB:256:hdd:de:100NN:100:nocache:best}
	\end{subfigure}
	\begin{subfigure}{0.18\textwidth}
		\centering
		\includegraphics[width=\textwidth]{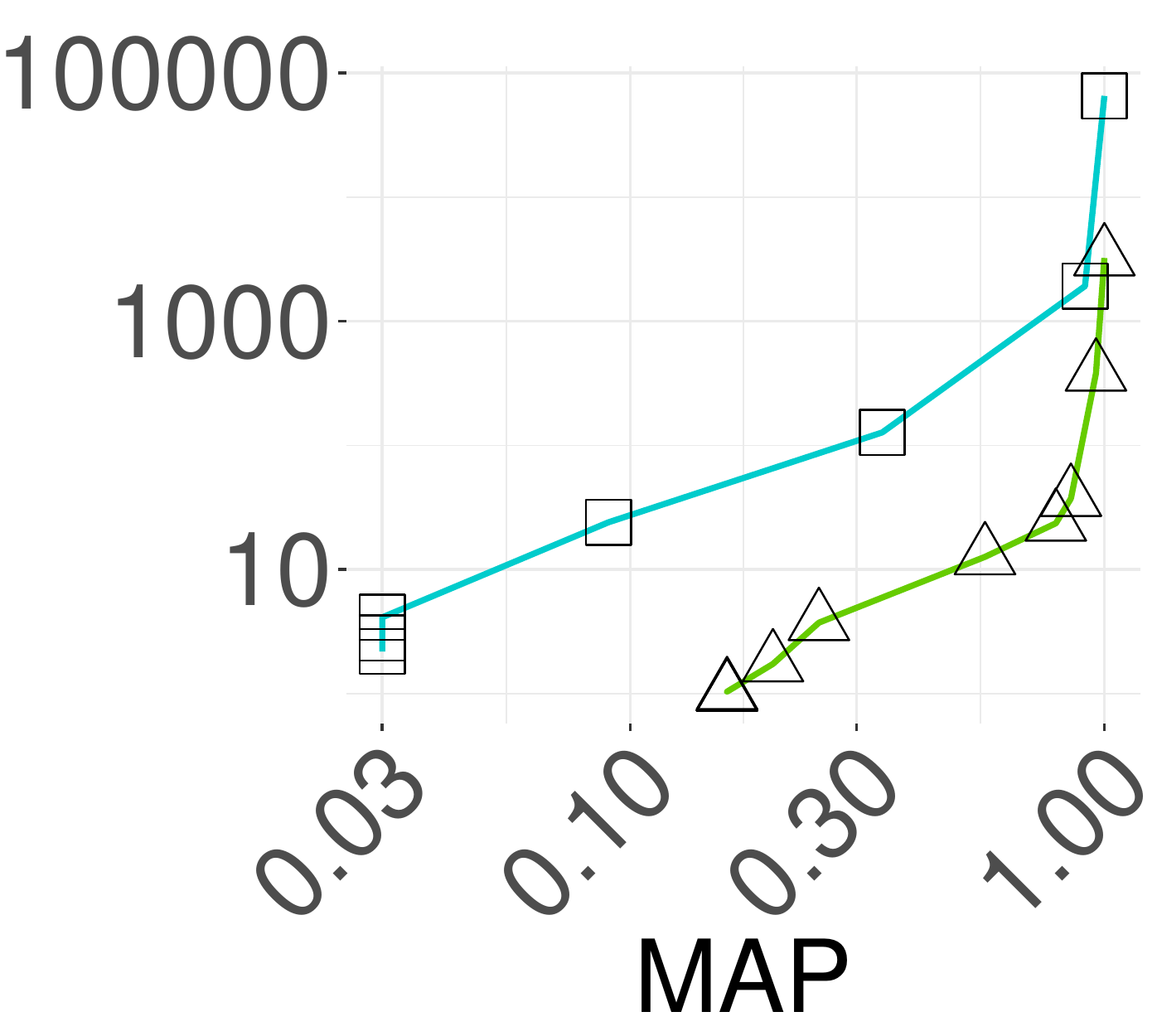}
		\caption{Sift250GB} 
		\label{fig:approx:accuracy:random:sift:250GB:128:hdd:de:100NN:100:nocache:best}
	\end{subfigure}
	\begin{subfigure}{0.18\textwidth}
		\centering
		\includegraphics[width=\textwidth]{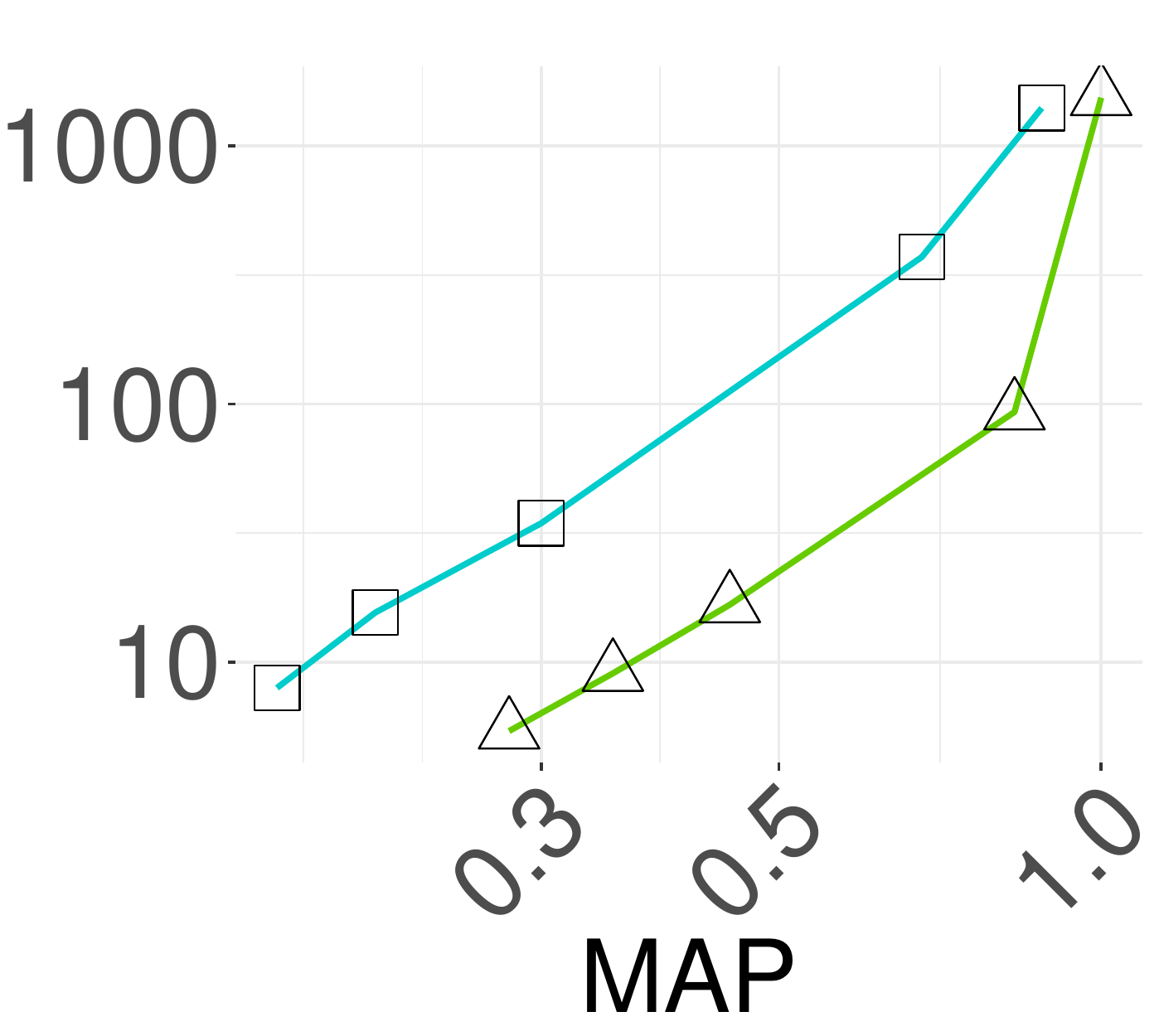}
		\caption{Deep250GB} 
		\label{fig:approx:accuracy:random:deep:250GB:96:hdd:de:100NN:100:nocache:best}
	\end{subfigure}
	\begin{subfigure}{0.18\textwidth}
		\centering
		\includegraphics[width=\textwidth]{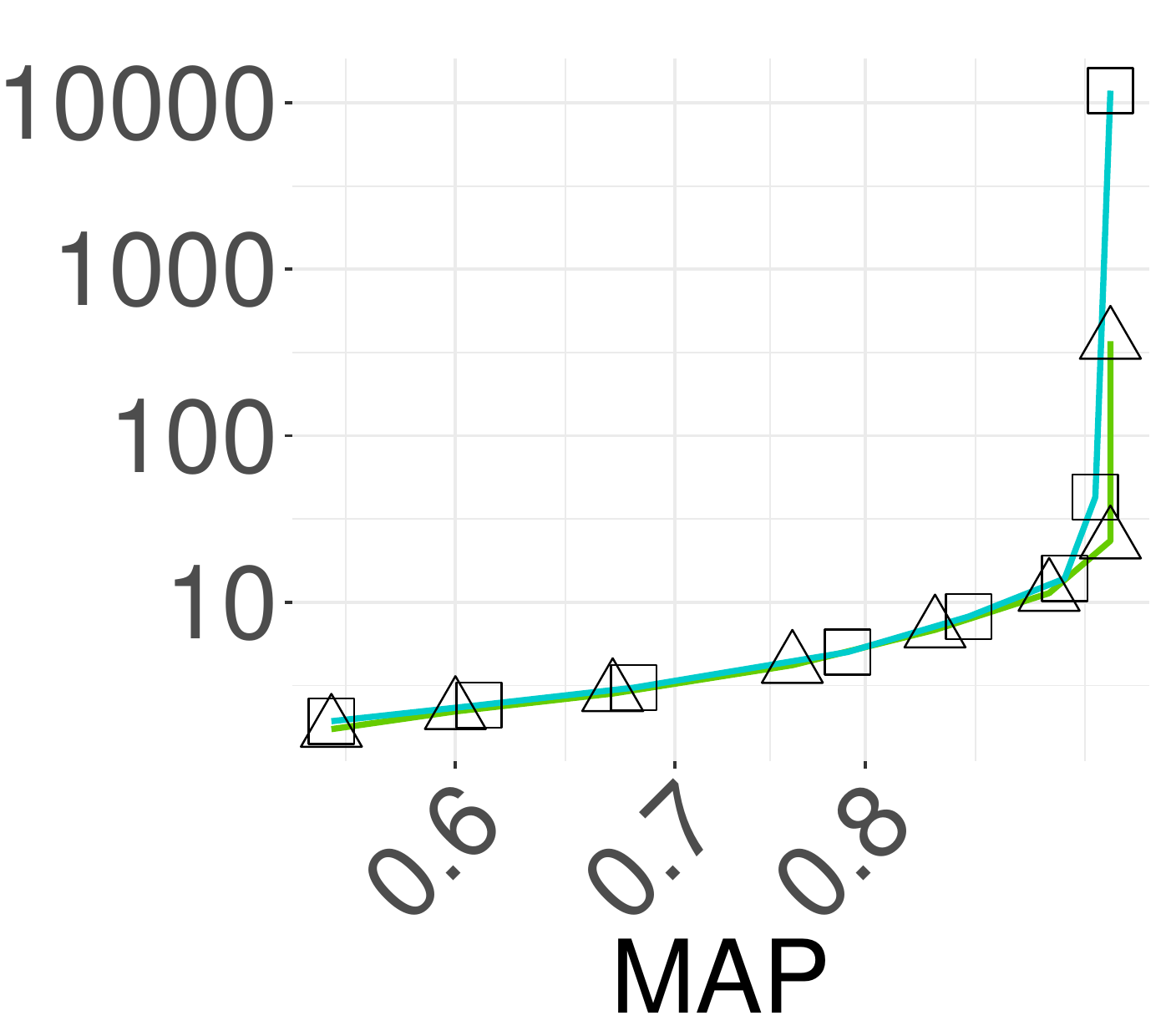}
		\caption{Sald100GB} 
		\label{fig:approx:accuracy:random:sald:100GB:128:hdd:de:100NN:100:nocache:best}
	\end{subfigure}
	\begin{subfigure}{0.18\textwidth}
		\centering
		\includegraphics[width=\textwidth]{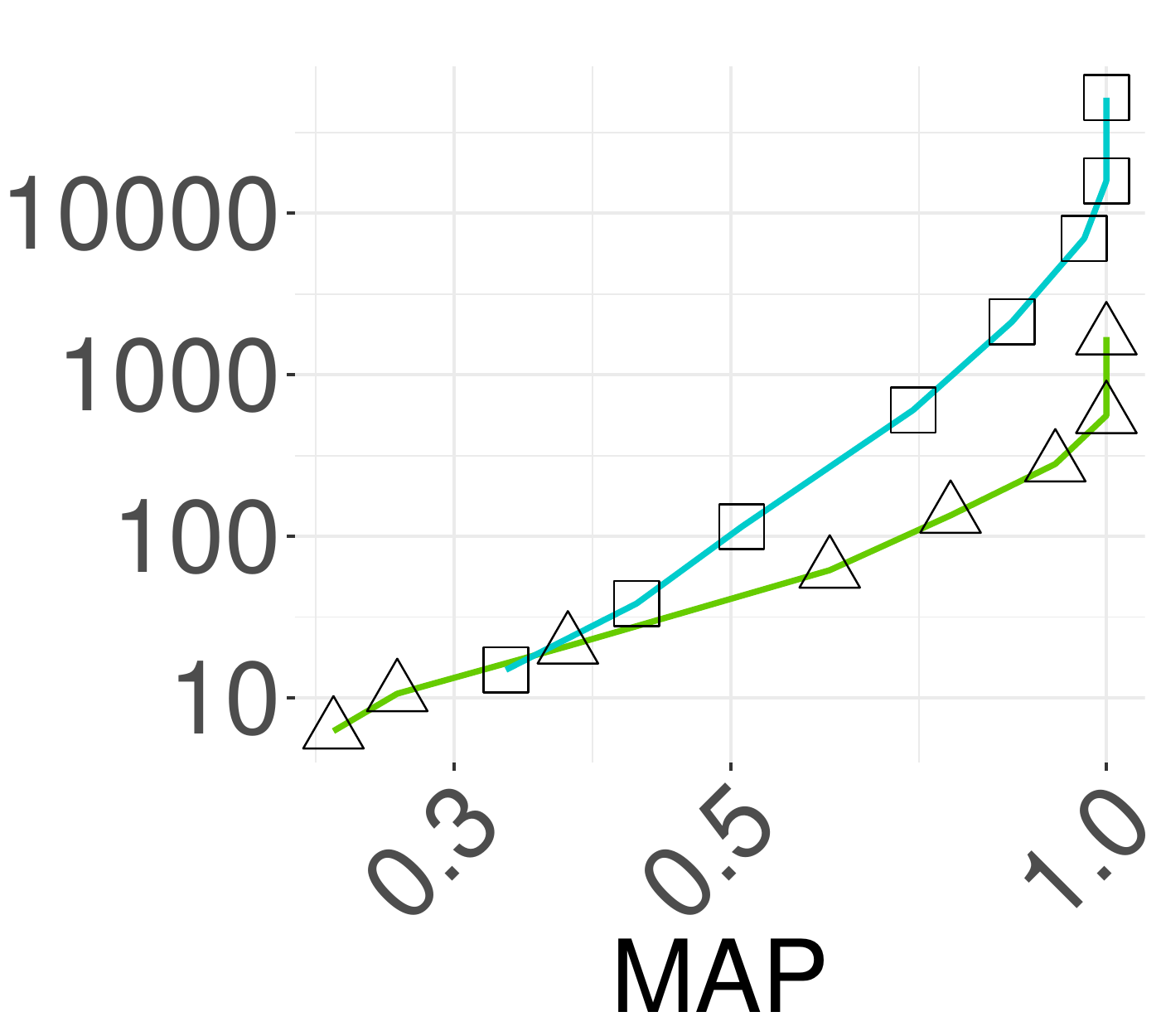}
		\caption{Seismic100GB} 
		\label{fig:approx:accuracy:random:seismic:100GB:256:hdd:de:100NN:100:nocache:best}
	\end{subfigure}
	\caption{{\color{black} Efficiency vs. accuracy for the best methods ($\bm{\epsilon}$-approximate)}}
	\label{fig:approx:accuracy:data:250GB:hdd:best}
\end{minipage}
\end{figure*}

\noindent\textbf{Amount of data accessed.} 
As expected, both DSTree and iSAX2+ need to access more data as the accuracy increases. 
Nevertheless, we observe that to achieve accuracies of almost 1, both methods access close to 100\% of the data for Sift250GB  (Figure~\ref{fig:approx:accuracy:data:sift:250GB:128:hdd:de:100NN:100:nocache:best}), Deep250GB (Figure~\ref{fig:approx:accuracy:data:deep:250GB:96:hdd:de:100NN:100:nocache:best}) and Seismic100GB (Figure~\ref{fig:approx:accuracy:data:seismic:100GB:256:hdd:de:100NN:100:nocache:best}), compared to 10\% of data accessed on Sald100GB (Figure~\ref{fig:approx:accuracy:data:sald:100GB:128:hdd:de:100NN:100:nocache:best}) and Rand250GB. (Figure~\ref{fig:approx:accuracy:data:synthetic:250GB:256:hdd:de:100NN:100:nocache:best}). 
The percentage of accessed data also varies among real datasets, Deep250GB and Sift250GB requiring the most. Note that for some datasets, a MAP of 1 is achievable with minimal data access. For instance DSTree needs to access about 1\% of the data to get a MAP of 1 on Sald100GB (Figure~\ref{fig:approx:accuracy:data:sald:100GB:128:hdd:de:100NN:100:nocache:best}).

\noindent\textbf{Number of Random I/Os.} 
To understand the nature of the data accesses discussed above, we report the number of random I/Os in Figure~\ref{fig:approx:accuracy:data:250GB:hdd:best} (bottom row). 
Overall, iSAX2+ incurs a higher number of random I/Os for all datasets. 
This is because iSAX2+ has a larger number of leaves, with a smaller fill factor than DSTree~\cite{journal/pvldb/echihabi2018}. 
For instance, the large number of random I/Os incurred by iSAX2+ (Figure~\ref{fig:approx:accuracy:random:seismic:100GB:256:hdd:de:100NN:100:nocache:best}) is what explains the faster runtime of DSTree on the Seismic100GB dataset (Figure~\ref{fig:approx:accuracy:throughput:seismic:100GB:256:hdd:ng:100NN:100:nocache:best}), even if DSTree accesses more data than iSAX2+ for higher MAP values (Figure~\ref{fig:approx:accuracy:data:seismic:100GB:256:hdd:de:100NN:100:nocache:best}). 
The Sald100GB dataset is an exception to this trend 
as iSAX2+ outperforms DSTree on all accuracies except for MAP is 1 (Figure~\ref{fig:approx:accuracy:throughput:sald:100GB:128:hdd:ng:100NN:100:nocache:best}), because it accesses less data incuring almost the same random I/O (Figures~\ref{fig:approx:accuracy:data:sald:100GB:128:hdd:de:100NN:100:nocache:best} and~\ref{fig:approx:accuracy:random:sald:100GB:128:hdd:de:100NN:100:nocache:best}).

\noindent\textbf{Effect of k.} Figure~\ref{fig:approx:efficiency:k:hdd} summarizes experiments varying k on different datasets in-memory and on-disk. 
We measure the total time required to complete a workload of 100 queries for each value of k. 
We observe that 
finding the first neighbor is the most costly operation, while finding the additional neighbors is much cheaper. 

\begin{figure}[tb]
	\captionsetup{justification=centering}
	\captionsetup[subfigure]{justification=centering}
	\begin{subfigure}{0.32\columnwidth}
		\centering
		\includegraphics[scale=0.23]{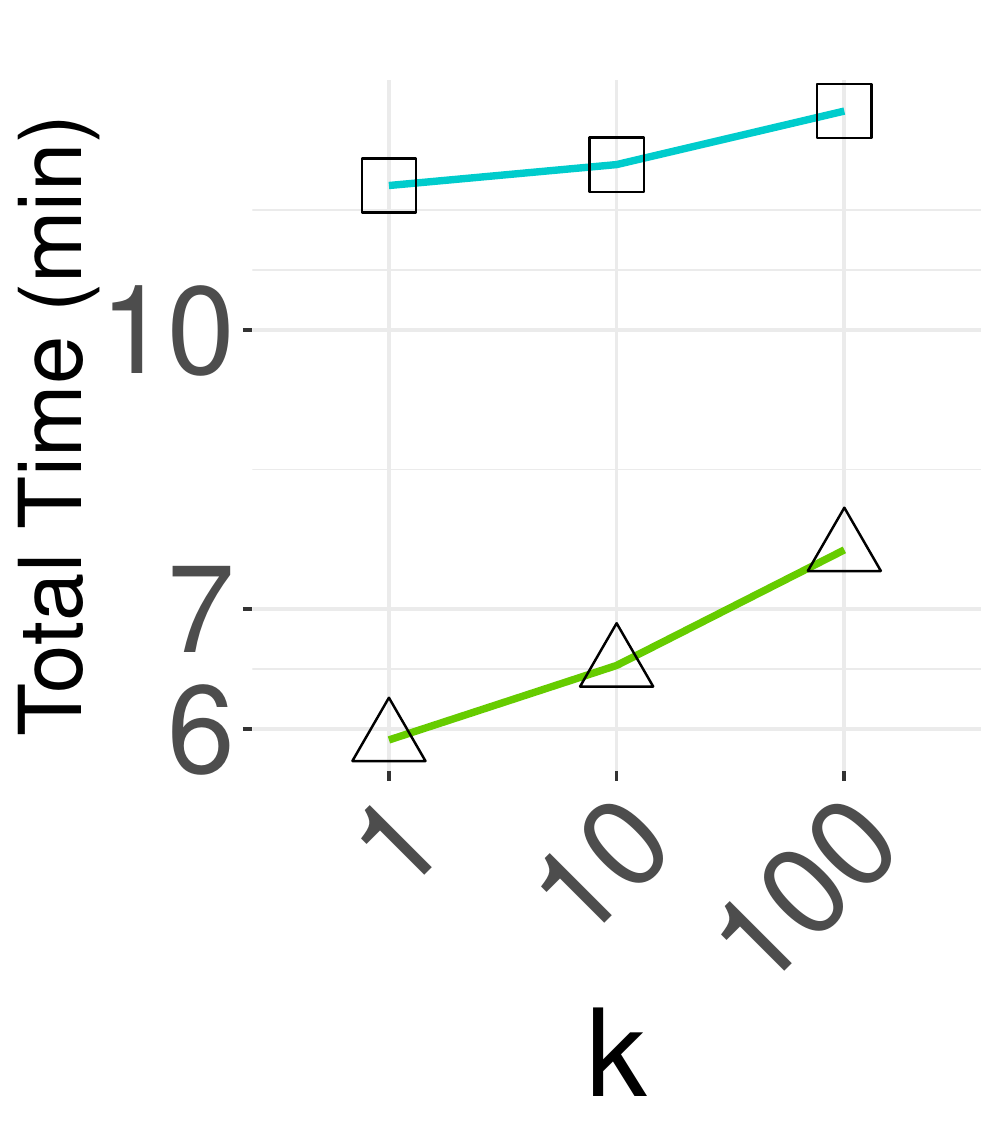}		\caption{Rand25GB}  
		\label{fig:approx:efficiency:k:hdd:rand:25GB}
	\end{subfigure}
	\begin{subfigure}{0.32\columnwidth}
		\centering
		\includegraphics[scale=0.23]{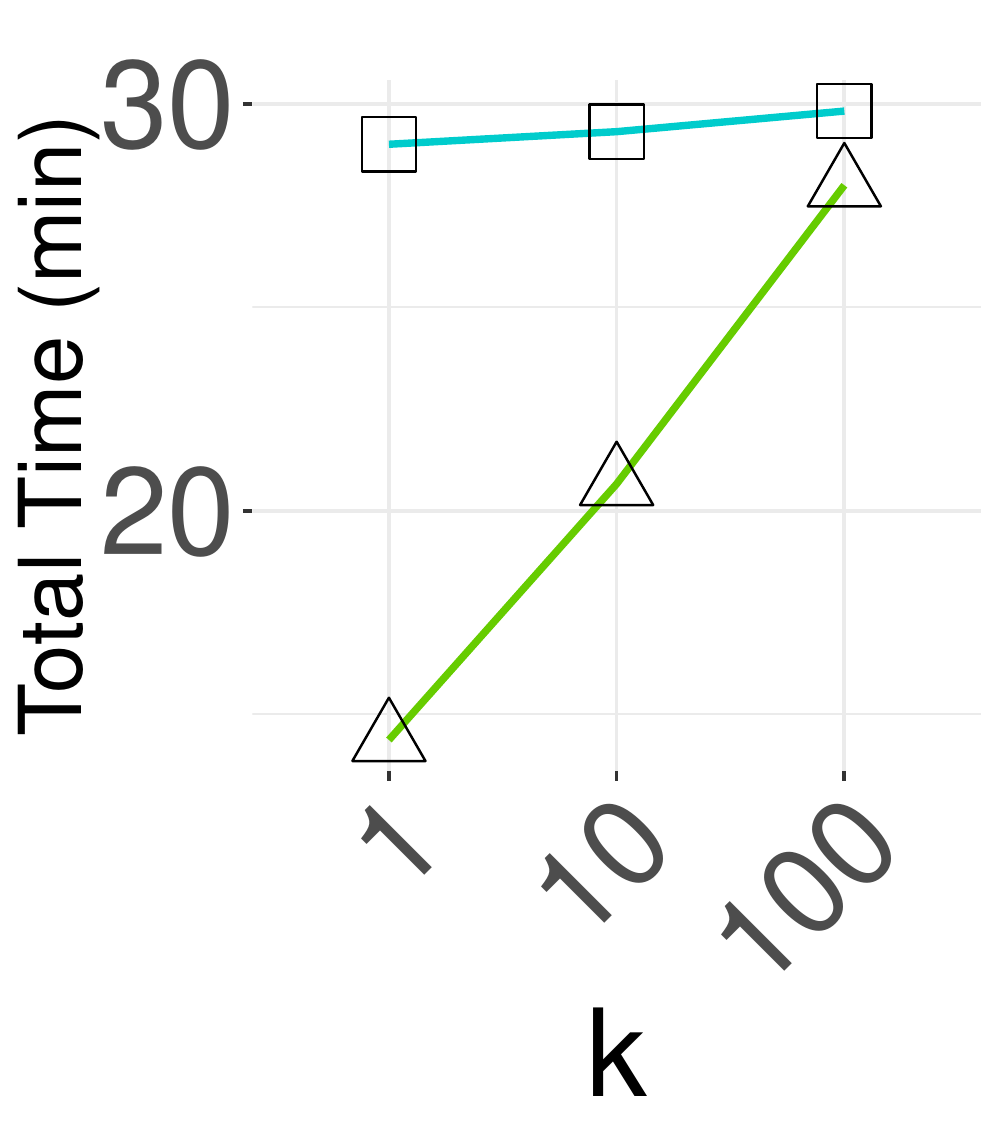}\\
		\caption{Sift25GB}  
		\label{fig:approx:efficiency:k:hdd:sift:25GB}
	\end{subfigure}
	\begin{subfigure}{0.32\columnwidth}
		\centering
		\includegraphics[scale=0.23]{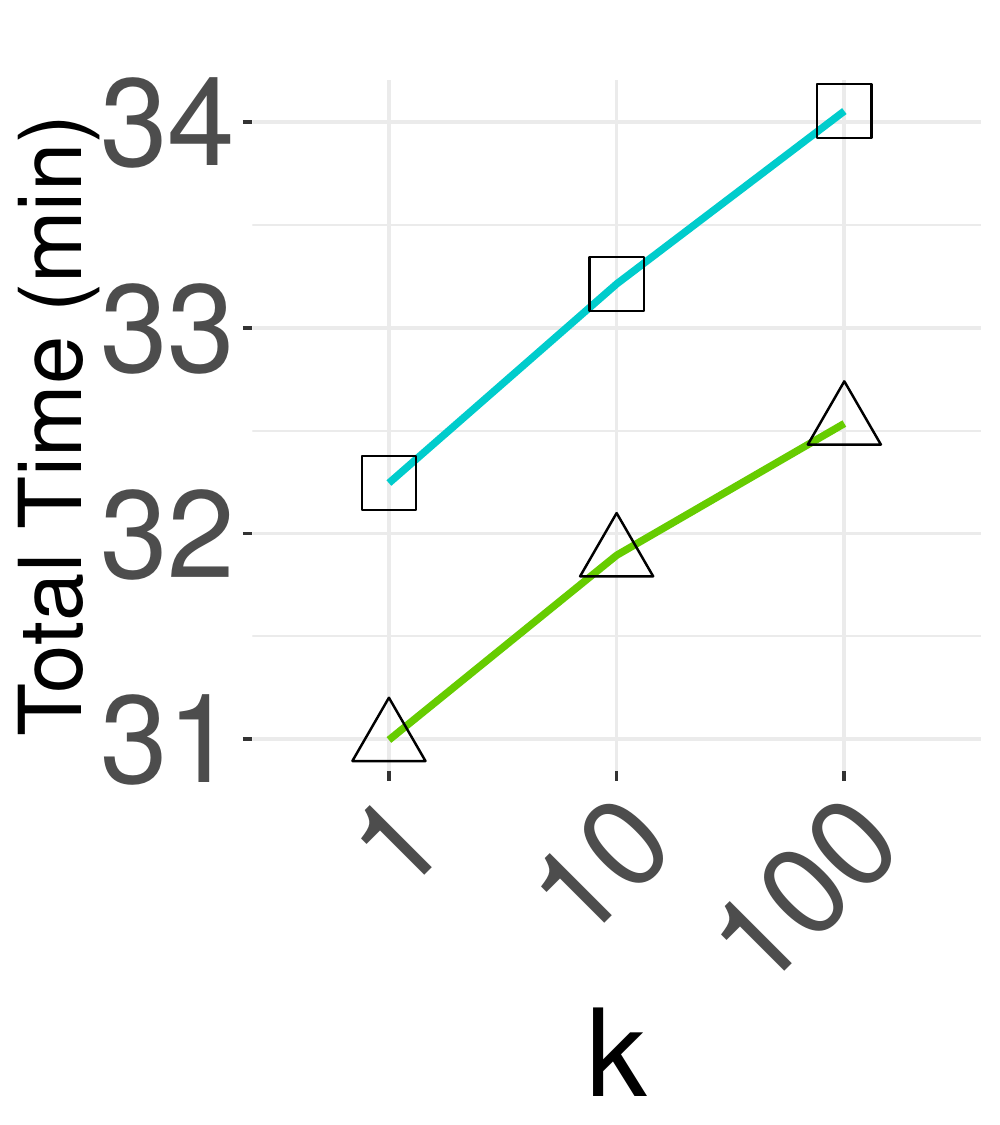}
		\caption{Deep25GB}  
		\label{fig:approx:efficiency:k:hdd:deep:25GB}
	\end{subfigure}\\
	\begin{subfigure}{0.32\columnwidth}
		\centering
		\includegraphics[scale=0.23]{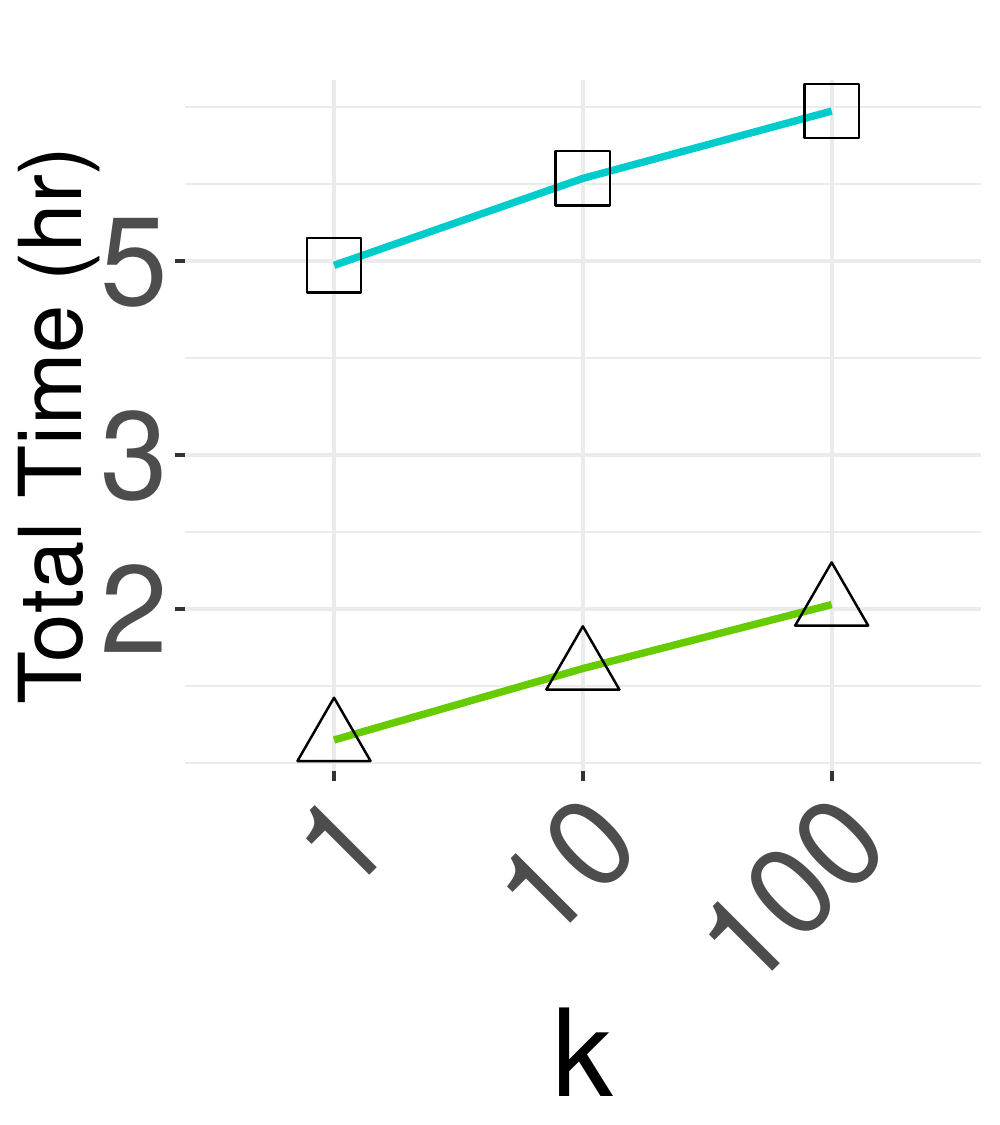}
		\caption{Rand250GB} 
		\label{fig:approx:efficiency:k:hdd:rand:250GB}
	\end{subfigure}
	\begin{subfigure}{0.32\columnwidth}
		\centering
		\includegraphics[scale=0.23]{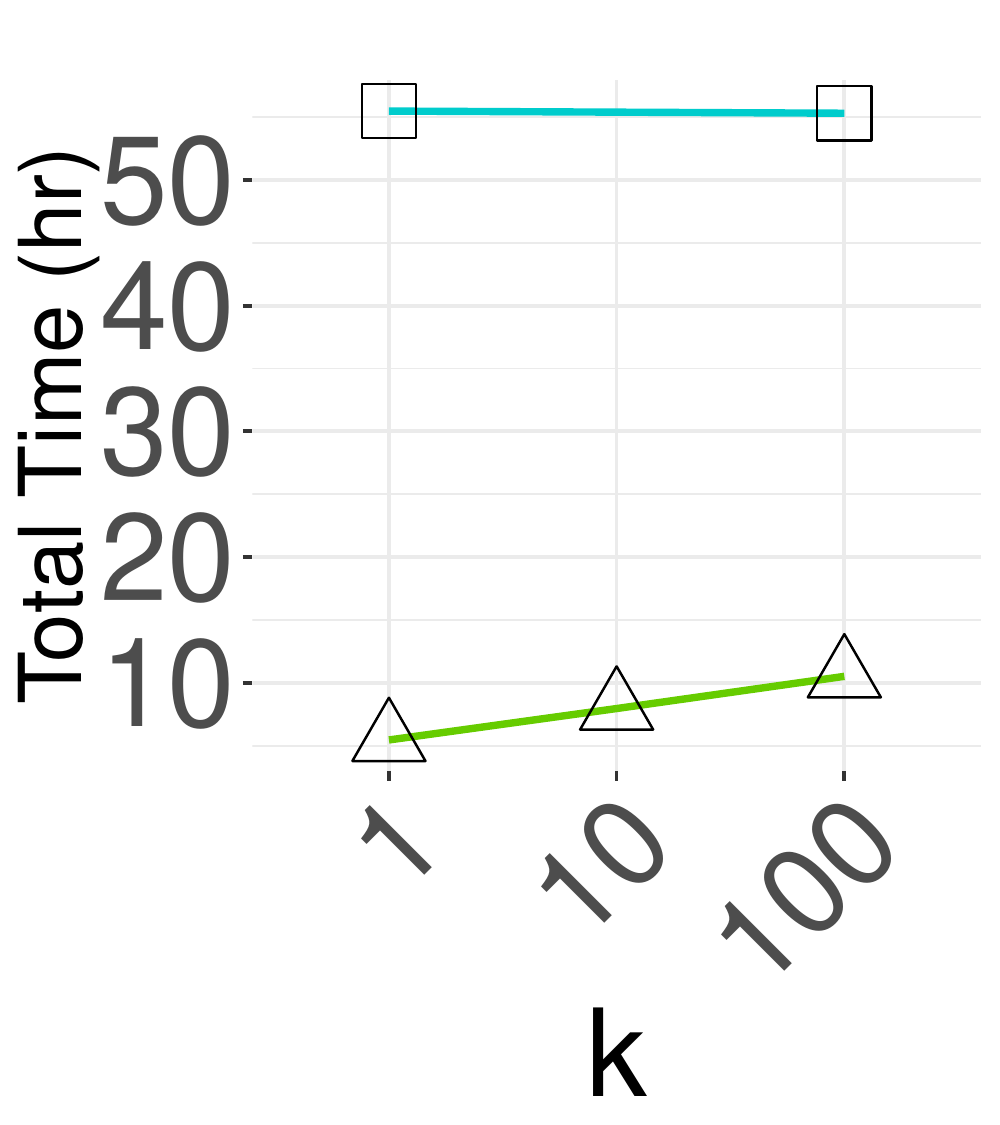}
		\caption{Sift250GB} 
		\label{fig:approx:efficiency:k:hdd:sift:250GB}
	\end{subfigure}
	\begin{subfigure}{0.32\columnwidth}
		\centering
		\includegraphics[scale=0.23]{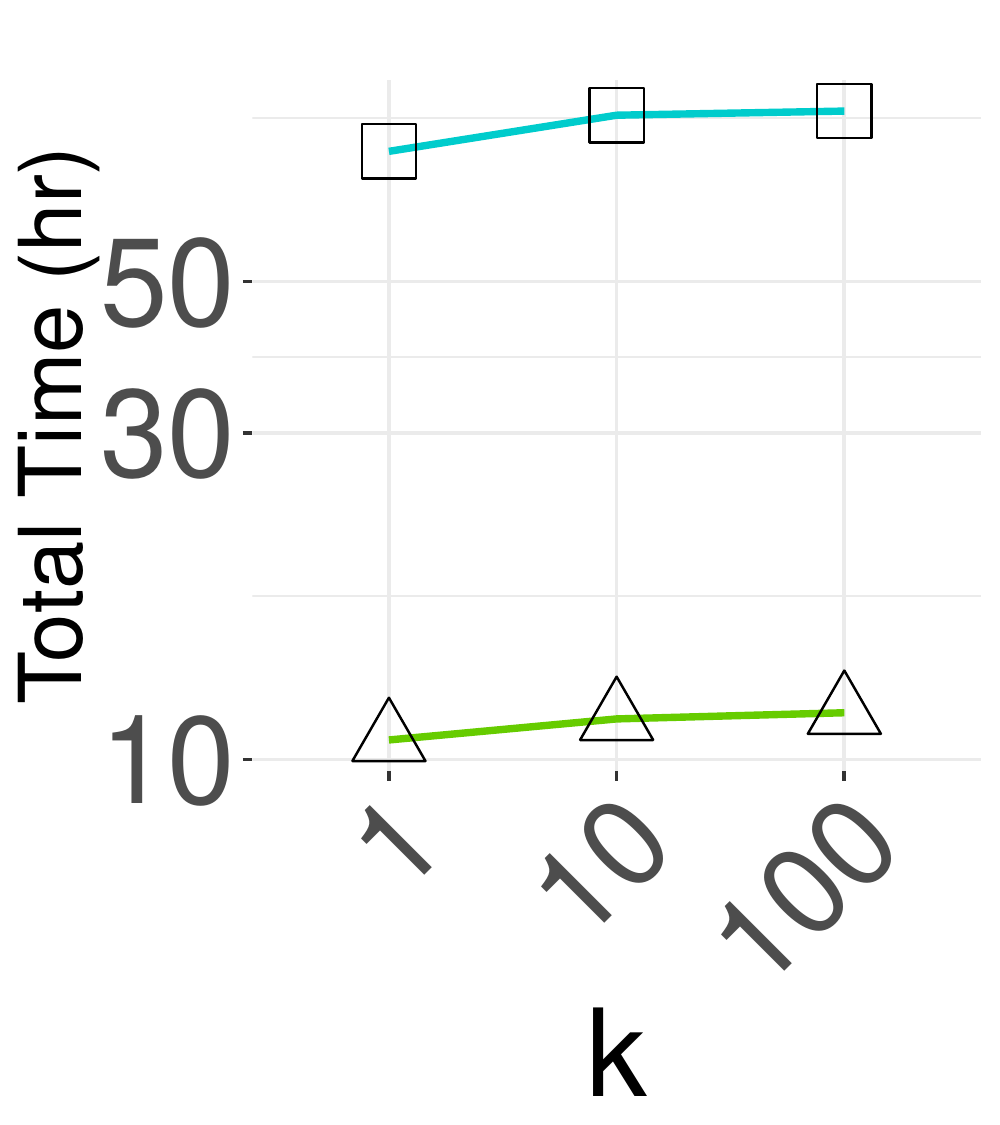}
		\caption{Deep250GB} 
		\label{fig:approx:efficiency:k:hdd:deep:250GB}
	\end{subfigure}
	\caption{Efficiency vs. k ($\bm{\epsilon}$-approximate)}	
	\label{fig:approx:efficiency:k:hdd}
	\vspace*{-0.5cm}
\end{figure}

\noindent\textbf{Effect of $\delta$ and $\epsilon$.} 
In Figure~\ref{fig:approx:accuracy_efficiency:delta:epsilon:synthetic:250GB:hdd}, we describe in more detail how varying $\delta$ and $\epsilon$ affects the performance of DSTree and iSAX2+. 
Figure~\ref{fig:approx:efficiency:epsilon:synthetic:250GB:hdd} shows that the throughput of both methods increases dramatically with increasing $\epsilon$. 
For example, a small value of $\epsilon = 5$ increases the throughput of iSAX2+ by two orders of magnitude, when compared to exact search ($\epsilon$ = 0). 
Moreover, note that both methods return the actual exact answers for small $\epsilon$ values, and accuracy drops only as $\epsilon$ goes beyond $2$ (Figure~\ref{fig:approx:accuracy:mapk:epsilon:synthetic:250GB:hdd}).
In addition, Figure~\ref{fig:approx:accuracy:mape:epsilon:synthetic:250GB:hdd} shows that the actual approximation error MRE is well below the user-tolerated threshold (represented by $\epsilon$), even for $\epsilon$ values well above $2$.
The above observations mean that these methods can be used in approximate mode, achieving very high throughput, while still returning answers that are exact (or very close to the exact).

As the probability $\delta$ increases, throughput stays constant and only plummets when search becomes exact ($\delta =1$ in Figure~\ref{fig:approx:efficiency:delta:synthetic:250GB:hdd}).
Similarly, accuracy also stays constant, then slightly increases (for a very high $\delta$ of 0.99), reaching 1 for exact search (Figure~\ref{fig:approx:accuracy:delta:synthetic:250GB:hdd}).
Accuracy plateaus as $\delta$ increases, because the first $ng$-approximate answer found by both algorithms is very close to the exact answer (Figures~\ref{fig:approx:accuracy:mapk:epsilon:synthetic:250GB:hdd} and~\ref{fig:approx:accuracy:mape:epsilon:synthetic:250GB:hdd}) and better than the approximation of $r_\delta$, thus the stopping condition is never triggered. 
When a high value of $\delta$ is used, the stopping condition takes effect for some queries, but the runtime is very close to that of the exact algorithm.

\begin{figure}[tb]
	\captionsetup{justification=centering}
	\captionsetup[subfigure]{justification=centering}
	\begin{subfigure}{0.30\columnwidth}
		\centering
		\includegraphics[width=\columnwidth]{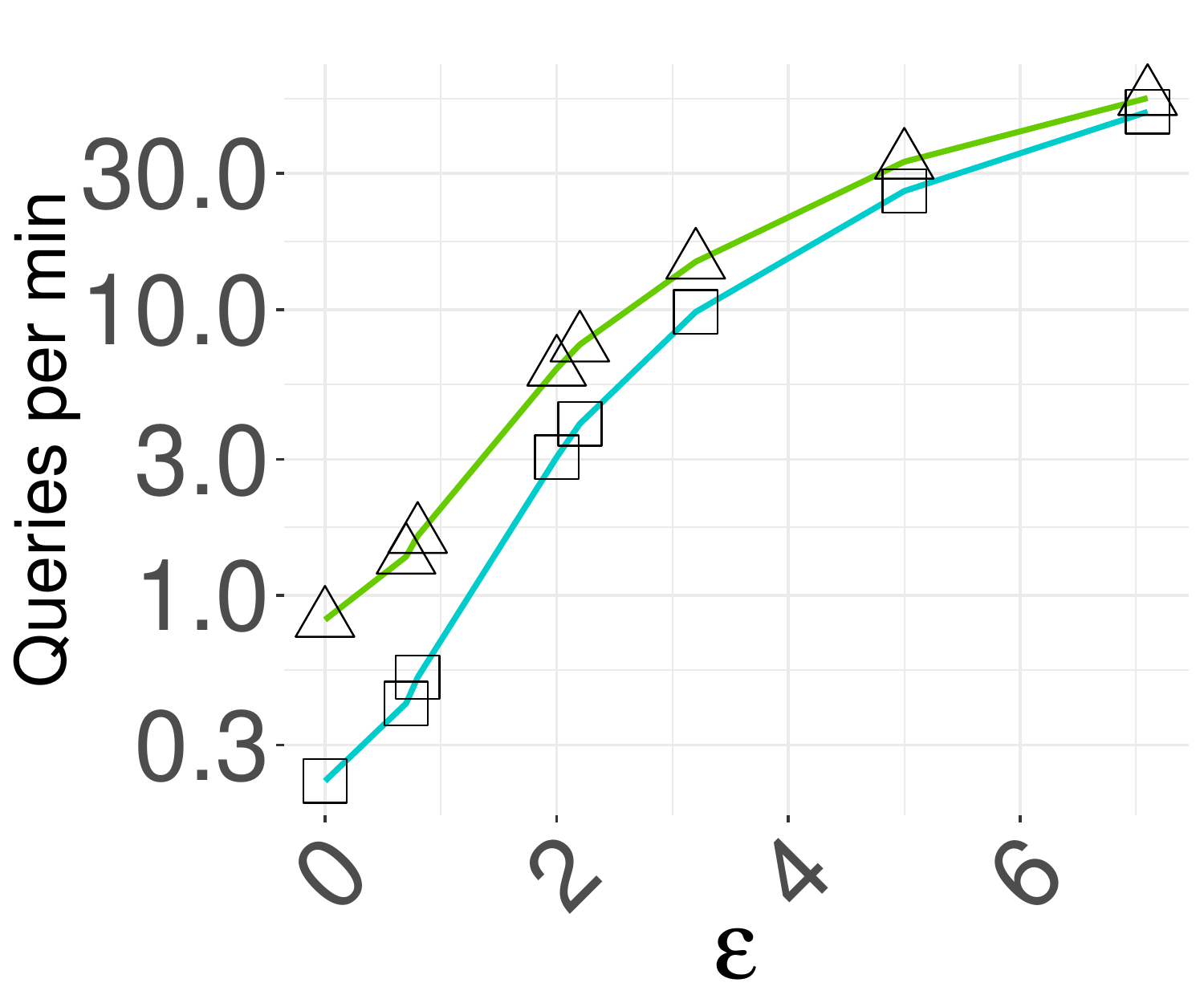}
		\caption{Time vs. $\bm{\epsilon}$ \\
			($\bm{\delta = 1}$)}  
		\label{fig:approx:efficiency:epsilon:synthetic:250GB:hdd}
	\end{subfigure}
	\begin{subfigure}{0.30\columnwidth}
		\centering
		\includegraphics[width=\columnwidth]{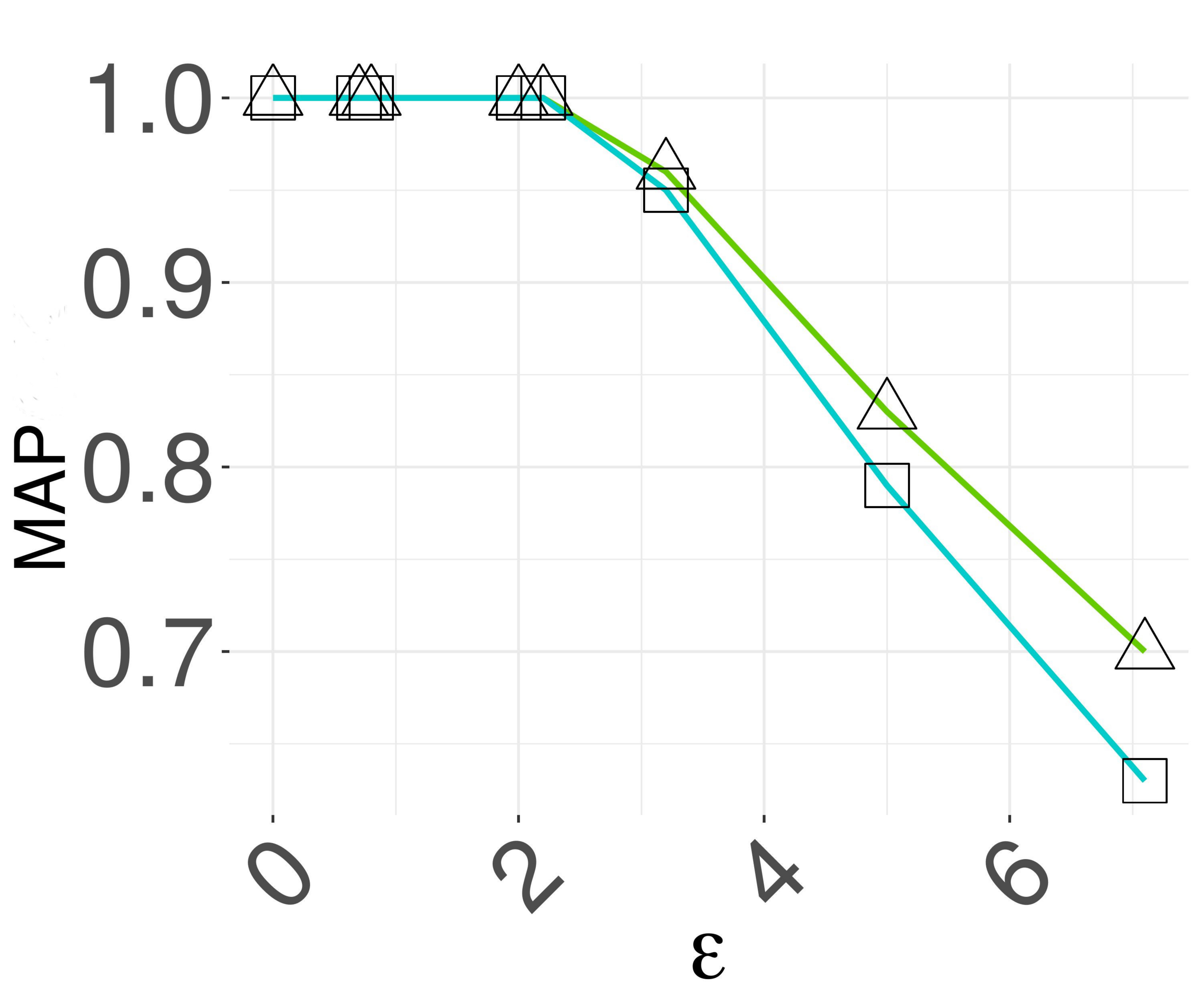}
		\caption{\color{black} MAP vs. $\bm{\epsilon}$ \\ ($\bm{\delta = 1}$)}	
		\label{fig:approx:accuracy:mapk:epsilon:synthetic:250GB:hdd}
	\end{subfigure}
	\begin{subfigure}{0.30\columnwidth}
		\centering
		\includegraphics[width=\columnwidth]{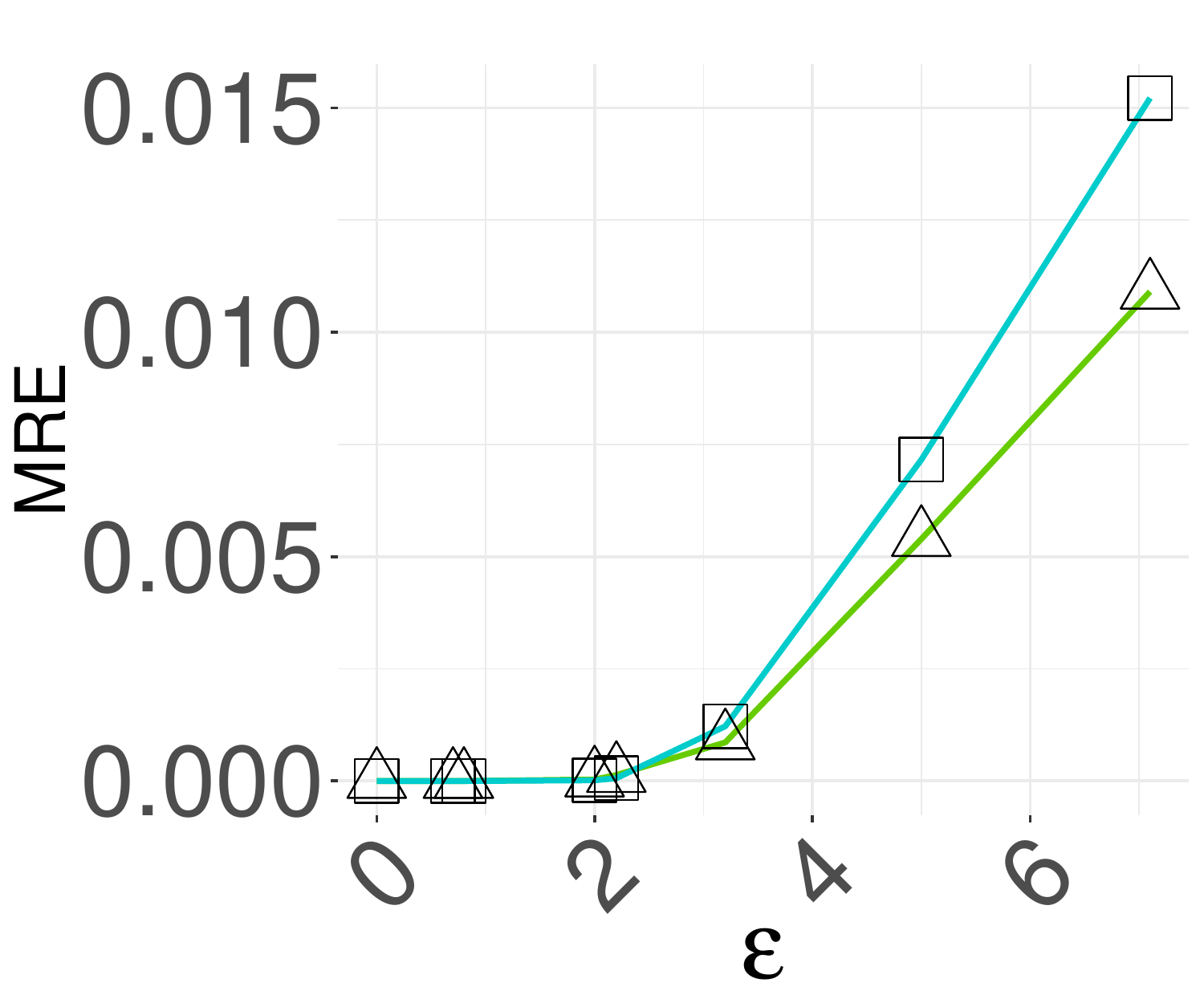}
		\caption{MRE vs. $\bm{\epsilon}$ \\
			($\bm{\delta = 1}$)}  
		\label{fig:approx:accuracy:mape:epsilon:synthetic:250GB:hdd}
	\end{subfigure}\\
	\begin{subfigure}{0.30\columnwidth}
		\centering
		\includegraphics[width=\columnwidth]{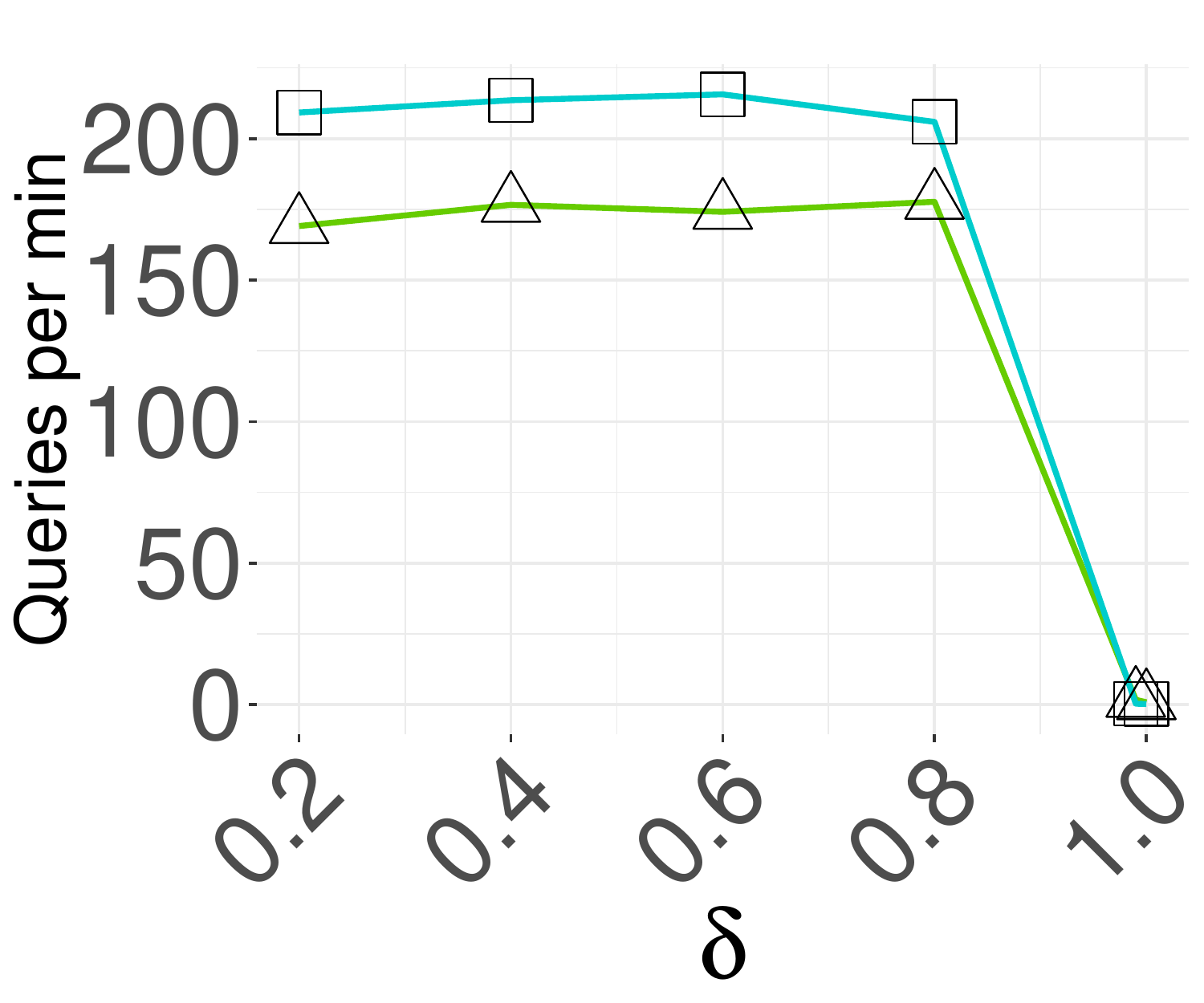}
		\caption{Time vs. $\bm{\delta}$ \\
			($\bm{\epsilon = 0}$)}  
		\label{fig:approx:efficiency:delta:synthetic:250GB:hdd}
	\end{subfigure}
	\begin{subfigure}{0.34\columnwidth}
		\centering
		\includegraphics[width=\columnwidth]{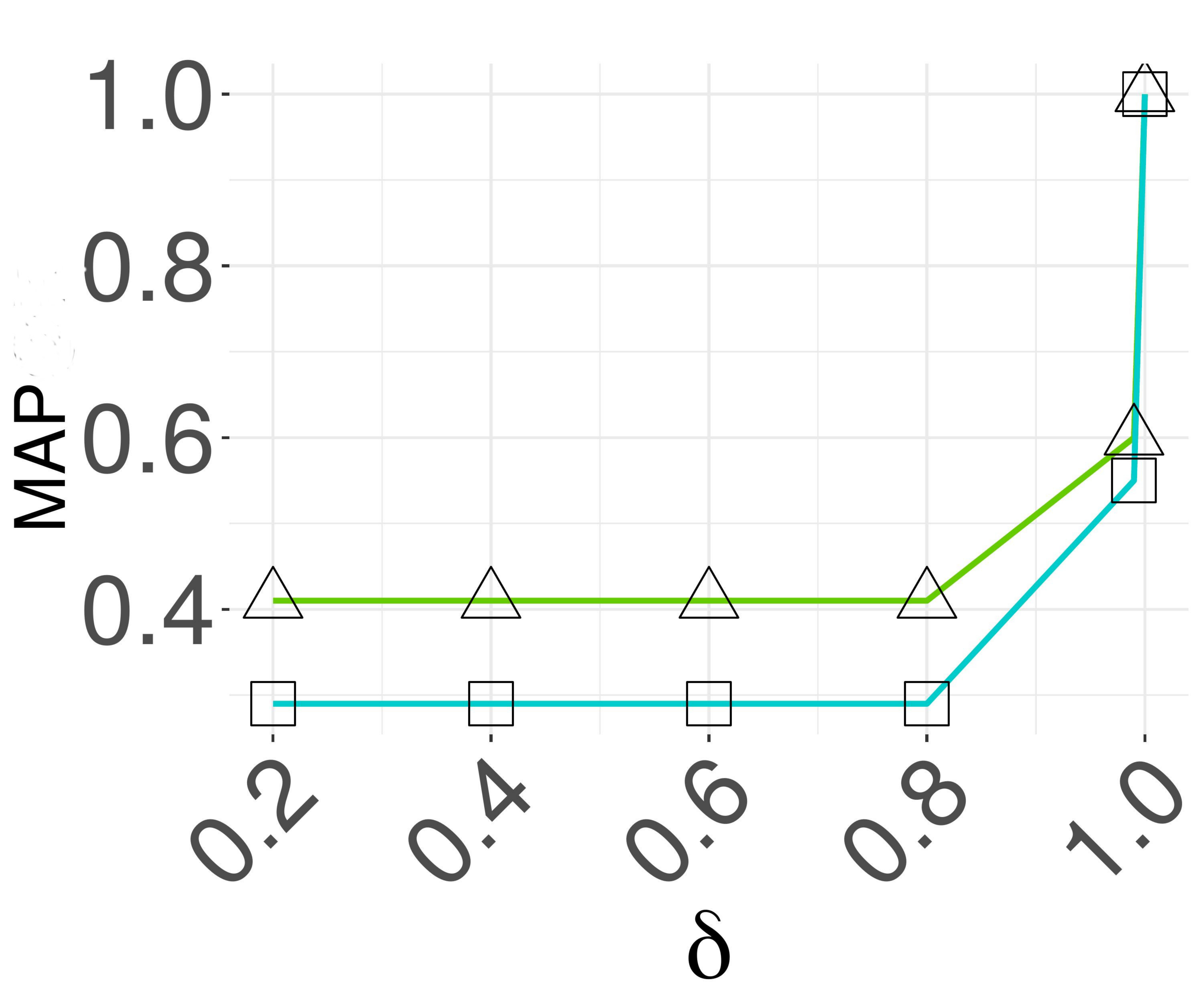}
		\caption{\color{black} MAP vs. $\bm{\delta}$ \\ ($\bm{\epsilon = 0}$)}
		\label{fig:approx:accuracy:delta:synthetic:250GB:hdd}
	\end{subfigure}
	\begin{subfigure}{0.30\columnwidth}
	\includegraphics[scale=0.15]{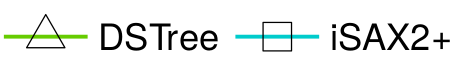}
	\end{subfigure}	
	\caption{Accuracy and efficiency vs. $\bm{\delta}$ and $\bm{\epsilon}$
	}	
	\vspace*{-0.4cm}
	\label{fig:approx:accuracy_efficiency:delta:epsilon:synthetic:250GB:hdd}
\end{figure}

}

{\color{black}
\section{Discussion}
\label{sec:discussion}

In the approximate NN search literature, experimental evaluations ignore the 
answering capabilities of data series methods. 
This is the first 
study that aims to fill 
this gap. 

\noindent{\bf Unexpected Results.}
Some of the results are surprising:
\\
(1) \emph{Effectiveness of $\delta$.} LSH techniques (like SRS and QALSH) exploit both $\delta$ and $\epsilon$ to tune the efficiency/accuracy tradeoff. We consider that they still fall short of expectations, because for a low $\epsilon$, high values of $\delta$ still produce low MAP and low values of $\delta$ still result in slow execution (Figure~\ref{fig:approx:accuracy:efficiency:synthetic:25GB:inmemory:hdd}). 
In the case of our extended methods, using $\epsilon$ yielded excellent empirical results, but introducing the probabilistic stop condition based on $\delta$ was ineffective (Figures~\ref{fig:approx:accuracy_efficiency:delta:epsilon:synthetic:250GB:hdd}-d,\ref{fig:approx:accuracy_efficiency:delta:epsilon:synthetic:250GB:hdd}-e). 
We believe that this is due to the inaccuracy of the (histogram-based) approximation of $r_{\delta}$. 
Therefore, 
improving the approximation of $r_{\delta}$, or devising novel approaches are interesting open research directions that will further improve the efficiency of these methods.  

(2) \emph{Approximate Query Answering with Data Series Indexes Performed Better than LSH.} 
Approximate query answering with DSTree and iSAX2+ outperfom SRS and QALSH (state-of-the-art LSH-based methods) both in space and time, while supporting better theoretical guarantees. 
This surprising result opens up exciting research opportunities, that is, devising efficient disk-based techniques that support both $ng$-approximate and $\delta$-$\epsilon$-approximate search with top performance~\cite{conf/vldb/echihabi19}. 
{\color{black} Note that data series indexes developed for distributed platforms~\cite{dpisax,conf/icde/zhang2019} also have the potential of outperforming LSH techniques~\cite{conf/cikm/bahmani2012,journal/pvdlb/sundaram2013} if extended following the ideas discussed in Section~\ref{sec:problem}.}

(3) \emph{Our results vs. the literature.}
Our results for the in-memory experiments are in-line with those reported in the literature, confirming that HNSW achieves the best accuracy/efficiency tradeoff when only query answering is considered~\cite{conf/sisap/martin17} (Figures~\ref{fig:approx:accuracy:qefficiency:synthetic:25GB:256:hdd:ng:100NN:100:nocache},~\ref{fig:approx:accuracy:qefficiency:sift:25GB:ng:hdd:100NN:100:nocache},~\ref{fig:approx:accuracy:qefficiency:deep:25GB:96:hdd:ng:100NN:100:nocache}). 
However, when indexing time is taken into account, HNSW loses its edge to iSAX2+/DSTree for both small  (Figures~\ref{fig:approx:accuracy:efficiency:synthetic:25GB:256:hdd:ng:100NN:100:nocache},~\ref{fig:approx:accuracy:efficiency:sift:25GB:ng:hdd:100NN:100:nocache},~\ref{fig:approx:accuracy:efficiency:deep:25GB:96:hdd:ng:100NN:100:nocache}) and large (Figures~\ref{fig:approx:accuracy:efficiency:synthetic:25GB:256:hdd:ng:100NN:10K:nocache},~\ref{fig:approx:accuracy:efficiency:sift:25GB:ng:hdd:100NN:10K:nocache},~\ref{fig:approx:accuracy:efficiency:deep:25GB:96:hdd:ng:100NN:10K:nocache}) query workloads.

Our results for IMI show a dramatic decrease in accuracy, in terms of MAP and Avg\_Recall for the Sift250GB and Deep250GB datasets, while high Avg\_Recall values have been reported in the literature for the full Sift1B and Deep1B datasets~\cite{conf/cvpr/yandex16,url/faiss}. 
We thoroughly investigated the reason behind such a discrepancy and ruled out the following factors: the Z-normalization of the Sift1B/Deep1B datasets, the size of the queries, and the number of NN. 
We believe that our results are different for the following reasons: 
(a) our queries return only the number of NN requested, while the smallest candidate list considered in~\cite{conf/cvpr/yandex16} is 10,000 for a 1-NN query; and (b) the results in~\cite{url/faiss} were obtained using training on a GPU with un-reported training sizes and times (we believe both were very large), while our focus was to evaluate methods on a CPU and account for training time. 
The difference in the accuracy results is most probably due to the fact that the training samples used in~\cite{url/faiss} were larger than the recommended numbers we used (1 million/4 million for the 25GB/250GB datasets, respectively). 
We tried to support this claim by running experiments with different training sizes: 
(i) we observed that increasing the training sizes for the smaller datasets improves the accuracy (the best results are reported in this study); 
(ii) we could not run experiments on the CPU with large training sizes for the 250GB datasets, because training was very slow: we stopped the execution after 48 hours; 
(iii) we tried a GPU-enabled server for training, but ran into a documented bug\footnote{ https://github.com/facebookresearch/faiss/issues/67}. 

\noindent{\bf Practicality of QALSH, IMI and HNSW.} 
{\color{black} Although QALSH provides better accuracy than SRS, it does so at a high cost: it needs to build a different index for each desired query accuracy. This is a serious drawback, while our extended methods offer a neat alternative since the index is built once and the desired accuracy is determined at query time.} Although LSH methods (such as SRS) provide guarantees on the accuracy of search results, they are expensive both in time and space. The $ng$-approximate methods overcome these limitations. IMI and HNSW are considered the state-of-the-art in this category, and while they deliver better speed-accuracy tradeoffs than QALSH and SRS, they suffer from two major limitations: (a) having no guarantees can lead them to return incomplete result sets, for instance retrieving only a subset of the neighbors for a k-NN query and returning null values for the others; (b) they are very difficult to tune, which hinders their practicality. 
In fact, the speed-accuracy tradeoff is not determined only at query time, but also during index building, which means that an index may need to be built many times using different parameters before finding the right speed-accuracy tradeoff. 
This means that the optimal settings may differ across datasets, and even for different dataset sizes of the same dataset. 
Moreover, if the analyst builds an index with a particular accuracy target, and then their needs change, they will have to rebuild the index from scratch and go through the same process of determining the right parameter values.

For example, we built the IMI index for the Deep250GB dataset 8 times. 
During each run that required over 42 hours, we varied the PQ encoding sizes, the number of centroids, and training sizes but still could not achieve the desired accuracy. 
Regarding HNSW, we tried three different combinations of parameters (M/efConstruction = 4/500, 16/500, 48/200) for each dataset before choosing the optimal one; each run took over 40 hours on the small  Deep25GB.
Overall, we observe that using IMI and HNSW in practice is cumbersome and time consuming. 
Developing auto-tuning methods for these techniques is both an interesting problem and a necessity.

\noindent{\bf Importance of guarantees}. 
In the approximate search literature, accuracy has been evaluated using recall, and approximation error. 
LSH techniques are considered the state-of-the-art in approximate search with theoretically proven sublinear time performance and probabilistic guarantees on accuracy (approximation error). 
Our results indicate that using the approximate search functionality of data series techniques provides tighter bounds than LSH (since $\delta$ can be equal to 1), and a much better performance in practice, with experimental accuracy levels well above the theoretical accuracy guarantees (Figure~\ref{fig:approx:accuracy_efficiency:delta:epsilon:synthetic:250GB:hdd}c). Note that LSH techniques can only provide probabilistic answers ($\delta < 1$), whereas our extended methods can also answer exact and $\epsilon$-approximate queries ($\delta = 1)$. 
A promising research direction is to improve the existing guarantees on these new methods, or establish additional ones: (1) by adding guarantees on query time performance; or (2) by developing probabilistic or deterministic guarantees on the recall or MAP value of a result set, instead of the commonly used distance approximation error. 
Remember that recall and MAP are better indicators of accuracy, because even small approximation errors may still result in low recall/MAP values (Figure~\ref{fig:approx:map:mre:sift:25GB:128:ng:hdd}). 

\noindent{\bf Improvement of ng-approximate methods.}  
Our results indicate that $ng$-approximate query answering with exact methods offers a viable alternative to existing methods, particularly because index building is much faster and query efficiency/accuracy tradeoffs can be determined at query time.
Besides, the performance of DSTree and iSAX2+ supporting {\color{black}ng-approximate and} $\delta$-$\epsilon$-approximate search can be greatly improved by exploiting modern hardware (including SIMD vectorization, multi-cores, multi-sockets, and GPUs).

\noindent{\bf Incremental approximate k-NN}. We established that, on some datasets, a kNN query incurs a much higher time cost as k increases. Therefore, a future research direction is to build $\delta$-$\epsilon$-approximate methods that support incremental search, i.e., returning the neighbors one by one as they are found. The current approaches return the k nearest neighbors all at once which impedes their interactivity. 

\noindent{\bf Progressive Query Answering.} The excellent empirical results with approximate search using exact methods paves the way for another very promising research direction: progressive query answering~\cite{DBLP:conf/edbt/GogolouTPB19}. 
New approaches can be devised to return intermediate results 
with increasing accuracy until the exact answers are found.

\noindent{\bf Recommendations.}
Choosing the best approach to answer an approximate similarity search query depends on a variety of factors including the accuracy desired, the dataset characteristics, the size of the query workload, the presence of an existing index and the hardware. 
Figure~\ref{fig:recommendations} illustrates a decision matrix that recommends the best technique to use for answering a query workload using an existing index. 
Overall, DSTree is the best performer, with the exceptions of ng-approximate queries, where iSAX2+ also exhibits excellent performance, and of in-memory datasets, where HNSW is the overall winner. 
Accounting for index construction time as well, DSTree becomes the method of choice across the board, except for small workloads, where iSAX2+ wins.

\begin{figure}[tb]
	\captionsetup{justification=centering}
	\includegraphics[width =\columnwidth]{{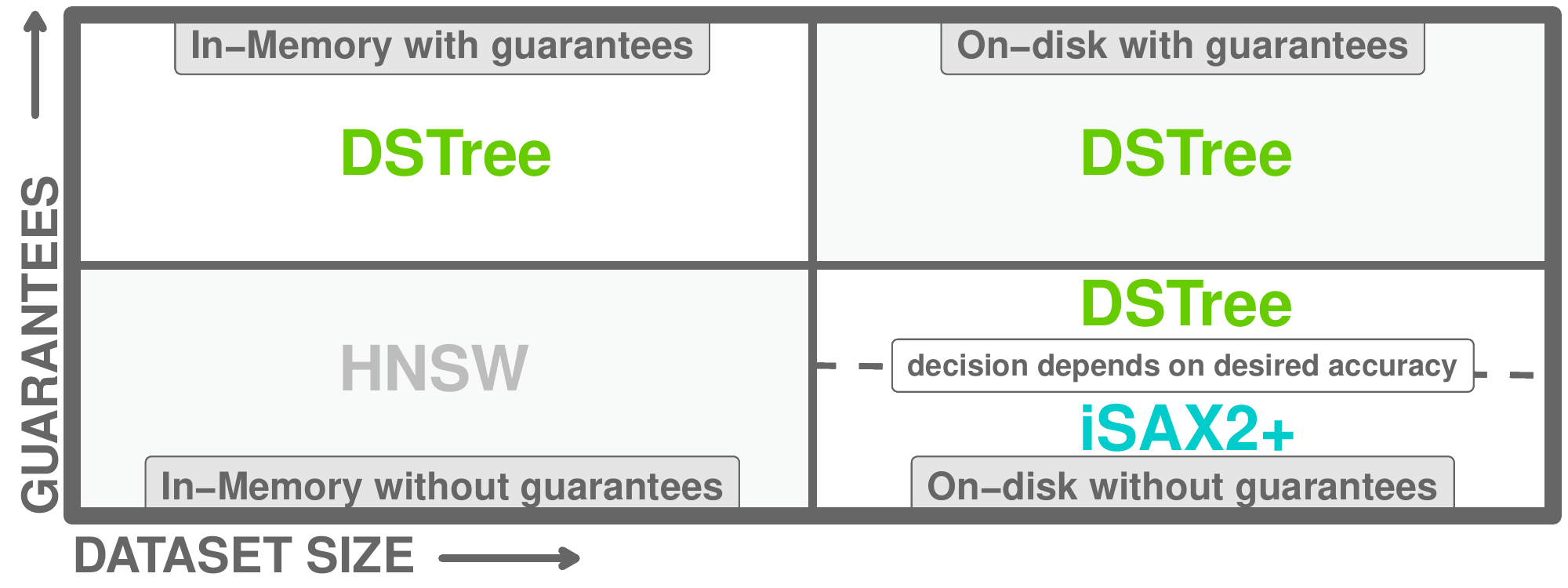}}
	\vspace*{-0.5cm}
	\caption{Recommendations (query answering).		
	\vspace*{-0.3cm}
		}
	\label{fig:recommendations}
\end{figure}

}

{\color {black}
\section{Conclusions}
\label{sec:conclusions}

We presented a taxonomy for data series similarity search techniques, proposed extensions of exact data series methods that can answer $\delta$-$\epsilon$-approximate queries, and conducted a thorough experimental evaluation of the state-of-the-art techniques from both the data series and vector indexing communities. 
The results reveal the pros and cons of the various methods, demonstrate the benefits and potential of approximate data series methods, and point to unexplored research directions in the approximate similarity search field. 

\vspace*{0.3cm}
\noindent{\bf Acknowledgments.} Work partially supported by program Investir l'Avenir and Univ. of Paris IDEX Emergence en Recherche ANR-18-IDEX-0001, EU project NESTOR (MSCA {\#}748945), and FMJH Program PGMO in conjunction with EDF-THALES.		

	\balance
	
	\bibliographystyle{abbrv}
	\def\thebibliography#1{
		\section*{References}
		\normalsize                  
		\list
		{[\arabic{enumi}]}
		{\settowidth\labelwidth{[#1]}
			\leftmargin\labelwidth
			\parsep 1pt                
			\itemsep 0.6pt               
			\advance\leftmargin\labelsep
			\usecounter{enumi}
		}
		\def\newblock{\hskip .11em plus .33em minus .07em}
		\sloppy\clubpenalty10000\widowpenalty10000
		\sfcode`\.=1000\relax
	}
	\bibliography{ref}  
	
\end{document}